\documentclass[aps,pra,reprint,superscriptaddress,nofootinbib]{revtex4-1}
\usepackage{afterpage}
\usepackage{amsmath}
\usepackage{amssymb}
\usepackage{amsfonts}
\usepackage{mathrsfs}
\usepackage{graphicx}
\usepackage[caption=false]{subfig}
\usepackage{enumitem}
\usepackage[normalem]{ulem}
\usepackage[percent]{overpic}
\usepackage{bm}
\usepackage{rotating}
\usepackage[usenames, dvipsnames]{color}
\usepackage{hyperref}
\usepackage{cancel}
\usepackage{verbatim}
\usepackage{mathtools}
\usepackage{tensor}
\usepackage{verbatim}
\usepackage{tikz}
\usepackage{booktabs}
\graphicspath{ {images/} }

\usepackage{braket}

\numberwithin{equation}{section}

\newcommand{\ii}{\mathrm{i}}

\renewcommand{\d}{\mathrm{d}}

\newcommand{\be}{\begin{equation}}
\newcommand{\bel}[1]{\begin{equation}\label{#1}}
\newcommand{\ee}{\end{equation}}

\begin{document}
\title{A Discrete Analog of General Covariance - Part 2:\\
Despite what you've heard, a perfectly Lorentzian lattice theory}
\author{Daniel Grimmer}
\email{daniel.grimmer@philosophy.ox.ac.uk}
\affiliation{Reuben College, University of Oxford, Oxford, OX2 6HW United Kingdom}
\affiliation{Faculty of Philosophy, University of Oxford, Oxford, OX2 6GG United Kingdom}
\affiliation{Barrio RQI, Waterloo, Ontario N2L 3G1, Canada}

\begin{abstract}
A crucial step in the history of General Relativity was Einstein's adoption of the principle of general covariance which demands a coordinate independent formulation for our spacetime theories. General covariance helps us to disentangle a theory's substantive content from its merely representational artifacts. It is an indispensable tool for a modern understanding of spacetime theories, especially regarding their background structures and symmetry. Motivated by quantum gravity, one may wish to extend these notions to quantum spacetime theories (whatever those are). Relatedly, one might want to extend these notions to discrete spacetime theories (i.e., lattice theories). In \cite{DiscreteGenCovPart1} I developed two discrete analogs of general covariance for non-Lorentzian lattice theories. This paper extends these results to a Lorentzian setting. 

In either setting these discrete analogs of general covariance reveal that lattice structure is rather less like a fixed background structure and rather more like a coordinate system, i.e., merely a representational artifact. These discrete analogs are built upon a rich analogy between the lattice structures appearing in our discrete spacetime theories and the coordinate systems appearing in our continuum spacetime theories. From this, in \cite{DiscreteGenCovPart1}, I argued that properly understood there are no such things as lattice-fundamental theories, rather there are only lattice-representable theories. It is well-noted by the causal set theory community that no theory on a fixed spacetime lattice is Lorentz invariant, however as I will discuss this is ultimately a problem of representation, not of physics. There is no need for the symmetries of our representational tools to latch onto the symmetries of the thing being represented. Nothing prevents us from using Cartesian coordinates to describe rotationally invariant states/dynamics. As this paper shows, the same is true of lattices in a Lorentzian setting: nothing prevents us from defining a perfectly Lorentzian lattice(-representable) theory.
\end{abstract}

\maketitle

\section{Introduction}\label{SecIntro}
A crucial step in the history of General Relativity (GR) was Einstein's adoption of the principle of general covariance which states that the form of our physical laws should be independent of any choice of coordinate systems. The conceptual benefits of writing a theory in a coordinate-free way are immense. A generally covariant formulation of a theory has at least two major benefits: 1) it more clearly exposes the theory's geometric background structure, and 2) it thereby helps clarify our understanding of the theory's symmetries (i.e., its structure/solution preserving transformations). It does both of these by disentangling the theory's substantive content from representational artifacts which arise in particular coordinate representations~\cite{Pooley2015,Norton1993,EarmanJohn1989Weas}. Thus, general covariance is an indispensable tool for a modern understanding of spacetime theories. 

Motivated by quantum gravity, one may wish to extend these notions to quantum spacetime theories (whatever those are). Relatedly, one might want to extend these notions to discrete spacetime theories (i.e., lattice theories\footnote{Given the results of this paper and of \cite{DiscreteGenCovPart1}, calling these ``lattice theories'' can be misleading. This would be analogous to referring to continuum spacetime theories as ``coordinate theories''. As I will discuss, in both cases the coordinate systems/lattice structure are merely representational artifacts and so do not deserve ``first billing'' so to speak. All lattice theories are best thought of as lattice-representable theories. Similarly, the term ``discrete spacetime theories'' ought to be here read as ``discretely-representable spacetime theories''. As discussed here (and in \cite{DiscreteGenCovPart1}), the defining feature of such theories is that they have a finite density of degrees of freedom, see the work of Achim Kempf~\cite{UnsharpKempf,Kempf2003,Kempf2004a,Kempf2004b,Kempf2006}}.). In \cite{DiscreteGenCovPart1} I developed two analogs of general covariance for such discrete spacetime theories in a non-Lorentzian setting. The aim of this paper is to extend these results to a Lorentzian setting. Indeed, the analysis provided here is nearly identical to the one carried out in \cite{DiscreteGenCovPart1}, although each paper is self-contained.

In either setting these discrete analogs of general covariance reveal that lattice structure is rather less like a fixed background structure or a fundamental part of some underlying manifold and rather more like a coordinate system, i.e., merely a representational artifact. Indeed, these discrete analogs are built upon a rich analogy between the lattice structures appearing in our discrete spacetime theories and the coordinate systems appearing in our continuum spacetime theories. 

This paper is largely inspired by the brilliant work of mathematical physicist Achim Kempf~\cite{Kempf_1997,UnsharpKempf,Kempf2000b,Kempf2003,Kempf2004a,Kempf2004b,Kempf2006,Martin2008,Kempf_2010,Kempf2013,Pye2015,Kempf2018} among others~\cite{PyeThesis,Pye2022,BEH_2020}. A key feature present both here and in Kempf's work is the sampling property of bandlimited function revealed by the Nyquist-Shannon sampling theory~\cite{GARCIA200263,SamplingTutorial,UnserM2000SyaS}. I review sampling theory in more detail in Sec.~\ref{SecSamplingTheory}, but let me overview here. Bandlimited functions are those with have a limited extent in Fourier space (i.e., compact support). Bandlimited functions have the following sampling property: they can be exactly reconstructed knowing only the values that they take on any sufficiently dense sample lattice. What ``sufficiently dense'' means is fixed in terms of the size of the function's support in Fourier space.

Nyquist-Shannon sampling theory was first discovered in the context of information processing as a way of converting between analog and digital signals (i.e., between continuous and discrete information). Sampling theory found its first application in fundamental spacetime physics with Kempf's \cite{Kempf_1997,UnsharpKempf}, ultimately leading to his thesis that ``Spacetime could be simultaneously continuous and discrete, in the same way that information can be'' \cite{Kempf_2010}. Kempf's thoughts on these topics is the primary inspiration for this paper and deserves wider appreciation by the philosophy of physics community. For an overview of Kempf's works on this topic see~\cite{Kempf2018}.

My thesis in \cite{DiscreteGenCovPart1} is in broad agreement with Kempf's with one crucial alteration. I stress that the sampling property of bandlimited functions indicates that bandlimited physics can be simultaneously \textit{represented as} continuous and discrete, (i.e., on a continuous or discrete spacetime). However, I further argue (both here and in \cite{DiscreteGenCovPart1}) that when one investigates these two representations one finds substantial issue with taking the discrete representation as fundamental. These issues stem from the rich analogy between the lattice structures and coordinate systems mentioned above.

This analogy is supported here (and in \cite{DiscreteGenCovPart1}) by the three lessons each of which tell against an intuitions one is likely to have regarding lattice structure. To motivate these (wrong) intuitions, consider the following situation.

\begin{figure}[t!]
\includegraphics[width=0.4\textwidth]{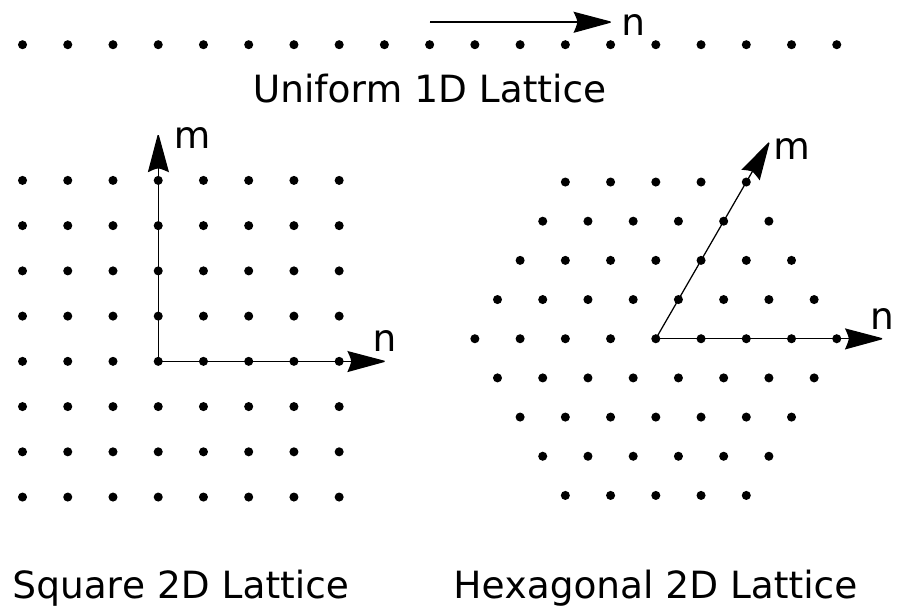}
\caption{Three of the lattice structures in space considered throughout this paper. The arrows indicate the indexing conventions for the lattice sites. Repeated in time, these give us something like a square 2D lattice, a cubic 3D lattice, and a hexagonal 3D lattice respectively. [\textit{Reproduced with permission from} \cite{DiscreteGenCovPart1}.]}\label{FigLat}
\end{figure}

Suppose that after substantial empirical investigation of our micro-physical reality we find what appear to be ``lattice artifacts''. For instance, we may find ourselves restricted to only quarter rotation, or one-sixth rotation symmetries. Intuitively, this would suggest that the world is fundamentally set on a lattice of the kind shown in Fig.~\ref{FigLat}, i.e., a square or hexagonal lattice.

Suppose that we have great predictive success when modeling the world as being set on (for instance) a square lattice with next-to-nearest neighbor interactions. Would this in any way prove that the world is fundamentally set on such a lattice? No, all this would prove is that the world can be \textit{faithfully represented} on such a lattice with such interactions, at least empirically. Anything can be faithfully represented in any number of ways, this is just mathematics. Some extra-empirical work must be done to know if we should take the lattice structure appearing in this representation seriously. That is, we must ask which parts of the theory are substantive and which parts are merely representational? The discrete analogs of general covariance developed here (and in \cite{DiscreteGenCovPart1}) answer this question: lattice structures are coordinate-like representational artifacts and so ultimately have no physical content.

To flesh out the contrary received position however, let us proceed without this analogy for the moment. We can ask: beyond merely appearing in our hypothetical empirically successful theory, what reason do we have to take the lattice structures which appear in this theory seriously? Well, intuitively the lattice structures appear to play a very substantial role in these theories, not merely a representational one. One likely has the following three interconnected first intuitions regarding the role that the lattice and lattice structure play in discrete spacetime theories:
\begin{itemize}
\item[1.] They restrict our possible symmetries. Taking the lattice structure to be a part of the theory's fixed background structure, our possible symmetries are limited to those which preserve this fixed structure. Intuitively a theory set on a square lattice can only have the symmetries of that lattice. Similarly for a hexagonal lattice, or even an unstructured lattice.
\item[2.] Differing lattice structures distinguishes our theories. Two theories with different lattice structures (e.g., square, hexagonal, irregular, etc.) cannot be identical. As suggested above they have different fixed background structures and as therefore have different symmetries.
\item[3.] The lattice is fundamentally ``baked-into'' the theory. Firstly, it is what the fundamental fields are defined over: they map lattice sites (and in \cite{DiscreteGenCovPart1} times) into some value space. Secondly, the bare lattice is what the lattice structure structures. Thirdly, it is what limits us to discrete permutation symmetries in advance of further limitations from the lattice structure.
\end{itemize}
These intuitions will be fleshed out and made more concrete in Sec.~\ref{SecKlein1}. However, as this paper demonstrates, each of the above intuitions are doubly wrong and overhasty.

What goes wrong with the above intuitions is that we attempted to directly transplant our notions of background structure and symmetry from continuous to discrete spacetime theories. This is an incautious way to proceed and is apt to lead us astray. Recall that, as discussed above, our notions of background structure and symmetry are best understood in light of general covariance. It is only once we understand what is substantial and what is merely representational in our theories that we have any hope of properly understanding them. Therefore, we ought to instead first transplant a notion of general covariance into our discrete spacetime theories and then see what conclusions we are led to regarding the role that the lattice and lattice structure play in our discrete spacetime theories. This transplant has been done in a non-Lorentzian setting in \cite{DiscreteGenCovPart1}. Here I extend these results to a Lorentzian setting.

This paper will teach us three lessons each of which negates one of the above intuitions about the role that lattice structure plays in discrete spacetime theories.

Firstly, as I will show, taking any lattice structure seriously as a fixed background structure systematically under predicts the symmetries that discrete theories can and do have. Indeed, as I will show neither the bare lattice itself nor its lattice structure in any way restrict a theory's possible symmetries. In \cite{DiscreteGenCovPart1}, for non-Lorentzian theories I have shown that there is no conceptual barrier to having a theory with continuous translation and rotation symmetries formulated on a discrete lattice. Indeed, in \cite{DiscreteGenCovPart1} I presented a perfectly rotation invariant lattice theory. As I discuss in \cite{DiscreteGenCovPart1}, this is analogous to the familiar fact that there is no conceptual barrier to having a continuum theory with rotational symmetry formulated on a Cartesian coordinate system. Here, I repeat this analysis in a Lorentzian context. In Sec.~\ref{PerfectLorentz}, I present a perfectly Lorentzian lattice theory. 

Secondly, as I will show, discrete theories which are initially presented to us with very different lattice structures (i.e., square vs. hexagonal) may nonetheless turn out to be completely equivalent theories or to be overlapping parts of some larger theory. Moreover, given any discrete theory with some lattice structure we can always re-describe it using a different lattice structure. As I will discuss, this is analogous to the familiar fact that our continuum theories can be described in different coordinates, and moreover we can switch between these coordinate systems freely.

Thirdly, as I will show, in addition to being able to switch between lattice structures, we can also reformulate any discrete theory in such a way that it has no lattice structure whatsoever. Indeed, we can always do away with the lattice altogether. As I will discuss, this is analogous to the familiar fact that any continuum theory can be written in a generally covariant (i.e., coordinate-free) way.

These three lessons combine to give us a rich analogy between lattice structures and coordinate systems. It is from this rich analogy that the central claims of this paper follow. Namely, from this analogy it follows that the lattice structure supposedly underlying any discrete ``lattice'' theory has the same level of physical import as coordinates do, i.e., none at all. Thus, as I argued in \cite{DiscreteGenCovPart1}, the world cannot be ``fundamentally set on a square lattice'' (or any other lattice) any more than it could be ``fundamentally set in a certain coordinate system''. Like coordinate systems, lattice structures are just not the sort of thing that can be fundamental; they are both thoroughly merely representational. Spacetime cannot be a lattice (even when it might be representable as such). Specifically, I claimed that properly understood, there are no such things as lattice-fundamental theories, rather there are only lattice-representable theories. This paper extends these conclusions to a Lorentzian context.

Once one begins thinking of lattices as a merely representational structure, a path opens for perfectly Lorentzian lattice theories. As proponents of causal set theory correctly point out, no single fixed spacetime lattice is Poincar\'e invariant. This (apparently) spells big trouble for any lattice-based Lorentzian theories. They, however, avoid this issue by considering instead a random Poisson sprinkling of lattice points which does not pick out any preferred direction and hence does not explicitly break Poincar\'e symmetry, at least on average. However, given the deflationary position this paper takes towards lattices, I claim there is no issue to be avoided. Like coordinate systems, lattice structures are just a representational tool for helping us express our theory. There is no need for the symmetries of our representational tools to latch onto the symmetries of the thing being represented. Cartesian coordinates are distorted under Lorentz boosts, but we can still use them to describe our Lorentzian theories without issue. The same is true of lattices. Indeed, in Sec.~\ref{PerfectLorentz} I will present a perfectly Lorentzian lattice theory.

\subsection{Outline of the Paper}
In Sec.~\ref{SecSevenKG}, I will introduce seven discrete Klein Gordon equations in an interpretation-neutral way and solve their dynamics. Then, in Sec.~\ref{SecKlein1}, I will make a first attempt at interpreting these theories. I will (ultimately wrongly) identify their underlying manifold, locality properties, and symmetries. Among other issues, a central problem with this first attempt is that it takes the lattice itself to be the underlying spacetime manifold and thereby unequivocally cannot support continuous translation and rotation symmetries. This systematically under predicts the symmetries that these theories can and do have.

In Sec.~\ref{SecKlein2}, I will provide a second attempt at interpreting these theories which fixes this issue (albeit in a slightly unsatisfying way). In particular, in this second attempt I deny that the lattice is the underlying spacetime manifold. Instead, I ``internalize'' it into the theory's value space. Fruitfully, this second interpretation does allow for continuous translation and rotation symmetries and even a (limited) Lorentz boost symmetry. However, the key move here of ``internalization'' has several unsatisfying consequences. For instance, the continuous symmetries we find here are all classified as internal (i.e, associated with the value space) whereas intuitively they ought to be external (i.e, associated with the manifold).

We thus will need a third attempt at interpreting these theories which externalizes these symmetries. Sec.~\ref{SecExtPart1} - Sec.~\ref{SecExtPart2} lay the groundwork for this third interpretation. In particular, they describe a principled way of 1) inventing a continuous spacetime manifold for our formerly discrete theories to live on and 2) embedding our theory's states/dynamics onto this manifold as a new dynamical field. In the middle of this, in Sec.~\ref{SecSamplingTheory}, I will provide an informal overview of the primary mathematical tools used in the latter half of this paper. Namely, I will review the basics of Nyquist-Shannon sampling theory and bandlimited functions. 

With this groundwork complete, in Sec.~\ref{SecKlein3} and Sec.~\ref{SecKlein3Extra} I will provide a third attempt at interpreting these seven theories which fixes all issues arising in the previous two interpretations. For instance, like in my second attempt, this third interpretation can support continuous translation and rotation symmetries as well as a (limited) Lorentz boost symmetry. However, unlike the second attempt it realizes them as external symmetries (i.e., associated with the underlying manifold, not the theory's value space).

In Sec.~\ref{SecDisGenCov}, I will review the lessons learned in these three attempts at interpretation. As I will discuss, the lessons learned combine to give us a rich analogy between lattice structures and coordinate systems. As I will discuss, there are actually two ways of fleshing out this analogy: one internal and one external. This section spells out these analogies in detail, each of which gives us a discrete analog of general covariance. I find reason to prefer the external notion, but either is likely to be fruitful for further investigation/use. Sec.~\ref{PerfectLorentz} provides us with a perfectly Lorentzian lattice theory as promised.

Finally, in Sec. \ref{SecConclusion} I will summarize the results of this paper, discuss its implications, and provide an outlook of future work.

For comments on what this means for the the dynamical vs geometrical spacetime debate~\cite{EarmanJohn1989Weas,TwiceOver,BelotGordon2000GaM,Menon2019,BrownPooley1999,Nonentity,HuggettNick2006TRAo,StevensSyman2014Tdat,DoratoMauro2007RTbS,Norton2008,Pooley2013} see \cite{DiscreteGenCovPart1}. Here I will focus on the implications this work has for quantum gravity especially causal set theory~\cite{Surya2019}.

\section{Seven Discrete Klein Gordon Equations}\label{SecSevenKG}
In this section I will introduce seven discrete Klein Gordon equations (KG1-KG7) in an interpretation-neutral way and solve their dynamics. These theories are all describable as being set on a lattice in both space and time. In each of these theories the lattice in space will simply repeat itself in time. I consider the following three cases for the lattice in space: a uniform 1D lattice, a square 2D lattice, and a hexagonal 2D lattice, see Fig.~\ref{FigLat}. Repeated in time, these give us something like a square 2D lattice, a cubic 3D lattice, and a hexagonal 3D lattice respectively.

As harmful as these choices seem to be to Lorentz invariance (as well as continuous translation and rotation invariance) as I will show they ultimately pose no barrier to our theories having these symmetries. As I will argue, these choices of lattice are ultimately merely choices of representation which have absolutely nothing to do with the thing being represented. In particular, there is no need for our representational structure to have the same symmetries as the thing being represented. There is no issue with using Cartesian coordinates to describe a rotationally invariant state/dynamics. I claim that analogously there is no issue with using a lattice to describe a state/dynamics with continuous translation and rotation invariance and even Lorentz invariance. This claim has already been demonstrated in \cite{DiscreteGenCovPart1} for states/dynamics with continuous translation and rotation invariance. This paper extends this claim about lattice structures to Lorentz invariance as well.

\subsection{Introducing KG1-KG7}
To begin, let us consider the continuum Klein Gordon equations in $1+1$ and $2+1$ dimensions:
\begin{align}\label{KG00}
&\text{\bf Continuum Klein Gordon Eq. (KG00):}\\
\nonumber
&\partial_{t}^2\varphi(t,x) = (\partial_x^2-M^2) \, \varphi(t,x)\\
\label{KG0}
&\text{\bf Continuum Klein Gordon Eq. (KG0):}\\
\nonumber
&\partial_{t}^2\varphi(t,x,y) = (\partial_x^2+\partial_y^2-M^2) \, \varphi(t,x,y)
\end{align}
with some mass $M\geq0$. For a generally covariant (i.e., coordinate-free) view of these theories, see Sec.~\ref{SecFullGenCov}. 

For our first three discrete Klein Gordon theories, let us consider the theories with only nearest-neighbor (N.N.) interactions on the above discussed lattices which best approximate KG0 and KG00. Namely:
\begin{align}\label{KG1Long}
&\text{\bf 1D N.N. Klein Gordon Eq. (KG1):}\\
\nonumber
&[\phi_{j-1,n}-2\phi_{j,n}+\phi_{j+1,n}]\\
\nonumber
&\quad=[\phi_{j,n-1}-2\phi_{j,n}+\phi_{j,n+1}]
-\mu^2 \phi_{j,n}\\
&\text{\bf Square N.N. Klein Gordon Eq. (KG4):}\label{KG4Long}\\
\nonumber
&[\phi_{j-1,n,m}-2\phi_{j,n,m}+\phi_{j+1,n,m}]\\
\nonumber
&\quad = [ \ \phi_{j,n-1,m}-2\phi_{j,n,m}+\phi_{j,n+1,m}\\
\nonumber
&\quad \ \ +\phi_{j,n,m-1}-2\phi_{j,n,m}+\phi_{j,n,m+1}]
-\mu^2 \phi_{j,n,m}\\
&\text{\bf Hexagonal N.N. Klein Gordon Eq. (KG5):}\label{KG5Long}\\
\nonumber
&[\phi_{j-1,n,m}-2\phi_{j,n,m}+\phi_{j+1,n,m}]\\
\nonumber
&\quad = \frac{2}{3}[ \ \phi_{j,n-1,m}-2\phi_{j,n,m}+\phi_{j,n+1,m}\\
\nonumber
&\qquad \quad +\phi_{j,n,m-1}\!-\!2\phi_{j,n,m}\!+\!\phi_{j,n,m+1}\\
\nonumber
&\qquad \quad +\phi_{j,n+1,m-1}-2\phi_{j,n,m}+\phi_{j,n-1,m+1}]
-\mu^2 \phi_{j,n,m}
\end{align}
with $j\in\mathbb{Z}$ indexing time and $n\in\mathbb{Z}$ and $m\in\mathbb{Z}$ indexing space. See Fig.~\ref{FigLat} for the indexing convention. Here $\mu\in\mathbb{R}$ is a dimensionless number playing the role of the field's mass. The terms in square brackets in the above expressions are the best possible approximations of the second derivative on each lattice which make use of only nearest neighbor interactions. These theories are named KG1, KG4, and KG5 in correspondence with the discrete heat equations considered in \cite{DiscreteGenCovPart1} and in anticipation of their further treatment later in this section. 

This section has promised to introduce these theories in an interpretation neutral way. As such, some of the above discussion needs to be hedged. In particular, in introducing these theories I have made casual comparison between parts of these theories' dynamical equations and various approximations of the second derivative. While, as I will discuss, such comparisons can be made, to do so immediately is unearned. It comes dangerously close to imagining the spacetime lattices discussed above as being embedded in a continuous manifold. This may be something we want to do later (see Sec.~\ref{SecExtPart1}), but it is a non-trivial interpretational move which ought not be done so casually. 

Crucially, in this paper I will begin by analyzing these theories as discrete-native theories. As such, it's important to think of these discrete spacetime theories as self-sufficient theories in their own right. We must not begin by thinking of them as various discretizations or bandlimitations of the continuum theories. While, as I will discuss, these discrete theories have some notable relationships to various continuum theories it is important to resist any temptation to see these continuum theories as ``where they came from''. Rather, let us pretend these theories ``came from nowhere'' and let us see what sense we can make of them.

Another bit of hedging: in introducing the above three theories I casually associated them with the lattice structures shown in Fig.~\ref{FigLat} (each repeated in time). Making such associations ab initio is unwarranted. While we may eventually associate these theories with those lattice structures we cannot do so immediately. Such an association would need to be made following careful consideration of the dynamics. (Such an exercise is carried out in Sec.~\ref{SecKlein1}.) Beginning here in an interpretation-neutral way these theories ought to be seen as being defined over a completely unstructured lattice. 

I will reflect this concern in my notation as follows. The labels for the lattice sites are presently too structured (e.g., $(j,n)\in\mathbb{Z}^2$ and $(j,n,m)\in\mathbb{Z}^3$). Instead we ought to think of the lattice sites as having labels $\ell\in L$ for some set $L$. Crucially, at this point the set of labels for the lattice sites, $L$, is just that, an unstructured set.

Up to isomorphism (here, generic bijections, i.e. generic relabelings), sets are uniquely specified by their cardinality. The set of labels for the lattice sites is here countable, $\ell\in L\cong\mathbb{Z}\cong\mathbb{Z}^2\cong\mathbb{Z}^3$. Reframed this way the above discussed theories each consider the same discrete variables $\phi_\ell\in\mathbb{R}$. In particular, KG1 considers variables $\phi_\ell$ which under some convenient relabeling of the lattice sites, $\ell\in L\mapsto (j,n)\in\mathbb{Z}^2$, satisfies Eq.~\eqref{KG1Long}. Similarly, KG4 and KG5 consider variables $\phi_\ell$ which under some convenient relabeling of the lattice sites, $\ell\in L\mapsto (j,n,m)\in\mathbb{Z}^3$, satisfy Eq.~\eqref{KG4Long} and Eq.~\eqref{KG5Long} respectively.

It's important to stress that the mere existence of these convenient relabelings by itself has no interpretative force. The fact that our labels $(j,n)\in\mathbb{Z}^2$ and $(j,n,m)\in\mathbb{Z}^3$ in some sense form a square 2D lattice and cubic 3D lattice in no way forces us to think of $L$ as being structured in this way (indeed, we might later like to think of $L$ as a hexagonal 3D lattice). In particular, the fact that these labels are in a sense equidistant from each other does not force us to think of the lattice sites as being equidistant from each other. Nor are we forced to think that ``the distance between lattice sites'' to be meaningful at all. Dynamical considerations may later push us in this direction, but the mere convenience of this labeling should not.

I have above introduced three out of seven discrete Klein Gordon theories. In order to introduce the other four theories, it is convenient (but not necessary) to first reformulate things. In particular, let us reorganize the $\phi_\ell$ variables into a vector, namely,
\begin{align}\label{PhiDef}
\bm{\Phi}= \sum_{\ell\in L} \phi_\ell \, \bm{b}_\ell. 
\end{align}
where $\bm{b}_\ell$ is a linearly-independent basis vector for each $\ell\in L$ and \mbox{$\bm{\Phi}$} is a vector in the vector space \mbox{$\mathbb{R}^L\coloneqq\text{span}(\{\bm{b}_\ell\}_{\ell\in L})$}. For later reference, it should be noted that $\phi_\ell$ is also a vector in a vector space: namely, $F_L$ the space of functions $f:L\to\mathbb{R}$. Note that Eq.~\eqref{PhiDef} is an vector space isomorphism between these vector spaces, $\mathbb{R}^L\cong F_L$. Everything which follows concerning $\bm{\Phi}\in\mathbb{R}^L$ has an isomorphic description in terms of $\phi_\ell\in F_L$.

Recall that for KG1 the lattice sites $\ell\in L$ have a convenient relabeling in terms of two integer indices, \mbox{$\ell\in L\mapsto (j,n)\in\mathbb{Z}^2$}. We can use this relabeling to grant the vector space a tensor product structure as \mbox{$\mathbb{R}^L\mapsto\mathbb{R}^\mathbb{Z}\otimes\mathbb{R}^\mathbb{Z}$} by taking \mbox{$\bm{b}_\ell\mapsto \bm{e}_j\otimes\bm{e}_n$} where 
\begin{align}
\bm{e}_m = (\dots,0,0,1,0,0,\dots)^\intercal\in\mathbb{R}^\mathbb{Z}
\end{align}
with the 1 in the $m^\text{th}$ position. Under this restructuring of KG1 we have, 
\begin{align}\label{PhiVec1}
\bm{\Phi}= \sum_{j,n\in\mathbb{Z}} \phi_{j,n} \, \bm{e}_j\otimes\bm{e}_n. 
\end{align}
In these terms the dynamics of KG1 is given by,
\begin{align}
\label{DKG1}
&\text{\bf Klein Gordon Equation 1 (KG1):}\\
\nonumber
&\Delta_{(1),\text{j}}^2\,\bm{\Phi}=\Delta_{(1),\text{n}}^2 \, \bm{\Phi}
-\mu^2\bm{\Phi}
\end{align}
where the notation $A_\text{j}\coloneqq A\otimes\openone$ and $A_\text{n}\coloneqq\openone\otimes A$ mean $A$ acts only on the first or second tensor space respectively. The linear operator $\Delta_{(1)}^2$ appearing twice in the above expression is the following bi-infinite Toeplitz matrix:
\begin{align}\label{Delta12}
\Delta_{(1)}^2=\{\Delta^+,\Delta^-\}
&=\text{Toeplitz}(1,\,-2,\,1)\\
\nonumber
\Delta^+&=\text{Toeplitz}(0,-1,\,1)\\
\nonumber
\Delta^-&=\text{Toeplitz}(-1,\,1,\,0)
\end{align}
where the curly brackets indicate the anticommutator, $\{A,B\}= \frac{1}{2}(A\,B + B\,A)$. Recall that Toeplitz matrices are so called diagonal-constant matrices with \mbox{$[A]_{i,j}=[A]_{i+1,j+1}$}. Thus, the values in the above expression give the matrix's values on either side of the main diagonal.

Although above I warned about thinking in terms of derivative approximations prematurely, a few comments are here warranted. Note that $\Delta^+$ is associated with the forward derivative approximation, $\Delta^-$ is be associated with the backwards derivative approximation, and $\Delta_{(1)}^2$ is associated with the nearest neighbor second derivative approximation,
\begin{align}
\nonumber
\Delta_{(1)}^2/\epsilon^2:\ \partial_x^2 f(x)
&\approx\frac{f(x+\epsilon)-2 f(x)+f(x-\epsilon)}{\epsilon^2}.
\end{align}
As stressed above, we ought to be cautious not to lean too heavily on these relationships when interpreting these discrete theories. 

In addition to KG1, I will also consider two more theories with ``improved derivative approximations''. Namely,
\begin{align}
\label{DKG2}
&\text{\bf Klein Gordon Equation 2 (KG2):}\\
\nonumber
&\Delta_{(2),\text{j}}^2\,\bm{\Phi}=\Delta_{(2),\text{n}}^2 \ \bm{\Phi}-\mu^2 \, \bm{\Phi}\\
\label{DKG3}
&\text{\bf Klein Gordon Equation 3 (KG3):}\\
\nonumber
&D_{\text{j}}^2\,\bm{\Phi}=D_{\text{n}}^2 \ \bm{\Phi}-\mu^2 \, \bm{\Phi}
\end{align}
where
\begin{align}\label{BigToeplitz}
\Delta_{(2)}^2&=\text{Toeplitz}(\frac{-1}{12},\,\frac{4}{3},\,\frac{-5}{2},\,\frac{4}{3},\,\frac{-1}{12})\\
\nonumber
D&=\text{Toeplitz}(\dots,\!\frac{-1}{5},\!\frac{1}{4},\!\frac{-1}{3},\!\frac{1}{2},\!-1,\!0,\!1,\!\frac{-1}{2},\!\frac{1}{3},\!\frac{-1}{4},\!\frac{1}{5},\!\dots)\\
\nonumber
D^2&=\text{Toeplitz}(\dots,\!\frac{-2}{16},\!\frac{2}{9},\!\frac{-2}{4},\!\frac{2}{1},\!\frac{-2\pi^2}{6},\!\frac{2}{1},\!\frac{-2}{4},\!\frac{2}{9},\!\frac{-2}{16},\!\dots).
\end{align}
Note that $\Delta_{(2)}^2$ is related to the next-to-nearest-neighbor approximation to the second derivative. Obviously, the longer range we make our derivative approximations the more accurate they can be. The infinite-range operator $D$ (and its square $D^2$) in some sense are the best discrete approximations to the derivative (and second derivative) possible. The defining property of $D$ is that it is diagonal in the (discrete) Fourier basis with spectrum,
\begin{align}\label{LambdaD}
\lambda_D(k)=-\ii\,\underline{k}    
\end{align}
where $\underline{k}=k$ for $k\in[-\pi,\pi]$ repeating itself cyclically with period $2\pi$ outside of this region. This is in tight connection with the continuum derivative operator $\partial_x$ which is diagonal in the (continuum) Fourier basis with spectrum \mbox{$\lambda_{\partial_x}(k)=-\ii \, k$} for $k\in[-\infty,\infty]$. 

Alternatively, one can construct $D^2$ in the following way: generalize $\Delta_{(1)}^2$ and $\Delta_{(2)}^2$ to $\Delta_{(n)}^2$ namely the best second derivative approximation which considers up to $n^\text{th}$ neighbors to either side. Taking the limit $n\to\infty$ gives $D^2=\lim_{n\to\infty}\Delta_{(n)}^2$. Other aspects of $D$ will be discussed in Sec.~\ref{SecSamplingTheory} (including its related derivative approximation Eq.~\eqref{ExactDerivative}) but enough has been said for now.

While these connections to derivative approximations allow us to export some intuitions from the continuum theories into these discrete theories, we must resist this (at least for now). In particular, I should stress again that we should not  be thinking of any of KG1, KG2 and KG3 as coming from the continuum theory under some approximation of the derivative.

Let's next reformulate KG4 and KG5 in terms of \mbox{$\bm{\Phi}\in\mathbb{R}^L$}. In these cases we have a convenient relabeling of the lattice sites in terms of three integer indices, $\ell\mapsto (j,n,m)$. As before we can use this relabeling to grant the vector space a tensor product structure as $\mathbb{R}^L\mapsto\mathbb{R}^\mathbb{Z}\otimes\mathbb{R}^\mathbb{Z}\otimes\mathbb{R}^\mathbb{Z}$ by taking by taking \mbox{$\bm{b}_\ell\mapsto \bm{e}_j\otimes\bm{e}_n\otimes\bm{e}_m$}. Under this restructuring we have,
\begin{align}\label{PhiVec2}
\bm{\Phi}= \sum_{j,n,m\in\mathbb{Z}} \phi_{j,n,m} \, \bm{e}_j\otimes\bm{e}_n\otimes\bm{e}_m. 
\end{align}
In these terms the dynamics of KG4 given above (namely, Eq.~\eqref{KG4Long}) is now given by,
\begin{align}\label{DKG4}
&\text{\bf Klein Gordon Equation 4 (KG4):}\\
\nonumber
&\Delta_{(1),\text{j}}^2\,\bm{\Phi}=(\Delta_{(1),\text{n}}^2+\Delta_{(1),\text{m}}^2) \ \bm{\Phi}-\mu^2 \, \bm{\Phi}
\end{align}
A similar treatment of the dynamics of KG5 (namely, Eq.~\eqref{KG5Long}) gives us,
\begin{align}\label{DKG5}
&\text{\bf Klein Gordon Equation 5 (KG5):}\\
\nonumber
&\Delta_{(1),\text{j}}^2\,\bm{\Phi} =\frac{2}{3} \, \Big[\Delta^2_{(1),\text{n}}+ \Delta^2_{(1),\text{m}}\\
\nonumber
&\qquad\qquad\quad \ +\big\{\Delta^+_\text{m}-\Delta^+_\text{n},\Delta^-_\text{m}-\Delta^-_\text{n}\big\}\Big] \, \bm{\Phi}-\mu^2 \, \bm{\Phi}
\end{align}
While the third term in the square brackets looks complicated, it is just the analog of $\Delta^2_{(1),\text{n}}$ and $\Delta^2_{(1),\text{m}}$ but in the $m-n$ direction. See Eq.~\eqref{Delta12}.

Finally, in addition to KG4 and KG5 I consider the following two theories:
\begin{align}
\label{DKG6}
&\text{\bf Klein Gordon Equation 6 (KG6):}\\
\nonumber
&D_{\text{j}}^2\,\bm{\Phi}=(D_{\text{n}}^2+D_{\text{m}}^2) \ \bm{\Phi}-\mu^2 \, \bm{\Phi}\\
\label{DKG7}
&\text{\bf Klein Gordon Equation 7 (KG7):}\\
\nonumber
&D_{\text{j}}^2\,\bm{\Phi}= \frac{2}{3}\left(D^2_\text{n}+D^2_\text{m}+(D_\text{m}-D_\text{n})^2\right) \, \bm{\Phi}-\mu^2 \, \bm{\Phi}
\end{align}
which resemble KG4 and KG5 but which make use of an infinite range coupling between lattice sites. Having introduced these seven theories, let us next solve their dynamics.

\subsection{Solving Their Dynamics}
Conveniently, each of KG1-KG7 admit planewave solutions. Moreover, in each case these planewave solutions form a complete basis of solutions.

Considering first KG1-KG3 we have solutions of the form,
\begin{align}\label{PlaneWave123}
\phi_{j,n}(\omega,k)=e^{-\ii\, \omega \, j - \ii\, k \, n}.
\end{align}
with $\omega,k\in\mathbb{R}^2$. It should be noted however, that outside of the range $\omega,k\in[-\pi,\pi]$ these planewaves repeat themselves with period $2\pi$ due to Euler's identity, $\exp(2\pi\ii)=1$. In terms of $\mathbb{R}^L\cong\mathbb{R}^\mathbb{Z}\otimes\mathbb{R}^\mathbb{Z}$ these planewaves are:
\begin{align}
\bm{\Phi}(\omega,k)=\sum_{j,n\in\mathbb{Z}} \phi_{j,n}(\omega,k) \, \bm{e}_j\otimes\bm{e}_n. 
\end{align}
From this planewave basis we can recover the $\bm{e}_j\otimes\bm{e}_n$ basis as:
\begin{align}
\bm{e}_j\otimes\bm{e}_n
&=\frac{1}{(2\pi)^2}\int\!\!\!\!\int_{-\pi}^\pi e^{\ii\,\omega \,j+\ii\, k\, n}\, \bm{\Phi}(\omega,k)\,\d\omega\d k. 
\end{align}
These planewaves are only a solution if $\omega$ and $k$ satisfy the theory's dispersion relation which can be straight-forwardly calculated from the theory's dynamics:
\begin{align}
\text{KG1:}& \quad \!\! 2-2\,\cos(\omega)= \mu^2+2-2\,\cos(k)\\
\text{KG2:}& \quad \!\! \frac{1}{6}\,(\cos(2\,\omega)-16\,\cos(\omega)+15)\\
\nonumber
&= \mu^2+\frac{1}{6}\,(\cos(2\,k)-16\,\cos(k)+15)\\
\text{KG3:}& \quad \!\! \underline{\omega}^2= \mu^2+\underline{k}^2
\end{align}
where $\underline{\omega}=\omega$ and $\underline{k}=k$ for $\omega,k\in[-\pi,\pi]$ repeating themselves cyclically with period $2\pi$ outside of this region. Note that the dispersion relation for KG3 follows from Eq.~\eqref{LambdaD}, essentially from the definition of $D$.

\begin{figure}[t!]
\includegraphics[width=0.45\textwidth]{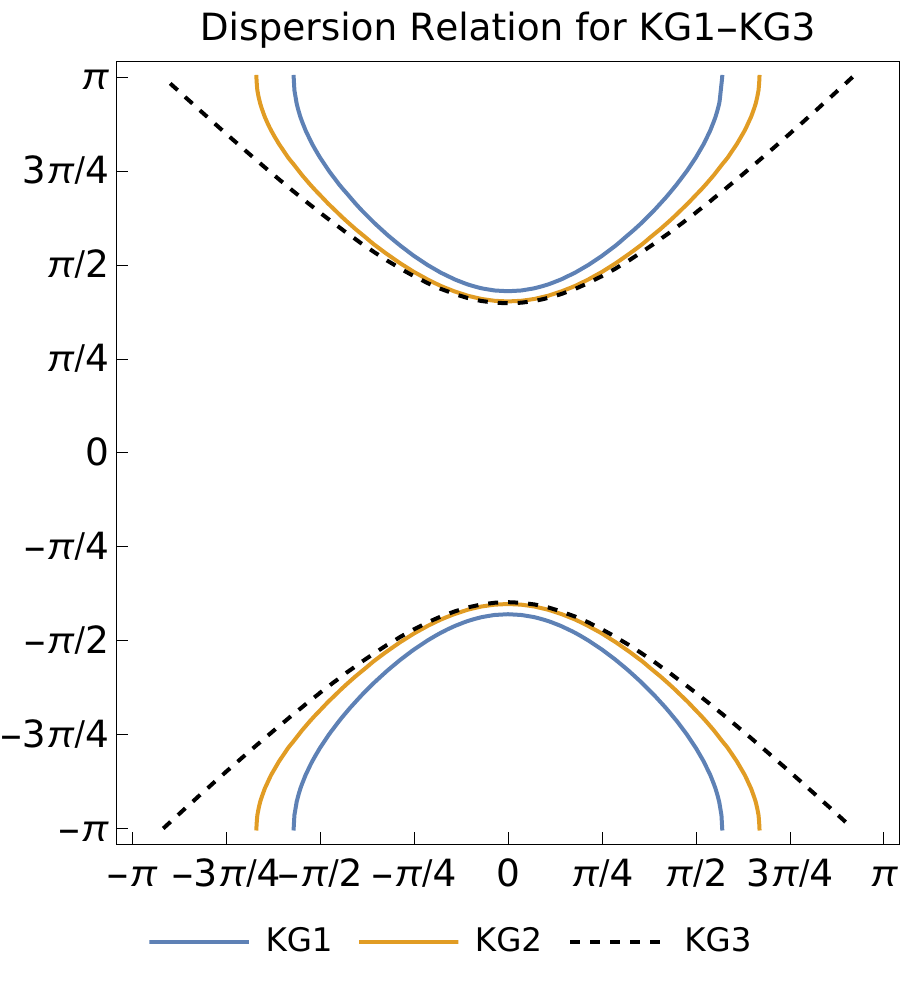}
\caption{The dispersion relations for the planewave solutions to the discrete Klein Gordon equations are plotted as a function of wavenumber for KG1, KG2 and KG3 with $\mu=1.25$.}\label{FigKleinDisp}
\end{figure}

Fig.~\ref{FigKleinDisp} shows these dispersion relations restricted to the region $\omega,k\in[-\pi,\pi]$ with a field mass of $\mu=1.25$. Qualitatively, KG1-KG3 all seem to agree with each other at low wavenumbers. They appear to mostly differ with respects to the rate at which high wavenumber planewaves oscillate. Let's investigate how these theories behave for planewaves with periods and wavelengths which span many lattice sites, that is with $\vert\omega\vert,\vert k\vert\ll\pi$.
\begin{align}
\nonumber
\text{KG1:}& \quad \!\! \omega^2= \mu^2+k^2+\frac{\omega^4-k^4}{12}+\mathcal{O}(\omega^6)+\mathcal{O}(k^6)\\
\nonumber
\text{KG2:}& \quad \!\! \omega^2= \mu^2+k^2-\frac{\omega^6-k^6}{90}+\mathcal{O}(\omega^8)+\mathcal{O}(k^8)\\
\text{KG3:}& \quad \!\! \omega^2 = \mu^2+k^2.
\end{align}
Note that the dispersion relation for KG3 exactly matches that of the continuum theory, not only within this regime but for all $\omega,k\in[-\pi,\pi]$. In the $\vert\omega\vert,\vert k\vert\ll\pi$ regime, KG2 gives a better approximation of the continuum theory than KG1 does. This is due to its longer range coupling giving a better approximation of the derivative. 

If we consider only solutions with all or most of their planewave support with $\vert\omega\vert,\vert k\vert\ll\pi$, we have an approximate one-to-one correspondence between the solutions to these theories. This is roughly why each of these theories have the same continuum limit, namely KG00 defined above. In terms of the rate at which these theories converge to the continuum theory in the continuum limit, one can expect KG3 to outpace KG2 which outpaces KG1. (As I discussed in \cite{DiscreteGenCovPart1}, this is in a way counter-intuitive: why does the most non-local discrete theory give the best approximation of our perfectly local continuum theory?)

However, while interesting in their own right, these relationships with the continuum theory are not directly helpful in helping us understand KG1-KG3 in their own terms as discrete-native theories.

Moving on to KG4-KG7, their planewave solutions are of the form, 
\begin{align}
\phi_{j,n,m}(\omega,k_1,k_2)=e^{-\ii\,\omega \,j-\ii\, k_1\, n-\ii\, k_2\, m}
\end{align}
with $\omega,k_1,k_2\in\mathbb{R}^2$. Again, it should be noted however, that outside of the range \mbox{$\omega,k_1,k_2\in[-\pi,\pi]$} these planewaves repeat themselves with period $2\pi$ due to Euler's identity, $\exp(2\pi\ii)=1$. In terms of \mbox{$\mathbb{R}^L\cong\mathbb{R}^\mathbb{Z}\otimes\mathbb{R}^\mathbb{Z}\otimes\mathbb{R}^\mathbb{Z}$} these planewaves are:
\begin{align}
\nonumber
\bm{\Phi}(\omega,k_1,k_2)=\sum_{j,n,m\in\mathbb{Z}} \phi_{j,n,m}(\omega,k_1,k_2) \, \bm{e}_j\otimes\bm{e}_n\otimes\bm{e}_m. 
\end{align}
From this planewave basis we can recover the $\bm{e}_j\otimes\bm{e}_n\otimes\bm{e}_m$ basis as:
\begin{align}
&\bm{e}_j\otimes\bm{e}_n\otimes\bm{e}_m\\
\nonumber
&=\frac{1}{(2\pi)^3}\int\!\!\!\!\int\!\!\!\!\int_{-\pi}^\pi e^{\ii\,\omega \,j+\ii\, k_1 n+\ii\, k_2 m}\,\bm{\Phi}(\omega,k_1,k_2)\,\d\omega\d k_1 \d k_2. 
\end{align}

The dispersion relation for each of these theories is given by:
\begin{align}
\text{KG4:}& \ 2-2\cos(\omega)= \mu^2\!+\!4\!-\!2\cos(k_1)\!-\!2\cos(k_2)\\
\nonumber
\text{KG5:}& \  2-2\,\cos(\omega)= \mu^2 \\
\nonumber
&+\frac{4}{3}\big[3-\cos(k_1)-\cos(k_1)-\cos(k_2-k_1)\big]\\
\nonumber
\text{KG6:}& \ \underline{\omega}^2= \mu^2 +\underline{k_1}^2 +\underline{k_2}^2\\
\nonumber
\text{KG7:}& \ \underline{\omega}^2= \mu^2 +\frac{2}{3}\big[\underline{k_1}^2 +\underline{k_2}^2 +(\underline{k_2}-\underline{k_1})^2\big].
\end{align}
Note that the dispersion relation for KG6 and KG7 follow from Eq.~\eqref{LambdaD}, essentially from the definition of $D$.

Unlike KG1-KG3, these theories do not all agree with each other in the small $\omega,k_1,k_2$ regime. KG4 and KG6 agree that for $\vert\omega\vert, \vert k_1\vert, \vert k_2\vert \ll \pi$ we have \mbox{$\omega^2= \mu^2 +k_1^2 +k_2^2$}. Moreover, KG5 and KG7 agree with each other in this regime, but not with KG4 and KG6. Do we have two different results in the continuum limit here?

Closer examination reveals that we do not. The key to realizing this is to note that under the transformation, 
\begin{align}\label{SkewKG7KG6}
\omega\mapsto \omega,\quad
k_1\mapsto k_1,\quad
k_2\mapsto \frac{1}{2}\, k_1+\frac{\sqrt{3}}{2}\,k_2.
\end{align}
we have the dispersion relation for KG7 mapping exactly onto the one for KG6. The inverse of this map is
\begin{align}\label{SkewKG6KG7}
\omega\mapsto \omega,\quad
k_1\mapsto k_1,\quad
k_2\mapsto \frac{2k_2-k_1}{\sqrt{3}}.
\end{align}
Technically, when acting on the planewaves $\bm{\Phi}(\omega,k_1,k_2)$ these transformations are only each other's inverses when we have \mbox{$\omega,k_1,k_2\in[-\pi,\pi]$} both before and after the transformation. This is due to the $2\pi$ periodicity of these planewaves. Fortunately however, all of KG7's planewave solutions with \mbox{$\omega,k_1,k_2\in[-\pi,\pi]$} remain in this region after applying Eq.~\eqref{SkewKG7KG6}. The same is true of KG6: its planewave solutions with \mbox{$\omega,k_1,k_2\in[-\pi,\pi]$} also remain in this region after applying Eq.~\eqref{SkewKG6KG7}. As I will soon discuss, this means we have an exact one-to-one correspondence between KG6 and KG7's solutions (much ado will be made about this later.) Applying this transformation to KG5 does not map it onto KG4, but it does bring their $\vert\omega\vert,\vert k_1\vert,\vert k_2\vert\ll\pi$ behavior into agreement. 

Thus, if we consider only solutions with all or most of their planewave support with $\vert\omega\vert,\vert k_1\vert,\vert k_2\vert\ll\pi$ (or the appropriately transformed regime for KG5 and KG7) then we have an approximate one-to-one correspondence between the solutions to these theories. Within this regime we can define their common continuum limit, KG0. Repeating our analysis of the convergence rates of KG1-KG3 here, we expect KG6 and KG7 to converge in the continuum limit faster than KG4 and KG5 do.

\vspace{0.25cm}

This paper will make three attempts at interpreting these seven discrete theories. Allow me to identify in advance three important points of comparison between these interpretations. 

The first important point of comparison is what sense they make of these different convergence rates in the continuum limit. As discussed above, in terms of this convergence rate we expect \mbox{$\text{KG3}>\text{KG2}>\text{KG1}$} and similarly \mbox{$\text{KG6}, \text{KG7} > \text{KG4}, \text{KG5}$} with higher rated theories converging more quickly. This is in tension with our intuitive sense of locality for these theories: judging locality by the number of lattice sites coupled together we have
\mbox{$\text{KG1} > \text{KG2} > \text{KG3}$} with higher rated theories being more local and similarly
\mbox{$\text{KG4},\text{KG5} > \text{KG6},\text{KG7}$}. How is it that our most non-local theories are somehow the nearest to our perfectly local continuum theory?

Regarding how these three interpretations deal with this tension, not much changes between the discrete heat equations considered in \cite{DiscreteGenCovPart1} and the discrete Klein Gordon equations considered here. As such, I will leave any detailed discussion of this issue to \cite{DiscreteGenCovPart1} and direct the interested reader there. Roughly, the second and third interpretations deal with this tension by negating or reversing all of the above intuitive locality judgements.

A second important point of comparison between these three interpretations will be what sense they make of the  above-noted exact one-to-one correspondence between KG6 and KG7's solutions. (More will be said about this in Sec.~\ref{SecKlein2}.) It is important to note that the mere existence of such a one-to-one correspondence does not automatically mean that these theories are identical or even equivalent; All it means technically is that their solution spaces have the same cardinality. As I will discuss, some of the coming interpretations recognize KG6 and KG7 as being equivalent whereas others do not.

A third important point of comparison between the coming interpretations will be what sense they make of these theories having continuous symmetries. For instance, the dispersion relation for KG6 appears to be in some sense rotation invariant and even Lorentz invariant (at least in Fourier space and staying inside of the region \mbox{$\omega,k_1,k_2\in[-\pi,\pi]$}). In a sense, KG7 might have these symmetries too: given the above-noted one-to-one correspondence between the solutions of KG6 and KG7, there may be some (skewed) sense in which KG7 is rotation invariant and Lorentz invariant as well. All of this will be made precise later on. As I will discuss, some interpretations consider KG6 and KG7 to have a rotation symmetry and even (limited) Lorentz boost symmetries whereas others do not. As I will discuss in Sec.~\ref{PerfectLorentz}, KG6 and KG7 can be seen as representationally-limited parts of a larger perfectly Lorentzian lattice theory.

Having introduced these theories and solved their dynamics in an interpretation-neutral way. We can now make a first (ultimately misled) attempt at interpreting them.

\section{A First Attempt at Interpreting KG1-KG7}\label{SecKlein1}
Now that we have introduced these seven discrete theories and solved their dynamics, let's get on to interpreting them. Let us begin by following our first intuitions and analyze these seven discrete theories concerning their underlying manifold, locality properties and symmetries. Ultimately however, as I will discuss later, much of the following is misled and will need to be revisited and revised later. Luckily, retracing where we went wrong here will be instructive later.

Let's start by taking the initial formulation of the above theories in terms of $\phi_\ell$ seriously, i.e. Eq.~\eqref{KG1Long}, Eq.~\eqref{KG4Long} and Eq.~\eqref{KG5Long}. Taken literally as written, what are these theories about? Intuitively these theories are about a field $\phi_\ell$ which maps lattice sites ($\ell\in L\cong\mathbb{Z}\cong\mathbb{Z}^2\cong\mathbb{Z}^3$) into field amplitudes ($\phi_\ell\in\mathbb{R}$). That is a field $\phi:Q\to \mathcal{V}$ with a discrete manifold $Q=L$ and value space $\mathcal{V}=\mathbb{R}$. Thus, taking $\phi:Q\to \mathcal{V}$ seriously as a fundamental field leads us to thinking of $Q=L$ as the theory's underlying manifold and $\mathcal{V}=\mathbb{R}$ as the theory's value space. It is important to note that here, $Q$ is the entire manifold, it is not being thought of as embedded in some larger manifold. (However, a view like this will be considered in Sec.~\ref{SecExtPart1}.) 

Taking $Q$ to be these theories' underlying manifold has consequences for our understanding of the locality of these theories. In a highly intuitive sense, theory KG1 is the most local in that it couples together the fewest lattice sites (only nearest neighbors). Following this KG2 is the next most local in the same sense: it couples only next-to-nearest neighbors. Finally, in this sense KG3 is the least local, it has an infinite range coupling. As mentioned above, there is some tension however with the rate we expect each of these theories to converge at in the continuum limit. How is it that our most non-local theories are somehow the nearest to our perfectly local continuum theory? This first interpretation can do little to resolve this tension, I refer the interested reader to \cite{DiscreteGenCovPart1} for further discussion.

\subsection{Intuitive Symmetries}
With this manifold $Q=L$ and value space $\mathcal{V}=\mathbb{R}$ picked out, what can we expect of these theories' symmetries? For any spacetime theory there are roughly three kinds of symmetries: 1) external symmetries associated with automorphisms of the manifold, here $\text{Auto}(Q)$, 2) internal symmetries associated with automorphisms of the value space, here $\text{Auto}(\mathcal{V})$, and gauge symmetries which result from allowing these internal symmetries to vary smoothly across the manifold. But what are the relevant notions of automorphism here?

Answering this question for $\text{Auto}(Q)$ will require us to distinguish what structures are ``built into'' $Q$ and what are ``built on top of'' $Q$. The analogous distinction in the continuum case is that we generally take the manifold's differentiable structure to be built into it while the Minkowski metric, for instance, is something additional built on top of the manifold. In this paper, I am officially agnostic on where we draw this line in the discrete case. However, for didactic purposes I will here be as conservative as possible giving $Q$ as little structure as is sensible. Note that the less structure we associate with $Q$ the larger the class of relevant automorphisms $\text{Auto}(Q)$ will be. Thus, I am taking $\text{Auto}(Q)$ to be as large as it can reasonably be. 

Here the minimal structure we can reasonably associate with $Q=L$ is that of a set. As such the largest $\text{Auto}(Q)$ could reasonably be is permutations of the lattice sites, \mbox{$\text{Auto}(Q)=\text{Perm}(L)$}. 

In addition to $\text{Auto}(Q)$ we might also have internal symmetries $\text{Auto}(\mathcal{V})$ and gauge symmetries. While in general there may be abundant internal or gauge symmetries, for the present cases there are not many. In particular, for all of the above-mentioned theories we only have $\mathcal{V}=\mathbb{R}$. As mentioned following Eq.~\eqref{PhiDef}, our (potentially off-shell) discrete fields are themselves vectors $\phi_\ell\in F_L$. Namely, they are closed under addition and scalar multiplication and hence form a vector space. This addition and scalar multiplication is carried out lattice-site-by-lattice-site. Thus, the field's value space $\mathcal{V}=\mathbb{R}$ is also structured like a vector space. 

The value space $\mathcal{V}=\mathbb{R}$ may additionally have more structure than this. However, as above, for didactic purposes I will here minimize the assumed structure in order to maximize possible symmetries. We can even drop the zero vector from our consideration taking $\mathcal{V}=\mathbb{R}$ to be an affine vector space.
Therefore, I will take \mbox{$\text{Auto}(\mathcal{V})=\text{Aff}(\mathbb{R})$} such that our internal symmetries are linear-affine rescalings of $\phi_\ell$, namely $\phi_\ell\mapsto c_1\phi_\ell+c_2$. We can find the theory's gauge symmetries by letting \mbox{$c_1,c_2\in\mathbb{R}$} vary smoothly across $Q$. That is, $\phi_\ell\mapsto c_{\ell,1}\phi_\ell+c_{\ell,2}$.

Thus, in total, for KG1-KG7 the widest scope of symmetry transformations available to us (at least on this interpretation) are:
\begin{align}\label{PermutationLong}
s:\quad \phi_\ell\mapsto c_{P(\ell),1} \phi_{P(\ell)}+c_{P(\ell),2}
\end{align}
for some permutation $P:L\to L$.

For later reference it will be convenient to translate these potential symmetry transformations in terms of the vector, $\bm{\Phi}\in\mathbb{R}^L$, as 
\begin{align}\label{Permutation}
s:\quad\bm{\Phi}\mapsto C_1\,P\,\bm{\Phi}+\bm{c}_2,
\end{align}
for some permutation matrix, $P$, a diagonal matrix $C_1$ and a vector $\bm{c}_2$. Here $P$ captures the theory's possible external symmetries:  the possibility of permuting lattice sites. The diagonal matrix $C_1$ and the vector $\bm{c}_2$ capture the theory's possible gauge symmetries: the possibility of linear-affine rescalings of $\phi_\ell$ which vary smoothly across $Q$.

I will next discuss which transformations of this form preserve the dynamics of KG1-KG7. It should be clear from the outset however that (at least on this interpretation) these theories cannot have continuous spacial translation and rotation, let alone Lorentzian boost symmetries. Indeed, I have been charitable considering the lattice sites structured only as a set (perhaps artificially) increasing the size of $\text{Auto}(Q)$. Given this, it would be highly surprising if we found KG1-KG7 to have symmetries outside of this set. (Such a surprise is coming in the Sec.~\ref{SecKlein2}.) 

As I will show in Sec.~\ref{SecKlein2}, this first interpretation of these theories systematically under predicts the symmetries that discrete spacetime theories can and do have. Fixing this issue will lead one to a discrete analog of general covariance. We here under-predict symmetries because we are taking these theories' lattice structures too seriously. Properly understood, they are merely a coordinate-like representational artifact and so do not limit our symmetries. Before that however, let's see the symmetries these theories have on this interpretation.

\subsubsection*{Symmetries of KG1-KG7: First Attempt}
What then are the symmetries of KG1-KG7 according to this interpretation? A technical investigation of the symmetries of KG1-KG7 on this interpretation is carried out in Appendix~\ref{AppA}, but the results are the following. For KG1-KG3 the dynamical symmetries of the form Eq.~\eqref{PermutationLong} are:
\begin{flushleft}\begin{enumerate}
    \item[1)] discrete shifts which map lattice site \mbox{$(j,n)\mapsto (j-d_1,n-d_2)$} for some integers $d_1,d_2\in\mathbb{Z}$,
    \item[2)] two negation symmetries which map lattice site $(j,n)\mapsto (-j,n)$ and $(j,n)\mapsto (j,-n)$ respectively,
    \item[3)] global linear rescaling which maps \mbox{$\phi_\ell\mapsto c_1\phi_\ell$} for some $c_1\in\mathbb{R}$,
    \item[4)] local affine rescaling which maps \mbox{$\phi_\ell\mapsto \phi_\ell + c_{2,\ell}(t)$} for some $c_{2,\ell}(t)$ which is also a solution of the dynamics.
\end{enumerate}\end{flushleft}
These are the symmetries of a uniform 1D lattices in space  and a uniform 1D lattice in time, \mbox{$z_{j,n}=(j,n)\in\mathbb{R}^2$} (plus linear-affine rescalings). These are two independent 1D lattices (rather than a single square 2D lattice) because we do not have quarter rotations between space and time among our symmetries. Previously I had warned against prematurely interpreting the lattice sites underlying KG1-KG3 as being organized into a square lattice. As it turns out, having investigated these theories' dynamical symmetries this warning was warranted.

What about KG4 and KG6? For KG4 and KG6 the dynamical symmetries of the form Eq.~\eqref{PermutationLong} are:
\begin{flushleft}\begin{enumerate}
    \item[1)] discrete shifts which map lattice site \mbox{$(j,n,m)\mapsto (j-d_1,n-d_2,m-d_3)$} for some integers $d_1,d_2,d_3\in\mathbb{Z}$,
    \item[2)] three negation symmetries which map lattice site $(j,n,m)\mapsto (-j,n,m)$ and $(j,n,m)\mapsto (j,-n,m)$ and $(j,n,m)\mapsto (j,n,-m)$ respectively,
    \item[3)] a 4-fold symmetry which maps lattice site \mbox{$(j,n,m)\mapsto (j,m,-n)$}, 
    \item[4)] global linear rescaling which maps \mbox{$\phi_\ell\mapsto c_1\phi_\ell$} for some $c_1\in\mathbb{R}$,
    \item[5)] local affine rescaling which maps \mbox{$\phi_\ell\mapsto \phi_\ell + c_{2,\ell}(t)$} for some $c_{2,\ell}(t)$ which is also a solution of the dynamics.
\end{enumerate}\end{flushleft}
These are the symmetries of a square 2D lattice in space and a uniform 1D lattice in time, \mbox{$z_{j,n,m}=(j,n,m)\in\mathbb{R}^3$} (plus linear-affine rescalings). The above 4-fold symmetry corresponds to quarter rotation in space. These are two independent lattices (rather than a single cubic 3D lattice) because we do not have quarter rotations between space and time among our symmetries. Previously I had warned against prematurely interpreting the lattice sites underlying KG4 and KG6 as being organized into a cubic lattice. As it turns out, having investigated these theories' dynamical symmetries this warning was warranted.

What about KG5 and KG7? For KG5 and KG7 the dynamical symmetries of the form Eq.~\eqref{PermutationLong} are:
\begin{flushleft}\begin{enumerate}
    \item[1)] discrete shifts which map lattice site \mbox{$(j,n,m)\mapsto (j-d_1,n-d_2,m-d_3)$} for some integers $d_1,d_2,d_3\in\mathbb{Z}$,
    \item[2)] an exchange symmetry which maps lattice site $(j,n,m)\mapsto (j,m,n)$,
    \item[3)] a 6-fold symmetry which maps lattice site \mbox{$(j,n,m)\mapsto (j,-m,n+m)$}. (Roughly, this permutes the three terms in the right hand side of Eq.~\eqref{DKG5} for KG5 and Eq.~\eqref{DKG7} for KG7.),
    \item[4)] global linear rescaling which maps \mbox{$\phi_\ell\mapsto c_1\phi_\ell$} for some $c_1\in\mathbb{R}$,
    \item[5)] local affine rescaling which maps \mbox{$\phi_\ell\mapsto \phi_\ell + c_{2,\ell}(t)$} for some $c_{2,\ell}(t)$ which is also a solution of the dynamics.
\end{enumerate}\end{flushleft}
These are the symmetries of a hexagonal 2D lattice in space and a uniform 1D lattice in time, \mbox{$z_{j,n,m}=(j,n+m/2,\sqrt{3}m/2)\in\mathbb{R}^3$} (plus linear-affine rescalings). The above 6-fold symmetry corresponds to one-sixth rotation in space. Previously I had warned against prematurely interpreting the lattice sites underlying KG5 and KG7 as being organized into a cubic 3D lattice, prompted by the convenient relabeling $\ell\mapsto (j,n,m)$. As it turns out, having investigated these theories' dynamical symmetries this warning was well warranted.

Thus, by investigating these theories' dynamical symmetries we were able to find what sort of lattice structure the assumed-to-be unstructured lattice $L$ actually has for each theory (e.g. in space a uniform 1D lattice, a square lattice, and a hexagonal lattice each together with a uniform 1D lattice in time). %Recall the discussion of matching dynamical symmetries with spacetime symmetries in Sec.~\ref{SecGenCov}. For continuum spacetime theories, we can only discover their fixed spacetime structures by investigating the dynamics. (Of course, we have no hope of discovering them directly through dynamical means, they are dynamically-fixed.) The same is true here, we started with a bare lattice, $L$, investigated the dynamics, and now we have good candidates for what lattice structures each of these theories have in addition to $L$.

\vspace{0.25cm}

Finally, in this interpretation what sense can be made of KG6 and KG7 having a nice one-to-one correspondence between their solutions discussed at the end of Sec.~\ref{SecSevenKG}? While this correspondence between solutions certainly exists, little sense can be made of it here in support of the equivalence of these theories. As the above discussion has revealed this interpretation associates very different symmetries to KG6 and KG7 and correspondingly very different lattice structures. While there is nothing technically wrong per se with this assessment our later interpretations will make better sense of this correspondence.

\vspace{0.25cm}

To summarize, this interpretation has the benefit of being highly intuitive. Taking the fields given to us, \mbox{$\phi:Q\to\mathbb{R}$}, seriously we identified the underlying manifold as $Q=L$. From this we got some intuitive notions of locality. Moreover, by finding these theories' dynamical symmetries we were able to grant some more structure to their lattice sites. By and large, the interpretation seems to validate all of the first intuitions laid out in Sec.~\ref{SecIntro}. On this interpretation, the lattice seems to play a substantive role in the theory: it seems to restrict our symmetries, it seems to distinguish our theories from one another, be essentially ``baked-into'' the formalism. (As I will discuss in the next section, none of this is right.)

However, there are three major issues with this interpretation which will become clear in light of our later interpretations. Firstly, our locality assessments are in tension with the rates at which these theories converge to the (perfectly local) continuum theory in the continuum limit, see \cite{DiscreteGenCovPart1} for further discussion. Secondly, despite the niceness of the one-to-one correspondence between the solutions to KG6 and KG7, this interpretation regards them as significantly different theories: with different lattice structures and (here consequently) different symmetries. The final issue (which will become clear in the next section) is that this interpretation drastically under predicts the kinds of symmetries which KG1-KG7 can and do have. In fact, each of KG1-KG7 have a hidden continuous translation symmetry. Moreover, KG6 and KG7 have a hidden continuous rotation symmetry. Moreover still, KG3, KG6, and KG7 all have a hidden (limited) Lorentzian boost symmetry.

As I will discuss, the root of all of these issues is taking the theory's lattice structure too seriously. As I will argue, when properly understood, they are merely a coordinate-like representational artifact. Indeed, as I will show in the next section they do not limit our theory's symmetries. Moreover, theories appearing initially with different lattice structures may nonetheless be equivalent. Finally, these theories can always be reformulated to refer to no lattice structure at all. These three points establish a strong analogy between the lattice structures appearing in our discrete spacetime theories and the coordinate systems appearing in our continuum theories. Ultimately, spelling out this analogy in detail in Sec.~\ref{SecDisGenCov} will give us a discrete analog of general covariance (now extended to a Lorentzian context). Indeed, this will lead us to a perfectly Lorentzian lattice theory in Sec.~\ref{PerfectLorentz}.

\section{A Second Attempt at Interpreting KG1-KG7}\label{SecKlein2}
In the previous section, I claimed that KG1-KG7 have hidden continuous translation and rotation symmetries and even (limited) Lorentzian boost symmetries. But how can this be? How can discrete spacetime theories have such continuous symmetries? As I discussed in the previous section, if we take our underlying manifold to be \mbox{$Q=L$} then these theories clearly cannot support continuous translation and rotation symmetries let alone Lorentzian boosts.

In order to avoid this conclusion we must deny the premise, $Q$ must not be the underlying manifold. What led us to believe $Q$ was the underlying manifold? We arrived at this conclusion by focusing on $\phi_\ell\in F_L$ and thereby taking the real scalar field $\phi:Q\to\mathcal{V}$ to be fundamental. $Q$ is the underlying manifold because it is where our fundamental field maps from. In order to avoid this conclusion we must deny the premise, the field $\phi:Q\to\mathcal{V}$ must not be fundamental. 

But if $\phi:Q\to\mathcal{V}$ is not fundamental then what is? Fortunately, our above discussion has already provided us with another object which we might take as fundamental. Namely, $\bm{\Phi}$ defined in Eq.~\eqref{PhiDef}. These vectors $\bm{\Phi}\in\mathbb{R}^L$ are in a one-to-one correspondence with the discrete fields $\phi_\ell\in F_L$, Moreover, these vector spaces are isomorphic $\mathbb{R}^L\cong F_L$ with Eq.~\eqref{PhiDef} being a vector space isomorphism between them.

On this second interpretation I will be taking the formulations of KG1-KG7 in terms of $\bm{\Phi}$ seriously: namely Eq.~\eqref{DKG1}, Eq.~\eqref{DKG2}, Eq.~\eqref{DKG3}, and Eqs.~\eqref{DKG4}-\eqref{DKG7}. Taken literally as written, what are these theories about? These theories are intuitively about an infinite dimensional vector ($\bm{\Phi}\in\mathbb{R}^L$) which satisfies some dynamical constraint. 

There are two ways in which one might try to make sense of $\bm{\Phi}$ as a field $\bm{\Phi}:\mathcal{M}\to\mathcal{V}$ with some manifold $\mathcal{M}$ and some value space $\mathcal{V}$. Firstly, one might consider a one-point manifold $\mathcal{M}=p$ with $\bm{\Phi}$ simply being the field value there $\bm{\Phi}\coloneqq \bm{\Phi}(p)$. Alternatively, one might try to make sense of $\bm{\Phi}$ as a field ``from nowhere'' with an empty manifold $\mathcal{M}=\emptyset$. On this view $\bm{\Phi}$ is a vector field which takes in no input and returns the vector $\bm{\Phi}$.

In any case, on this interpretation $Q$ is no longer the underlying manifold for KG1-KG7. Indeed, on this interpretation the lattice sites, $L$, are no longer our spacetime manifold (there may not even be a manifold). Rather, they have been ``internalized'' into the value space $\mathcal{V}=\mathbb{R}^L$. In particular, in defining this vector space we have associated with each lattice site $\ell\in L$ a basis vector $\bm{b}_\ell$. See Eq.~\eqref{PhiDef}. However, as I will discuss, these particular basis vectors play no special role in these theories. Indeed, looking back at the dynamics for each of KG1-KG7 written in terms of $\bm{\Phi}$, one can see that in each case it can be made basis-independent.

Let's see what consequences this interpretive stance has for these theories' locality and symmetry. To preview: this second interpretation either dissolves or resolves all of our issues with the first interpretation. Firstly, the tension is dissolved between our theories' differences in locality and their differences in convergence rate in the continuum limit. With no spacetime manifold we no longer have access to any spaciotemporal notion of locality. There simply are no differences in locality anymore. I refer the interested reader to \cite{DiscreteGenCovPart1} for further discussion.

Secondly, KG6 and KG7 are seen to be equivalent in a stronger sense. And thirdly, perhaps most importantly, this interpretation reveals KG1-KG7's hidden continuous translation and rotation symmetries and even (limited) Lorentzian boost symmetries. However, as I will discuss, this interpretation has some issues of its own which will ultimately require us to make a third attempt at interpreting these theories in Sec.~\ref{SecKlein3}.

\subsection{Internalized Symmetries}
How does this internalization move affect our theory's capacity for symmetry? How can we now have continuous translation and rotation symmetries as well as a Lorentzian boost symmetry? At first glance, this may appear to have made things worse. Without a manifold (or even if the manifold is a single point) we no longer have any possibility for external symmetries. However, while there are certainly less possible external symmetries, we are now open to a wider range of internal symmetries. It is among these internal symmetries that we will find KG1-KG7's hidden continuous translation and rotation symmetries and even (limited) Lorentzian boost symmetries. As I will argue these symmetries can reasonably be given these names despite being internal symmetries. (In Sec.~\ref{SecKlein3} I will present a third attempt at interpreting these theories which ``externalizes'' these symmetries, making them genuinely spacial translations, rotations, and Lorentzian boosts.)

With our focus now on $\bm{\Phi}\in\mathbb{R}^L$, let us consider its possibilities for symmetries. As discussed above, we have no possibility of external symmetries associated with the manifold. However, we do have possible internal symmetries associated with the value space (i.e., an infinite dimensional vector space) we now have the full range of invertible linear-affine transformations over $\mathbb{R}^L$, namely $\text{Auto}(\mathcal{V})=\text{Aff}(\mathbb{R}^L)$. There are no gauge symmetries here as there is no longer any manifold for them to smoothly vary across. Thus, in total the possibly symmetries for our theories under this interpretation are,
\begin{align}\label{GaugeVR}
s:\quad\bm{\Phi}\mapsto \Lambda\,\bm{\Phi}+\bm{c}
\end{align}
for some invertible linear transformation $\Lambda\in\text{GL}(\mathbb{R}^L)$ and some vector $\bm{c}\in \mathbb{R}^L$. 

Contrast this with the symmetries available to us on our first interpretation, namely Eq.~\eqref{Permutation}. We can role this new class of possible symmetries back onto our first interpretation as follows: Note that because $\mathbb{R}^L \cong F_L$ we have $\text{Aff}(\mathbb{R}^L)\cong \text{Aff}(F_L)$. The set of transformations $\text{Aff}(F_L)$ acting on $\phi_\ell\in F_L$ is much larger than what we previously considered: namely, \mbox{$\text{Auto}(\mathbb{V})=\text{Aff}(\mathbb{R})$} varying smoothly over \mbox{$\text{Auto}(Q)=\text{Perm}(L)$}. Indeed, the present interpretation has a significantly wider class of symmetries than before. 

Moving back to our second interpretation, our previous class of transformations (i.e, \mbox{$\text{Auto}(\mathbb{V})=\text{Aff}(\mathbb{R})$} varying smoothly over \mbox{$\text{Auto}(Q)=\text{Perm}(L)$}) corresponds to only a subset of our present consideration: $\text{Aff}(\mathbb{R}^L)$. The difference is that before we could only apply a permutation matrix $P$ followed by a diagonal matrix $C_1$ whereas now we are allowed a general linear transformation $\Lambda$. 

Note that permutation and diagonal matrices are basis-dependent notions. Our first interpretation took the lattice sites $\ell\in L$ seriously as a part of the manifold $Q=L$ and this is reflected in its conception of symmetries. Converted into $\mathbb{R}^L$ this first conception of these theories' possible symmetries gives special treatment to the basis associated with the lattice sites, namely $\{\bm{b}_\ell\}_{\ell\in L}$. In particular, on our first interpretation, our possible symmetries are those of the form Eq.~\eqref{GaugeVR} which \textit{additionally} preserve this basis (up to rescaling, and reordering).

This basis receives no special treatment on this second interpretation. While it is true that $\{\bm{b}_\ell\}_{\ell\in L}$ were used in the initial construction of $\bm{\Phi}$, after this they no longer play any special role. We are always free to redescribe $\bm{\Phi}$ in a different basis if we wish. Indeed, here any change of basis transformation is of the form Eq.~\eqref{GaugeVR} and hence a symmetry.

With this attachment to the basis $\{\bm{b}_\ell\}_{\ell\in L}$ dropped, we now have a wider class of symmetries. Indeed, everything which was previously considered a symmetry will be here as well and possibly more. Perhaps among this more general class of symmetries we may find new continuous symmetries. Let's see.

\subsubsection*{Symmetries of KG1-KG7: Second Attempt}
Which of the above transformations are symmetries for KG1-KG7? A non-exhaustive  investigation of the symmetries of KG1-KG7 on this interpretation is carried out in Appendix~\ref{AppA}, but the results are the following. For KG1 and KG2 the dynamical symmetries of the form Eq.~\eqref{GaugeVR} include:
\begin{flushleft}\begin{enumerate}
    \item[1)]  action by $T^\epsilon_\text{j}$ sending $\bm{\Phi}\mapsto T^\epsilon_\text{j} \bm{\Phi}$ where \mbox{$T^\epsilon_\text{j}=T^\epsilon\otimes\openone$} with $T^\epsilon$ defined below. Similarly for $T^\epsilon_\text{n}=\openone\otimes T^\epsilon$,
    \item[2)] two negation symmetries which map basis vectors as  $\bm{e}_j\otimes\bm{e}_n\mapsto \bm{e}_{-j}\otimes\bm{e}_n$ and $\bm{e}_j\otimes\bm{e}_n\mapsto \bm{e}_j\otimes\bm{e}_{-n}$ respectively,
    \item[3)] a local Fourier rescaling symmetry which maps \mbox{$\bm{\Phi}(\omega,k)\mapsto \tilde{f}(\omega,k)\,\bm{\Phi}(\omega,k)$} for some non-zero complex function $\tilde{f}(\omega,k)\in\mathbb{C}$ with $\omega,k\in[-\pi,\pi]$,
    \item[4)] local affine rescaling which maps \mbox{$\bm{\Phi}\mapsto \bm{\Phi} + \bm{c}_2$} for some $\bm{c}_2$ which is also a solution of the dynamics.
\end{enumerate}\end{flushleft}
These are exactly the same symmetries that we found on the previous interpretation with two differences: Firstly, global rescaling $\phi_\ell\mapsto c_1 \phi_\ell$ has been refined to a local Fourier rescaling. Note that the discrete Fourier transform itself is in $\text{Aff}(\mathbb{R}^L)$ and so is in the class of potential symmetries considered here.

Secondly, discrete shifts have been replaced with action by 
\begin{align}\label{TDef}
T^\epsilon\coloneqq\text{exp}(-\epsilon D)
\end{align}
with $\epsilon\in\mathbb{R}$ acting on each tensor factor. A straight-forward calculation shows that $T^\epsilon$ acts on the planewave basis \mbox{$\bm{\Phi}(k)\coloneqq \sum_{n\in\mathbb{Z}}e^{-\ii k n}\bm{e}_n$} with $k\in[-\pi,\pi]$ as
\begin{align}
\nonumber
T^\epsilon:\bm{\Phi}(k)\mapsto \exp(\ii k \epsilon)\,\bm{\Phi}(k).
\end{align}
Using \mbox{$\bm{e}_m
=\frac{1}{2\pi}\int_{-\pi}^\pi e^{\ii\, k\, m}\, \bm{\Phi}(k)\,\d k$} we can recover how $T^\epsilon$ acts on the basis $\bm{e}_m$ as
\begin{align}
T^\epsilon: \bm{e}_m \mapsto \sum_{b\in\mathbb{Z}} S_m(b+\epsilon) \, \bm{e}_b
\end{align}
where 
\begin{align}\label{SincDef}
S(y)=\frac{\sin(\pi y)}{\pi y}, \quad\text{and}\quad
S_m(y)=S(y-m),       
\end{align}
are the normalized and shifted sinc functions. Note that $S_n(m)=\delta_{nm}$ for integers $n$ and $m$.

As I will now discuss, $T^\epsilon$ can be thought of as a continuous translation operator for three reasons despite it being here classified as an internal symmetry. Note that none of these reasons depend on $T^\epsilon$ being a symmetry of the dynamics.

First note that $T^\epsilon$ is a generalization of the discrete shift operation in the sense that taking $\epsilon=d_1\in\mathbb{Z}$ reduces action by $T^\epsilon$ to the map $T^{d_1}:\bm{e}_m\mapsto \bm{e}_{m-d_1}$ on basis vectors and relatedly the map $m\mapsto m-d_1$ on lattice sites. 

Second note that $T^\epsilon$ is additive in the sense that $T^{\epsilon_1}\,T^{\epsilon_2}=T^{\epsilon_1+\epsilon_2}$. In the language or representation theory $T^\epsilon$ is a representation of the translation group on the vector space $\mathbb{R}^\mathbb{Z}\cong\mathbb{R}^L$. In particular, this means \mbox{$T^{1/2}\,T^{1/2}=T^1$}: there is something we can do twice to move one space forward. The same is true for all fractions adding to one. 

Third, recall from the discussion following Eq.~\eqref{LambdaD} that $D$ is closely related to the continuum derivative operator $\partial_x$, exactly matching its spectrum for $k\in[-\pi,\pi]$. Recall also that the derivative is the generator of translation, i.e. $h(x-\epsilon)=\text{exp}(-\epsilon\, \partial_x) h(x)$. In this sense also \mbox{$T^\epsilon=\text{exp}(-\epsilon D)$} is a translation operator. More will be said about $T^\epsilon$ in Sec.~\ref{SecSamplingTheory}.

Thus we have our first big lesson: despite the fact that KG1-KG2 can be represented on a lattice, they nonetheless have a continuous translation symmetry. This continuous translation symmetry was hidden from us on our first interpretation because we there took the lattice to be hard-wired in as a part of the manifold. Here, we do not take the lattice structure so seriously. We have internalized it into the value space where it then disappears from view as just one basis among many.

Before KG3 had the same symmetries as KG1 and KG2, now it does not. However, allow me to skip over KG3 temporarily. In the previous interpretation the symmetries of KG4 and KG6 matched each other, both being associated with a 2D square lattice. Moreover, the symmetries of KG5 and KG7 matched each other, both being associated with a hexagonal 2D lattice. Here however, these pairings are broken up and a new matching pair is formed between KG6 and KG7. More will be said about this momentarily.

Let's consider KG4 first. For KG4 the dynamical symmetries of the form Eq.~\eqref{GaugeVR} include:
\begin{flushleft}\begin{enumerate}
    \item[1)] action by $T^\epsilon_\text{j}$ sending $\bm{\Phi}\mapsto T^\epsilon_\text{j} \bm{\Phi}$. Similarly for $T^\epsilon_\text{n}$ and $T^\epsilon_\text{m}$,
    \item[2)] three negation symmetries which map basis vectors as \mbox{$\bm{e}_{j}\otimes\bm{e}_{n}\otimes\bm{e}_{m}\mapsto \bm{e}_{-j}\otimes\bm{e}_{n}\otimes\bm{e}_{m}$}, and \mbox{$\bm{e}_{j}\otimes\bm{e}_{n}\otimes\bm{e}_{m}\mapsto \bm{e}_{j}\otimes\bm{e}_{-n}\otimes\bm{e}_{m}$}, and \mbox{$\bm{e}_{j}\otimes\bm{e}_{n}\otimes\bm{e}_{m}\mapsto \bm{e}_{j}\otimes\bm{e}_{n}\otimes\bm{e}_{-m}$} respectively,
    \item[3)] a 4-fold symmetry which maps basis vectors as \mbox{$\bm{e}_{j}\otimes\bm{e}_{n}\otimes\bm{e}_{m}\mapsto \bm{e}_{j}\otimes\bm{e}_{m}\otimes\bm{e}_{-n}$},
    \item[4)] a local Fourier rescaling symmetry which maps \mbox{$\bm{\Phi}(\omega,k_1,k_2)\mapsto \tilde{f}(\omega,k_1,k_2)\,\bm{\Phi}(\omega,k_1,k_2)$} for some non-zero complex function $\tilde{f}(\omega,k_1,k_2)\in\mathbb{C}$ with $\omega,k_1,k_2\in[-\pi,\pi]$,
    \item[5)] local affine rescaling which maps \mbox{$\bm{\Phi}\mapsto \bm{\Phi} + \bm{c}_2$} for some $\bm{c}_2$ which is also a solution of the dynamics.
\end{enumerate}\end{flushleft}
These are exactly the symmetries which we found on our first interpretation (plus local Fourier rescaling) but with action by $T^\epsilon_\text{j}$, $T^\epsilon_\text{n}$ and $T^\epsilon_\text{m}$ replacing the discrete shifts. The same discussion following Eq.~\eqref{TDef} applies here, justifying us calling these continuous translation operations. Thus, KG4 has three continuous translation symmetries despite being initially represented on a lattice, Eq.~\eqref{KG4Long}.

Let's next consider KG5. For KG5 the dynamical symmetries of the form Eq.~\eqref{GaugeVR} include:
\begin{flushleft}\begin{enumerate}
    \item[1)] action by $T^\epsilon_\text{j}$ sending $\bm{\Phi}\mapsto T^\epsilon_\text{j} \bm{\Phi}$. Similarly for $T^\epsilon_\text{n}$ and $T^\epsilon_\text{m}$,
    \item[2)] a negation symmetry which maps basis vectors as \mbox{$\bm{e}_{j}\otimes\bm{e}_{n}\otimes\bm{e}_{m}\mapsto \bm{e}_{-j}\otimes\bm{e}_{n}\otimes\bm{e}_{m}$} and an exchange symmetry which maps basis vectors as \mbox{$\bm{e}_{j}\otimes\bm{e}_{n}\otimes\bm{e}_{m}\mapsto \bm{e}_{j}\otimes\bm{e}_{m}\otimes\bm{e}_{n}$},
    \item[3)] a 6-fold symmetry which maps basis vectors as \mbox{$\bm{e}_{j}\otimes\bm{e}_{n}\otimes\bm{e}_{m}\mapsto \bm{e}_{j}\otimes\bm{e}_{-m}\otimes\bm{e}_{n+m}$}. (Roughly, this permutes the three terms in Eq.~\eqref{DKG5}),
    \item[4)] a local Fourier rescaling symmetry which maps \mbox{$\bm{\Phi}(\omega,k_1,k_2)\mapsto \tilde{f}(\omega,k_1,k_2)\,\bm{\Phi}(\omega,k_1,k_2)$} for some non-zero complex function $\tilde{f}(\omega,k_1,k_2)\in\mathbb{C}$ with $t\in\mathbb{R}$ and $\omega,k_1,k_2\in[-\pi,\pi]$,
    \item[5)] local affine rescaling which maps \mbox{$\bm{\Phi}\mapsto \bm{\Phi} + \bm{c}_2$} for some $\bm{c}_2$ which is also a solution of the dynamics.
\end{enumerate}\end{flushleft}
These are exactly the symmetries which we found on our first interpretation (plus local Fourier rescaling) but with action by $T^\epsilon_\text{j}$, $T^\epsilon_\text{n}$ and $T^\epsilon_\text{m}$ replacing the discrete shifts. The same discussion following Eq.~\eqref{TDef} applies here, justifying us calling these continuous translation operations. Thus, KG5 has three continuous translation symmetries despite being initially represented on a lattice, Eq.~\eqref{KG5Long}.

Before moving on to analyze the symmetries of KG6 and KG7, let's first see what this interpretation has to say about them being equivalent to one another. As noted at the end of Sec.~\ref{SecSevenKG}, KG6 and KG7 have a nice one-to-one correspondence between their solutions. Allow me to spell this out in detail now.

Before this, however, it is worth briefly noting a rather weak sense in which each of KG4-KG7 are equivalent to each other. As noted following Eq.~\eqref{SkewKG7KG6} and Eq.~\eqref{SkewKG6KG7} there is an approximate one-to-one correspondence between each of these theories in the $\vert \omega\vert,\vert k_1\vert,\vert k_2\vert\ll\pi$ regime as they approach their common continuum limit, KG0. By contrast,  as I will show, KG6 and KG7 have an \textit{exact} one-to-one correspondence over \textit{the whole of} $\sqrt{k_1^2+k_2^2}<\pi$ and indeed more. This includes all of their solutions but not all of $\omega,k_1,k_2\in[-\pi,\pi]$.

This one-to-one correspondence is mediated by the transformations Eq.~\eqref{SkewKG7KG6} and Eq.~\eqref{SkewKG6KG7}. Let's first rewrite these in terms of \mbox{$\bm{\Phi}\in\mathbb{R}^L\cong\mathbb{R}^\mathbb{Z}\otimes\mathbb{R}^\mathbb{Z}\otimes\mathbb{R}^\mathbb{Z}$} as follows. Consider first the transformation which maps the dispersion relation for KG7 onto the one for KG6 (namely, Eq.~\eqref{SkewKG7KG6}). Consider its action on the planewave basis 
$\bm{\Phi}(\omega,k_1,k_2)$ with \mbox{$\omega,k_1,k_2\in[-\pi,\pi]$}, namely
\begin{align}
\nonumber
\Lambda_{\text{KG7}\to\text{KG6}}:\bm{\Phi}(\omega,k_1,k_2)\mapsto \bm{\Phi}\left(\omega,k_1,\frac{1}{2}k_1+\frac{\sqrt{3}}{2}k_2\right)  
\end{align}
A straight-forward calculation shows this acts on the \mbox{$\bm{e}_j\otimes\bm{e}_n\otimes\bm{e}_m$} basis as:
\begin{align}\label{SkewKG7KG6Basis}
&\Lambda_{\text{KG7}\to\text{KG6}}:\bm{e}_j\otimes\bm{e}_n\otimes\bm{e}_m\mapsto \\
\nonumber
&\bm{e}_j\otimes\left(\sum_{b_1,b_2\in\mathbb{Z}}S_n(b_1+b_2/2) \,S_m(\sqrt{3}\, b_2/2)\, \bm{e}_{b_1}\otimes\bm{e}_{b_2}\right).  
\end{align}
Consider also the transformation which maps the dispersion relation for KG6 onto the one for KG7 (namely, Eq.~\eqref{SkewKG6KG7}). Consider its action on the planewave basis 
$\bm{\Phi}(\omega,k_1,k_2)$ with \mbox{$\omega,k_1,k_2\in[-\pi,\pi]$}, namely as
\begin{align}
\nonumber
\Lambda_{\text{KG6}\to\text{KG7}}:\bm{\Phi}(\omega,k_1,k_2)\mapsto \bm{\Phi}\left(\omega,k_1,\frac{2k_2-k_1}{\sqrt{3}}\right)  
\end{align}
A straight-forward calculation shows this acts on the \mbox{$\bm{e}_j\otimes\bm{e}_n\otimes\bm{e}_m$} basis as:
\begin{align}
&\Lambda_{\text{KG6}\to\text{KG7}}:\bm{e}_j\otimes\bm{e}_n\otimes\bm{e}_m\mapsto \\
\nonumber
&\bm{e}_j\otimes\left(\sum_{b_1,b_2\in\mathbb{Z}}S_n(b_1-b_2/\sqrt{3}) \,S_m(2\, b_2/\sqrt{3})\, \bm{e}_{b_1}\otimes\bm{e}_{b_2}\right).  
\end{align}

It should be noted however that despite the fact that Eq.~\eqref{SkewKG7KG6} and Eq.~\eqref{SkewKG6KG7} are each other's inverses, $\Lambda_{\text{KG7}\to\text{KG6}}$ and $\Lambda_{\text{KG6}\to\text{KG7}}$ are not each other's inverses (at least not on the whole of \mbox{$\mathbb{R}^L\cong\mathbb{R}^\mathbb{Z}\otimes\mathbb{R}^\mathbb{Z}\otimes\mathbb{R}^\mathbb{Z}$}). This is due to the $2\pi$ periodic identification of the planewaves $\bm{\Phi}(\omega,k_1,k_2)$. Indeed, when viewed as acting on $\mathcal{V}=\mathbb{R}^L$, the transformation $\Lambda_{\text{KG7}\to\text{KG6}}$ is not even invertible. They are only each other's inverses when we have \mbox{$\omega,k_1,k_2\in[-\pi,\pi]$} both before and after these transformations.

For these reasons we need to consider the following two subspaces of \mbox{$\mathbb{R}^L\cong\mathbb{R}^\mathbb{Z}\otimes\mathbb{R}^\mathbb{Z}\otimes\mathbb{R}^\mathbb{Z}$}: 
\begin{align}
\mathbb{R}^L_\text{KG7}\!\coloneqq\text{span}(&\bm{\Phi}(\omega,k_1,k_2)\vert\text{with }\omega,k_1,k_2\in[-\pi,\pi]\\
\nonumber
&\text{ before and after applying Eq.~\eqref{SkewKG7KG6}})\\
\mathbb{R}^L_\text{KG6}\!\coloneqq\text{span}(&\bm{\Phi}(\omega,k_1,k_2)\vert\text{with }\omega,k_1,k_2\in[-\pi,\pi]\\
\nonumber
&\text{ before and after applying Eq.~\eqref{SkewKG6KG7}}).
\end{align}
For later reference it should be noted that the rotation invariant subspace,
\begin{align}\label{RLrotinv}
\mathbb{R}^L_\text{rot.inv}\coloneqq\text{span}\left(\bm{\Phi}(\omega,k_1,k_2)\Big\vert\,\sqrt{k_1^2+k_2^2}<\pi\right)
\end{align}
is a subspace of $\mathbb{R}^L_\text{KG6}$, that is $\mathbb{R}^L_\text{rot.inv}\subset \mathbb{R}^L_\text{KG6}$.

Restricted to $\mathbb{R}^L_\text{KG6}$ and $\mathbb{R}^L_\text{KG7}$ these transformations are invertible and indeed are each other's inverses. Fortunately, all of KG6's solutions are in $\mathbb{R}^L_\text{KG6}$ (and moreover they are in $\mathbb{R}^L_\text{rot.inv}$ as well). Similarly, all of KG7's solutions are in $\mathbb{R}^L_\text{KG7}$. Thus, $\Lambda_{\text{KG7}\to\text{KG6}}$ maps generic solutions to KG7 onto generic solutions for KG6 in an invertible way. Therefore, $\Lambda_{\text{KG6}\to\text{KG7}}$ and $\Lambda_{\text{KG7}\to\text{KG6}}$ give us not only a one-to-one correspondence between the solutions to KG6 and KG7 but a solution-preserving vector-space isomorphism between KG6 and KG7. One can gloss this situation saying: the KPMs of KG6 and KG7 are not isomorphic, but their DPMs are.

The fact that this is solution-preserving vector-space isomorphism rather than merely a one-to-one correspondence has substantial consequences for these theories' symmetries. Namely, this forces KG6 and KG7 to have the same symmetries. This is because these transformations are both of the form Eq.~\eqref{GaugeVR} (but notably not of the form Eq.~\eqref{Permutation}) for any symmetry transformation for KG6 there is a corresponding symmetry transformation for KG7 and vice versa. 

This is in strong contrast to the results of our previous analysis in Sec.~\ref{SecKlein1}. There KG6 was seen to have symmetries associated with a square 2D lattice and KG7 was seen to have the symmetries associated with a hexagonal 2D lattice. By contrast, in the present interpretation KG6 and KG7 are thoroughly equivalent: We have a solution-preserving vector-space isomorphism between them. Thus, on this interpretation the only difference between KG6 and KG7 is a change of basis.

Thus we have our second big lesson: despite the fact that KG6 and KG7 can be represented with very different lattice structures (i.e., a square lattice versus a hexagonal lattice) they have nonetheless turned out to be completely equivalent to one another. This equivalence was hidden from us on our first interpretation because we there took the lattice too seriously. As I will now discuss, this reduced their continuous rotation symmetries down to quarter rotations and one-sixth rotations respectively and thereby made them inequivalent. Here, we do not take the lattice structure so seriously. We have here internalized it into the value space where it subsequently disappears from view as just one basis among many.

In addition to switching between lattice structures, in this interpretation we can also do away with them altogether. In this interpretation, a lattice structure is associated with a choice of basis for $\mathcal{V}=\mathbb{R}^L$. A choice of basis (like a choice of coordinates) is ultimately a merely representational choice, reflecting no physics. We can always choose, if we like, to work within a basis-free formulation of these theories. That is, ultimately, a lattice-free formulation of these theories. Thus we have our third big lesson: given a discrete spacetime theory with some lattice structure we can always reformulate it in such a way that it has no lattice structure whatsoever.

In the rest of this subsection I will only discuss the symmetries KG6; analogous conclusions are true for KG7 after applying $\Lambda_{\text{KG6}\to\text{KG7}}$. For KG6 the dynamical symmetries of the form Eq.~\eqref{GaugeVR} include:
\begin{flushleft}\begin{enumerate}
    \item[1)] action by $T^\epsilon_\text{j}$ sending $\bm{\Phi}\mapsto T^\epsilon_\text{j} \bm{\Phi}$. Similarly for $T^\epsilon_\text{n}$ and $T^\epsilon_\text{m}$,
    \item[2)] three negation symmetries which map basis vectors as \mbox{$\bm{e}_{j}\otimes\bm{e}_{n}\otimes\bm{e}_{m}\mapsto \bm{e}_{-j}\otimes\bm{e}_{n}\otimes\bm{e}_{m}$}, and \mbox{$\bm{e}_{j}\otimes\bm{e}_{n}\otimes\bm{e}_{m}\mapsto \bm{e}_{j}\otimes\bm{e}_{-n}\otimes\bm{e}_{m}$}, and \mbox{$\bm{e}_{j}\otimes\bm{e}_{n}\otimes\bm{e}_{m}\mapsto \bm{e}_{j}\otimes\bm{e}_{n}\otimes\bm{e}_{-m}$} respectively,
    \item[3)] action by $R^\theta$  sending $\bm{\Phi}\mapsto R^\theta \bm{\Phi}$ with $R^\theta$ defined below. (This being a symmetry requires some qualification as I will discuss below.),
    \item[4)] action by $\Lambda^w_\text{j,n}$ sending $\bm{\Phi}\mapsto \Lambda^w_\text{j,n} \bm{\Phi}$ with $\Lambda^w_\text{j,n}$ defined below. Similarly for $\Lambda^w_\text{j,m}$ which is defined below as well. (This being a symmetry requires some qualification as I will discuss below.),
    \item[5)] a local Fourier rescaling symmetry which maps \mbox{$\bm{\Phi}(\omega,k_1,k_2)\mapsto \tilde{f}(\omega,k_1,k_2)\,\bm{\Phi}(\omega,k_1,k_2)$} for some non-zero complex function $\tilde{f}(\omega,k_1,k_2)\in\mathbb{C}$ with $\omega,k_1,k_2\in[-\pi,\pi]$,
    \item[6)] local affine rescaling which maps \mbox{$\bm{\Phi}\mapsto \bm{\Phi} + \bm{c}_2$} for some $\bm{c}_2$ which is also a solution of the dynamics.
\end{enumerate}\end{flushleft}
As with KG4 and KG5, we have here gained local Fourier rescaling and action by $T^\epsilon_\text{j}$, $T^\epsilon_\text{n}$ and $T^\epsilon_\text{m}$ has replaced the discrete shifts from before. The same discussion following Eq.~\eqref{TDef} applies here, justifying us calling these continuous translation operations. Thus, KG6 (and KG7) have three continuous translation symmetries despite being initially represented on a lattice.

Additionally, we have the quarter rotation symmetry from our first interpretation replaced with action by $R^\theta$, which as I will argue is essentially a continuous rotation transformation. Before that, it is worth noting a rather weak sense in which each of KG4-KG7 are rotation invariant. Each of these theories is approximately rotation invariant in the $\vert \omega\vert,\vert k_1\vert,\vert k_2\vert\ll\pi$ regime as they approach the continuum limit. By contrast, as I will show, KG6 is \textit{exactly} rotation invariant over \textit{the whole of} $\sqrt{k_1^2+k_2^2}<\pi$, that is the whole of $\mathbb{R}^L_\text{rot.inv.}$. Since all of KG6's solutions lie inside of $\mathbb{R}^L_\text{rot.inv.}$, $R^\theta$ will always map its solutions to solutions in an invertible way, and will hence be a symmetry.

This alleged continuous rotation transformation $R^\theta$ is given by 
\begin{align}\label{RthetaDef}
R^\theta \coloneqq \exp(-\theta (N_\text{n} D_\text{m}-N_\text{m} D_\text{n}))
\end{align}
with $\theta\in\mathbb{R}$ and where $N$ is the ``position operator'' which acts on the basis $\bm{e}_m$ as $N\bm{e}_m=m\,\bm{e}_m$ for $m\in\mathbb{Z}$. Thus \mbox{$N_\text{n}=\openone\otimes N\otimes\openone$} and \mbox{$N_\text{m}=\openone\otimes \openone\otimes N$} return the second and third index respectively when acting on $\bm{e}_j\otimes \bm{e}_n\otimes \bm{e}_m$.

A straight-forward calculation shows that $R^\theta$ acts on the planewave basis $\bm{\Phi}(\omega,k_1,k_2)$ with \mbox{$\omega,k_1,k_2\in[-\pi,\pi]$} as
\begin{align}
R^\theta:\,&\bm{\Phi}(\omega,k_1,k_2) \mapsto\\
\nonumber
&\bm{\Phi}(\omega,\cos(\theta)k_1-\sin(\theta)k_2,\sin(\theta)k_1+\cos(\theta)k_2).
\end{align}
and acts on the basis $\bm{e}_j\otimes \bm{e}_n\otimes \bm{e}_m$ as
\begin{align}
&R^\theta: \bm{e}_j\!\otimes\bm{e}_n\!\otimes\bm{e}_m \mapsto\bm{e}_{j}\otimes\!\!\!\!\sum_{b_1,b_2\in\mathbb{Z}} R_{nm}^{b_1 b_2}(\theta)\,
\bm{e}_{b_1}\!\!\otimes\bm{e}_{b_2}\\
\nonumber
&R_{nm}^{b_1 b_2}\!(\theta)\!=\!S_n(\cos(\theta) b_1 \!-\! \sin(\theta) b_2) S_m(\sin(\theta) b_1\!+\! \cos(\theta)b_2).
\end{align}

It should be noted that $R^\theta$ is not invertible (at least not on the whole of \mbox{$\mathbb{R}^L\cong\mathbb{R}^\mathbb{Z}\otimes\mathbb{R}^\mathbb{Z}\otimes\mathbb{R}^\mathbb{Z}$}). As I will soon discuss, this is due to the $2\pi$ periodic identification of the planewaves $\bm{\Phi}(\omega,k_1,k_2)$. However, $R^\theta$ is invertible over $\mathbb{R}^L_\text{rot.inv.}$ which contains all of KG6's solutions. Thus, for KG6 $R^\theta$ always maps solutions to solutions in an invertible way, and is hence a symmetry.

To see why $R^\theta$ is not invertible over all of $\mathcal{V}=\mathbb{R}^L$ note that $R^{\pi/4}$ maps two different planewaves to the same place: Firstly note,
\begin{align}
R^{\pi/4}\bm{\Phi}(\omega,\pi,\pi)&=\bm{\Phi}(\omega,0,\sqrt{2}\pi)\\
\nonumber
&=\bm{\Phi}(\omega,0,\sqrt{2}\pi-2\pi)
\end{align}
since the planewaves repeat themselves with period $2\pi$. Secondly note, 
\begin{align}
\nonumber
R^{\pi/4}\bm{\Phi}(\omega,\pi-\sqrt{2}\pi,\pi-\sqrt{2}\pi)&=\bm{\Phi}(\omega,0,\sqrt{2}\pi-2\pi).
\end{align}
Such issues do not arise when $\sqrt{k_1^2+k_2^2}<\pi$. Thus, when we restrict our attention to $\mathbb{R}^L_\text{rot.inv}$ (which contains all of KG6's solutions) then $R^\theta$ is invertible and indeed a symmetry. One can gloss this situation saying: rotation does not map the KPMs of KG6 onto themselves in an invertible way, but it does for the DPMs of KG6. If we cut the KPMs of KG6 down to $\mathbb{R}^L_\text{rot.inv}$ this minor issue is fixed.

As I will now discuss, $R^\theta$ can be thought of as a continuous rotation operator for three reasons despite it being here an internal symmetry. First note that $R^\theta$ is a generalization of quarter rotation operation in the sense that taking $\theta=\pi/2$ reduces action by $R^\theta$ to the map \mbox{$R^{\pi/2}:\bm{e}_{j}\otimes\bm{e}_{n}\otimes\bm{e}_{m}\mapsto \bm{e}_{j}\otimes\bm{e}_{m}\otimes\bm{e}_{-n}$} on basis vectors and relatedly the map $(j,n,m)\mapsto (j,m,-n)$ on lattice sites.

Second, note that restricted to $\mathbb{R}^L_\text{rot.inv.}$, $R^\theta$ is cyclically additive in the sense that $R^{\theta_1}\,R^{\theta_2}=R^{\theta_1+\theta_2}$ with $R^{2\pi}=\openone$. In the language of representation theory, $R^\theta$ is a representation of the rotation group on the vector space $\mathbb{R}^L_\text{rot.inv.}$. In particular, this means \mbox{$R^{\pi/4}\,R^{\pi/4}=R^{\pi/2}$}.  There is something we can do twice to make a quarter rotation. Similarly for all fractional rotations. Moreover, note that together with the above discussed translations, these constitute a representation of the Euclidean group over $\mathbb{R}^L_\text{rot.inv.}$.

Third, recall from the discussion following Eq.~\eqref{LambdaD} that $D$ is closely related to the continuum derivative operator $\partial_x$, exactly matching its spectrum for $k\in[-\pi,\pi]$. Recall also that rotations are generated through the derivative as \mbox{$h(R^\theta(x,y))= \exp(-\theta (x \partial_y-y \partial_x))h(x,y)$}. In this sense also $R^\theta$ is a rotation operator. More will be said about $R^\theta$ in Appendix~\ref{AppA}.

This adds to our first big lesson: despite the fact that KG6 and KG7 can be represented on a cubic 3D lattice and a hexagonal 3D lattice respectively, they nonetheless both have a continuous rotation symmetry. This, in addition to their continuous translation symmetries. These continuous translation and rotations symmetries were hidden from us on our first interpretation because we there took the lattice representations too seriously. Here, we do not take the lattice structure so seriously. Instead, we have internalized it into the value space where it then disappears from view as just one basis among many.

Let's next discuss KG6's (limited) Lorentzian boost symmetry. In addition to generalizing the 4-fold symmetry into to a continuous rotation symmetry, KG6 also has a brand new symmetry on this interpretation, namely ``action by $\Lambda^w_\text{j,n}$ and/or $\Lambda^w_\text{j,m}$''. As I will argue these are essentially Lorentz boost transformations. Before that, it is worth noting a rather weak sense in which each of KG4-KG7 are Lorentz invariant. Each of these theories is approximately Lorentz invariant as they approach the continuum limit regime $\vert \omega\vert,\vert k_1\vert,\vert k_2\vert\ll\pi$ at least boosts parameters $w$ which keep them in this regime. By contrast, as I will show, KG6 is \textit{exactly} Lorentz invariant over \textit{a finite-sized region} around $\omega,k_1,k_2=0$ and boost parameter $w=0$.

This alleged Lorentz boost transformations are given by
\begin{align}\label{LambdaWDef}
\Lambda^w_\text{j,n}\coloneqq\exp(-w (&N_\text{j} D_\text{n} + N_\text{n} D_\text{j})),\\
\nonumber
\Lambda^w_\text{j,m}\coloneqq\exp(-w (&N_\text{j} D_\text{m} + N_\text{m} D_\text{j}))
\end{align}
with $w\in\mathbb{R}$. Note that $\Lambda^w_\text{j,n}$ acts only on the first and second tensor factor, whereas $\Lambda^w_\text{j,m}$ acts only on the first and third factors. In what follows I will focus on $\Lambda^w_\text{j,n}$, with similar results following for $\Lambda^w_\text{j,m}$.

A straight-forward calculation shows that $\Lambda^w_\text{j,n}$ acts on the planewave basis $\bm{\Phi}(\omega,k_1,k_2)$ with \mbox{$\omega,k_1,k_2\in[-\pi,\pi]$} as
\begin{align}
\Lambda^w_\text{j,n}:\,&\bm{\Phi}(\omega,k_1,k_2) \mapsto\\
\nonumber
&\bm{\Phi}(\cosh(w)\omega+\sinh(w)k_1,\sinh(w)\omega+\cosh(w)k_1,k_2).
\end{align}
This is a Lorentz boost in discrete Fourier space. It follows from this that $\Lambda^w_\text{j,n}$ acts on the basis $\bm{e}_j\otimes \bm{e}_n\otimes\bm{e}_m$ as
\begin{align}
\nonumber
&\Lambda^w_\text{j,n}: \bm{e}_j\otimes\!\bm{e}_n\otimes\!\bm{e}_m\mapsto\!\!\!\!\sum_{b_1,b_2\in\mathbb{Z}} \Lambda_{jn}^{b_1 b_2}(w)\,
\bm{e}_{b_1}\otimes\!\bm{e}_{b_2}\otimes\!\bm{e}_m\\
&\Lambda_{jn}^{b_1 b_2}(w)=S_j(\cosh(w) b_1 + \sinh(w) b_2)\\ 
\nonumber
&\qquad\qquad\times S_n(\sinh(w) b_1+\cosh(w) b_2).
\end{align}
Like with $R^\theta$ discussed above, $\Lambda^w_\text{j,n}$ is a symmetries of the dynamics in a qualified sense: namely, $\Lambda^w_\text{j,n}$ is not invertible (at least not on the whole of \mbox{$\mathbb{R}^L\cong\mathbb{R}^\mathbb{Z}\otimes\mathbb{R}^\mathbb{Z}\otimes\mathbb{R}^\mathbb{Z}$}). Indeed, each of these transformations is only invertible for a subspace of \mbox{$\mathbb{R}^L\cong\mathbb{R}^\mathbb{Z}\otimes\mathbb{R}^\mathbb{Z}\otimes\mathbb{R}^\mathbb{Z}$}. As before, this is due to the $2\pi$ periodic identification of the planewaves $\bm{\Phi}(\omega,k_1,k_2)$. To see this note that $\Lambda^{\ln(2)}_\text{j,n}$ maps two different planewaves to the same place: Firstly note,
\begin{align}
\Lambda^{\ln(2)}_\text{j,n}\bm{\Phi}(\pi,\pi,k_2)&=\bm{\Phi}(2\pi,2\pi,k_2)\\
\nonumber
&=\bm{\Phi}(0,0,k_2)
\end{align}
since the planewaves repeat themselves with period $2\pi$. Secondly note, 
\begin{align}
\nonumber
\Lambda^{\ln(2)}_\text{j,n}\bm{\Phi}(0,0,k_2)&=\bm{\Phi}(0,0,k_2)
\end{align}
Such issues do not arise when we have \mbox{$\omega,k_1,k_2\in[-\pi,\pi]$} both before and after the Lorentz transformation. 

Thus, $\Lambda^{w}_\text{j,n}$ and $\Lambda^{w}_\text{j,m}$ are invertible over the portion of \mbox{$\mathbb{R}^L\cong\mathbb{R}^\mathbb{Z}\otimes\mathbb{R}^\mathbb{Z}\otimes\mathbb{R}^\mathbb{Z}$} spanned by planewaves $\bm{\Phi}(\omega,k_1,k_2)$ which satisfy the following two conditions:
\begin{align}\label{BoostCond}
\vert\cosh(w)\,\omega+\sinh(w)\, k_1\,\vert<&\pi\\
\nonumber
\vert\sinh(w)\,\omega+\cosh(w)\, k_2\,\vert<&\pi.
\end{align}

Unfortunately, unlike with $R^\theta$ this issue cannot be so easily contained: the region where both $\Lambda^{w}_\text{j,n}$ and $\Lambda^{w}_\text{j,m}$ are invertible depends on $w$ and indeed, shrinks to nothing as $w\to\pm\infty$. When boosted enough, any planewave except $\omega=k_1=k_2=0$ will leave the region \mbox{$\omega,k_1,k_2\in[-\pi,\pi]$}. Moreover, for any $w$ there is some planewave in \mbox{$\omega,k_1,k_2\in[-\pi,\pi]$} which when boosted by $w$ leaves this region. However, despite this, for every planewave in $\omega,k_1,k_2\in(-\pi,\pi)$ there is some small enough boost $w\neq0$ which keeps it in \mbox{$\omega,k_1,k_2\in[-\pi,\pi]$}. Moreover, for every \mbox{$w$} there is some neighborhood around \mbox{$\omega,k_1,k_2=0$} where boosting by $w$ leaves us in \mbox{$\omega,k_1,k_2\in[-\pi,\pi]$}. When restricted to acting on the span of these planewaves $\Lambda^{w}_\text{j,n}$ and $\Lambda^{w}_\text{j,n}$ are both invertible.

It is in the following sense that $\Lambda^{w}_\text{j,n}$ and $\Lambda^{w}_\text{j,m}$ are symmetries of KG6. Consider solutions to KG6 which are supported only over planewaves in some finite neighborhood of $\omega,k_1,k_2=0$. Consider $\Lambda^{w}_\text{j,n}$ and $\Lambda^{w}_\text{j,m}$ with $w$ in some finite neighborhood of $w=0$. Choose the neighborhoods such that Eq.~\eqref{BoostCond} is satisfied throughout. Restricting our attention to this regime, all of these transformations map exactly map solutions to solutions in an invertible way. As I claimed above, KG6 is \textit{exactly} Lorentz invariant over \textit{a finite-sized region} around $\omega,k_1,k_2=0$ and boost parameter $w=0$.

Recall as mentioned above that for all planewaves in $\omega,k_1,k_2\in(-\pi,\pi)$ there is some small enough boost \mbox{$w\neq0$} which keeps it in \mbox{$\omega,k_1,k_2\in[-\pi,\pi]$}. Thus, KG6 has a differential Lorentz boost invariance over all of \mbox{$\omega,k_1,k_2\in(-\pi,\pi)$}. Note that the same is true of translations and rotations. Including translations and rotations, KG6 has a differential Poincar\'e invariance over $\omega,k_1,k_2\in(-\pi,\pi)$. Concretely, taken all together $\Lambda^w_\text{j,n}$, $\Lambda^w_\text{j,m}$, $R^\theta$, $T^\epsilon_\text{j}$, $T^\epsilon_\text{n}$ and $T^\epsilon_\text{m}$ form a representation of the Poincar\'e algebra over the space spanned by \mbox{$\omega,k_1,k_2\in(-\pi,\pi)$}. In particular, this means that for every algebraic fact about the differential Poincar\'e transformations there is an analogous fact here with the group action being replaced by matrix multiplication. Exponentiating this representation of the Poincar\'e algebra we recover a finite part of the Poincar\'e group. Namely, the finite-sized collection of states and transformations satisfying:
\begin{align}\label{PoincareCond}
\vert\cosh(w)\,\omega+\sinh(w)\,\vert k\vert\,\vert<&\pi\\
\nonumber
\vert\sinh(w)\,\omega+\cosh(w)\,\vert k\vert\,\vert<&\pi\\
\nonumber
\vert\cos(\theta)\,k_1-\sin(\theta) k_2\vert<&\pi\\
\nonumber
\vert\sin(\theta)\,k_1+\cos(\theta) k_2\vert<&\pi.
\end{align}
Indeed, KG6 is \textit{exactly} Poincar\'e invariant over \textit{a finite-sized region} around $\omega,k_1,k_2=0$ and $w=\theta=0$.

As I will now discuss, in this regime $\Lambda^w_\text{j,m}$ and $\Lambda^w_\text{j,n}$ can be thought of as implementing Lorentzian boosts for two reasons despite being here categorized as an internal symmetry. Firstly, recall the close relationship noted above between $D$ and $\partial_x$. Recall also that Lorentz boosts are generated through the derivative as \mbox{$h(\Lambda^w(t,x))= \exp(-w (x \partial_t + t \partial_x))h(t,x)$}. 

Secondly, as I have already mentioned, together with our above discussed translations and rotations,  $\Lambda^w_\text{j,m}$ and $\Lambda^w_\text{j,n}$ give us a representation of the Poincar\'e algebra and even a finite portion of the Poincar\'e group.

Thus we have yet another addendum to our first big lesson: despite the fact that KG6 and KG7 can be represented on a cubic 3D lattice and a hexagonal 3D lattice respectively, they nonetheless both have a Lorentzian boost symmetries (in a finite but limited regime). This, in addition to their continuous translation and rotation symmetries. As impressive as this admittedly limited Lorentz symmetry is, we can do better: In Sec.~\ref{PerfectLorentz} I will provide a perfectly Lorentzian lattice theory.

Finally, let's consider KG3. For KG3 the dynamical symmetries of the form Eq.~\eqref{GaugeVR} include:
\begin{flushleft}\begin{enumerate}
    \item[1)] action by $T^\epsilon_\text{j}$ sending $\bm{\Phi}\mapsto T^\epsilon_\text{j} \bm{\Phi}$. Similarly for $T^\epsilon_\text{n}$.
    \item[2)] two negation symmetries which map basis vectors as \mbox{$\bm{e}_{j}\otimes\bm{e}_{n}\mapsto \bm{e}_{-j}\otimes\bm{e}_{n}$} and \mbox{$\bm{e}_{j}\otimes\bm{e}_{n}\mapsto \bm{e}_{j}\otimes\bm{e}_{-n}$} respectively,
    \item[3)] action by $\Lambda^w_\text{jn}$ sending $\bm{\Phi}\mapsto \Lambda^w_\text{jn} \bm{\Phi}$ where \mbox{$\Lambda^w_\text{jn}$} is defined above,
    \item[4)] a local Fourier rescaling symmetry which maps \mbox{$\bm{\Phi}(\omega,k)\mapsto \tilde{f}(\omega,k)\,\bm{\Phi}(\omega,k)$} for some non-zero complex function $\tilde{f}(\omega,k)\in\mathbb{C}$ with $\omega,k\in[-\pi,\pi]$,
    \item[5)] local affine rescaling which maps \mbox{$\bm{\Phi}\mapsto \bm{\Phi} + \bm{c}_2$} for some $\bm{c}_2$ which is also a solution of the dynamics.
\end{enumerate}\end{flushleft}
As with KG1 and KG2 we have gained local Fourier rescaling and discrete shifts have been replaced with action by $T^\epsilon$. As discussed above we are justified in calling $T^\epsilon$ a continuous translation operation. For KG3 we also have a new symmetry, namely action by $\Lambda^w_\text{jn}$. For the reasons discussed above, this can be thought of as a (limited) Lorentzian boost symmetry despite it being here classified as an internal symmetry.

\vspace{0.25cm}
To summarize: this second attempt at interpreting KG1-KG7 has fixed all of the issues with our previous interpretation. Firstly, there is no longer any tension between these theories' differing locality properties and the rates at which they converge to the (perfectly local) continuum theory in the continuum limit. (There are no longer any differences in locality.) See \cite{DiscreteGenCovPart1} for further discussion. Secondly, the fact that we have a nice one-to-one correspondence between the solutions to KG6 and KG7 is now more satisfyingly reflected in their matching symmetries. Finally, this interpretation has exposed the fact that KG1-KG7 have hidden continuous translation and rotation symmetries as well as a (limited) Lorentzian boost symmetries.

By and large, the interpretation invalidates all of the first intuitions laid out in Sec.~\ref{SecIntro}. As this interpretation has revealed, the lattice seems to play a merely representational role in the theory: it does not restrict our symmetries. Moreover, theories initially appearing with different lattice structures may nonetheless turn out to be completely equivalent. The process for switching between lattice structures is here a change of basis in the value space. Indeed, we have a third lesson: there is no sense in which these lattice structures are essentially ``baked-into'' these theories; on this interpretation our theories make no reference to any lattice structure if we work in a basis-independent way. No basis is dynamically favored. 

As I discussed in \cite{DiscreteGenCovPart1} these three lessons lay the foundation for a strong analogy between the lattice structures which appear in our discrete spacetime theories and the coordinate systems which appear in our continuum spacetime theories. This analogy here extended to Lorentzian theories.

These are substantial lessons, but ultimately this interpretation has a few issues of its own. Firstly, the way that the tension is dissolved between locality and convergence in the continuum limit is unsatisfying. Intuitively, we ought to be able to extract intuitions about locality from the lattice sites. 

Moreover, while this interpretation has indeed exposed KG1-KG7's hidden symmetries, the way it classifies them seems wrong. They are here classified as internal symmetries (i.e., symmetries on the value space) whereas intuitively they should be external symmetries (i.e., symmetries on the manifold).

The root of all of these issues is taking the theory's lattice structure to be internalized into the theory's value space. Our third attempt at interpreting these theories will fix this by externalizing these symmetries. As I will discuss, this gives us access to a perspective within which we can find a perfectly Lorentzian lattice theory (i.e., not limited as the above theories are).

\section{Externalizing these theories - Part 1}\label{SecExtPart1}
In the previous section it was revealed that KG1-KG7 have hidden continuous symmetry transformations which intuitively correspond to spacial translations and rotation and even (limited) Lorentzian boosts. In our first attempt at interpreting KG1-KG7 the possibility of such symmetries were outright denied, see Sec.~\ref{SecKlein1}. In our second attempt, these hidden symmetries were exposed, but they were classified (unintuitively) as internal symmetries, see Sec.~\ref{SecKlein2}. This is due to an ``internalization'' move made in our second interpretation. This move also had the unfortunate consequence of undercutting our ability to use the lattice sites to reason about locality.

In this section I will show how we can externalize these symmetries by in a principled way 1) inventing a continuous manifold for our formerly discrete theories to live on and 2) embedding our theory's states/dynamics onto this manifold as a new dynamical field.

\subsection{A Principled Choice of Spacetime Manifold}\label{SecKlein3A}
If we are going to externalize these symmetries then we need to have a big enough manifold on which to do the job. What spacetime manifold $\mathcal{M}$ might be up to the task? The first thing we must do is pick out which of our theory's symmetries we would like to externalize (there may be some symmetries we want to keep internal). For KG1-KG7 we want to externalize the following symmetries: continuous translations, mirror reflections, as well as  discrete rotations for KG4 and KG5 and continuous rotations for KG6 and KG7. For each theory we can collect these dynamical symmetries together in a group $G^\text{dym}_\text{to-be-ext}$. Clearly, our choice of spacetime manifold $\mathcal{M}$ needs to be big enough to have $G^\text{dym}_\text{to-be-ext}$ as a subgroup of $\text{Diff}(\mathcal{M})$. Let us call this the symmetry-fitting constraint\footnote{As I will soon discuss, there may be transformations which we want to externalize even when they are not symmetries.}. 

Of course, symmetry-fitting alone doesn't uniquely specify which manifold we ought to use. If $\mathcal{M}$ works, then so does any larger $\mathcal{M}'$ with $\mathcal{M}$ as a sub-manifold. For standard Occamistic reasons, it is natural to go with the smallest manifold which gets the job done. The larger the gap between the groups $G^\text{dym}_\text{to-be-ext}$ and $\text{Diff}(\mathcal{M})$ the more fixed spacetime structures will need to be introduced later on (see Sec.~\ref{SecFullGenCov}).

In principle, we are free to pick any large-enough manifold which we like to embed KG1-KG7 onto. However, perhaps surprisingly, if we make natural choices about how the translation operations we have already identified (see Eq.~\eqref{TDef}) fit onto the new spacetime manifold then our choice of $\mathcal{M}$ is actually fixed up to diffeomorphism. In particular, I demand the following: Certain translation operations on $\mathbb{R}^L$ are to correspond (perhaps in a complicated way) to parallel transport on the new spacetime manifold $\mathcal{M}$. For KG1-KG3 these to-be-externalized translation operations are $T^{\epsilon_1}_\text{j}$ and $T^{\epsilon_2}_\text{n}$. For KG4-KG7 these to-be-externalized translation operations are $T^{\epsilon_1}_\text{j}$, $T^{\epsilon_2}_\text{n}$, and 
$T^{\epsilon_3}_\text{m}$. Let us call this the translation-matching constraint. As I will soon show, this constraint fixes the new spacetime manifold $\mathcal{M}$ up to diffeomorphism.

Before fleshing this out, however, it's worth reflecting on two questions focusing on KG4-KG7: What exactly makes $T^{\epsilon_1}_\text{j}$, $T^{\epsilon_2}_\text{n}$, and $T^{\epsilon_3}_\text{m}$ translation operations? Moreover, what motivation do we have to externalize these particular translation operations? To answer: firstly, these can be thought of as translation operations for the reasons discussed following Eq.~\eqref{TDef}. Note these reasons are unrelated to the fact that these are dynamical symmetries of KG4-KG7. Suppose the dynamics of KG4 given by Eq.~\eqref{KG4Long} was modified to have explicit dependence on the index $n$. In this case, $T^{\epsilon_2}_\text{n}$ would no longer be a dynamical symmetry but it would still be a translation operation (and moreover, one worth externalizing).

Secondly, why externalize these translation operations in particular? As I will now discuss, any motivation for externalizing these particular translation operations must come from the dynamics. Forgoing any dynamical considerations, all we can say about KG1-KG7 is that they concern vectors $\bm{\Phi}\in\mathbb{R}^L$ (or alternatively functions \mbox{$\phi:L\to\mathbb{R}$} in $F_L$). Recall that pre-dynamics the set of labels for lattice sites $L$ is uncountable but otherwise unstructured; We might index it using any number of indices we like, \mbox{$L\cong\mathbb{Z}\cong\mathbb{Z}^2\cong\dots\cong\mathbb{Z}^{17}\cong\dots$}. Using each of these re-indexings we can grant $\mathbb{R}^L$ different tensor product structures, \mbox{$\mathbb{R}^L\cong\mathbb{R}^Z\cong\mathbb{R}^Z\otimes\mathbb{R}^Z\cong\dots$}. In each tensor factor we can define a translation operation $T^\epsilon$ as in Eq.~\eqref{TDef}. Thus, the vector space $\mathbb{R}^L$ supports representations of: the 1D translation group, the 2D translation group, $\dots$, the 17D translation group, etc. Given all these possibilities, why externalize $T^{\epsilon_1}_\text{j}$, $T^{\epsilon_2}_\text{n}$, and $T^{\epsilon_3}_\text{m}$ in particular? The answer must come from the dynamics.

One good reason to consider $T^{\epsilon_1}_\text{j}$, $T^{\epsilon_2}_\text{n}$, and $T^{\epsilon_3}_\text{m}$ worthy of externalization is that they are dynamical symmetries of KG4-KG7. However, as mentioned above, something might be a translation operation (and moreover, one worthy of externalizing) even if it's not a dynamical symmetry. Unfortunately, I don't have a perfect rule for how to identify such cases. The best I can offer is to check whether its associated derivative (or some function thereof) appears in the dynamical equations.

Regardless, its sufficiently clear for KG4-KG7 that $T^{\epsilon_1}_\text{j}$, $T^{\epsilon_2}_\text{n}$, and 
$T^{\epsilon_3}_\text{m}$ are among the translation operations worthy of externalizing. Our translation-matching constraint suggests that these ought to correspond (perhaps in a complicated way) to parallel transport on the new spacetime manifold $\mathcal{M}$. I claim that (at least in this case) this constraint fixes the spacetime manifold $\mathcal{M}$ up to diffeomorphism.

To show this I will first pick out at each point \mbox{$p\in\mathcal{M}$} on the manifold three independent directions in the tangent space at $p$. To realize the translation-matching constraint, I demand that differential translation by $T^{\epsilon_1}_\text{j}$, $T^{\epsilon_2}_\text{n}$, and $T^{\epsilon_3}_\text{m}$ is then to be carried out on the manifold as parallel transport in these three directions. Note that this already implies that the dimension of $\mathcal{M}$ is at least three.

Indeed, in this case we have good reason to take $\mathcal{M}$ to be exactly three dimensional. To see this, note that 
\begin{align}\label{RLCover}
\mathbb{R}^L&\cong \mathbb{R}^\mathbb{Z}\otimes\mathbb{R}^\mathbb{Z}\otimes\mathbb{R}^\mathbb{Z}\\
&=\text{span}(T^{\epsilon_1}_\text{j}  T^{\epsilon_2}_\text{n}  T^{\epsilon_3}_\text{m} \, \bm{e}_0\!\otimes\bm{e}_0\!\otimes\bm{e}_0\vert (\epsilon_1,\epsilon_2,\epsilon_3)\in\mathbb{R}^3)
\end{align}
That is, beginning from $\bm{e}_0\otimes\bm{e}_0\otimes\bm{e}_0$ these translations cover all of $\mathbb{R}^L\cong\mathbb{R}^\mathbb{Z}\otimes\mathbb{R}^\mathbb{Z}\otimes\mathbb{R}^\mathbb{Z}$. That is, they cover all of our state space in the second interpretation. The three translations $T^{\epsilon_1}_\text{j}$, $T^{\epsilon_2}_\text{n}$, and $T^{\epsilon_3}_\text{m}$ are thus already enough to cover every kinematic possibility. We need no further dimensions and so by Occam's razor we may then take the spacetime manifold $\mathcal{M}$ to be three dimensional. 

But how does the translation-matching constraint (which I have yet to state technically) fix $\mathcal{M}$ up to diffeomorphism? As I will now discuss, (at least in this case) $\mathcal{M}$ it is fixed up to diffeomorphism by the group theoretic properties of $T^{\epsilon_1}_\text{j}$, $T^{\epsilon_2}_\text{n}$, and $T^{\epsilon_3}_\text{m}$. Let $G_\text{trans.}$ be the group formed by all compositions of $T^{\epsilon_1}_\text{j}$, $T^{\epsilon_2}_\text{n}$, and $T^{\epsilon_3}_\text{m}$. Note that the group $G_\text{trans.}$ is a Lie group, and hence is also a differentiable manifold.

Recall that so far the translation-matching constraint associated to each of our differential translations, a direction in the tangent space of each point $p\in\mathcal{M}$. Since $G_\text{trans.}$ contains nothing but these translations, this gives us a one-to-one correspondence between the algebra $g_\text{trans.}$ of $G_\text{trans.}$ and (at least a subspace of) the of the tangent space of each point on the manifold. Let's take these tangent vectors to very smoothly across the manifold. Let $\mathcal{M}_\text{trans.}(p)$ be the submanifold of points on the spacetime manifold reachable from $p\in\mathcal{M}$ by following these tangent vectors. The algebra $g_\text{trans.}$ is related to the group $G_\text{trans.}$ in the same way that this tangent space it related to $\mathcal{M}_\text{trans.}(p)$, namely by repeated application of the exponential map. Hence, I take translation-matching to demand that $\mathcal{M}_\text{trans.}(p)\cong G_\text{trans.}$ as differentiable manifolds for every $p\in\mathcal{M}$. 

As discussed above, we have reason to take $\mathcal{M}$ to be three dimensional. Hence (assuming that $\mathcal{M}$ is connected) any three dimensional submanifold (e.g. $\mathcal{M}_\text{trans.}(p)$) must in fact be the entire manifold. Therefore we have $\mathcal{M}\cong G_\text{trans.}$ as differentiable manifolds.

But what can we say about this translation group $G_\text{trans.}$ viewed as a manifold? Notice that these three continuous translation operations comprising $G_\text{trans.}$ all commute with each other. As such for any combination of them $T_\text{generic}\in G_\text{trans.}$ we always have a unique factorization of the form,
\begin{align}\label{Tgeneric}
T_\text{generic} = T^{\epsilon_1}_\text{j} \ T^{\epsilon_2}_\text{n} \ T^{\epsilon_3}_\text{m}
\end{align}
Moreover, this factorization represents all of $G_\text{trans.}$ without redundancy. Indeed, we can smoothly parameterize all of $G_\text{trans.}$ using parameters \mbox{$(\epsilon_1,\epsilon_2,\epsilon_3)\in\mathbb{R}^3$} in a one-to-one way. That is, $G_\text{trans.}\cong\mathbb{R}^3$ as differentiable manifolds. Therefore, we have $\mathcal{M}\cong\mathbb{R}^3$ as well. Note this means that we have access to a global coordinate system for $\mathcal{M}$.

Thus, from the translation-matching constraint alone we are forced to take $\mathcal{M}\cong\mathbb{R}^3$ for KG4-KG7. For KG1-KG3 the same translation-matching constraint forces us to take $\mathcal{M}\cong\mathbb{R}^2$. Note that in both cases these manifolds satisfy the symmetry-fitting constraint: $\text{Diff}(\mathcal{M})$ is large enough to contain $G^\text{dyn}_\text{to-be-ext}$.

\subsection{A Principled Choice of Embedding}
Now that we have a spacetime manifold selected, we need to somehow embed $\bm{\Phi}$ (or equivalently $\phi_\ell$) into it. Here I will begin with $\bm{\Phi}$ with the approach from $\phi_\ell$ being addressed in Sec.~\ref{SecExtPart2}. The goal in either case is to construct in a principled way from either $\bm{\Phi}$ or $\phi_\ell$ a new field \mbox{$\phi:\mathcal{M}\to\mathbb{R}$} defined over this manifold. 

Before this, however, recall that our move from the first to the second interpretation was mediated by means of a vector-space isomorphism $F_L\cong\mathbb{R}^L$ namely Eq.~\eqref{PhiDef}. Here too, our reinterpretation will be mediated by a vector-space isomorphism.

Let $F(\mathcal{M})$ be the set of all real scalar functions \mbox{$f:\mathcal{M}\to\mathbb{R}$}. Notice that this is a vector space as it is closed under addition and scalar multiplication. Our goal here is to in a principle way construct an isomorphism between \mbox{$\mathbb{R}^L$} and some vector subspace of \mbox{$F(\mathcal{M})$}. Indeed, $F(\mathcal{M})$ is much larger than $\mathbb{R}^L$ (having an uncountably infinite dimension). As such we will only ever map $\mathbb{R}^L$ onto some subset of it, $F\subset F(\mathcal{M})$. In order for us to have a vector space isomorphism between $\mathbb{R}^L$ and $F$ the following conditions must be satisfied. $F$ must be: 1) closed under addition and scalar multiplication and hence a vector space, and 2) countably infinite dimensional and hence isomorphic to $\mathbb{R}^L$, i.e., \mbox{$F\cong\mathbb{R}^\mathbb{Z}\cong\mathbb{R}^L$}. With the target subset characterized, the embedding of $\bm{\Phi}$ onto $\mathcal{M}$ is then accomplished by picking a vector-space isomorphism $E:\mathbb{R}^L\to F$. The embedded field is then gives to us as $\phi\coloneqq E(\bm{\Phi})$.

It appears we have a great deal of freedom here. Suppose that, in line with the translation-matching constraint discussed above, we take $\mathcal{M}\cong\mathbb{R}^2$ for KG1-KG3 and $\mathcal{M}\cong\mathbb{R}^3$ for KG4-KG7. In this case, we still have great freedom in picking both the target vector space $F\subset F(\mathcal{M})$ and the isomorphism $E:\mathbb{R}^L\to F$. However, as I will now discuss, translation-matching also drastically limits these choices as well. Allow me to demonstrate with KG4-KG7.

First, allow me to explicitly define a coordinate system for $\mathcal{M}\cong G_\text{trans.}\cong\mathbb{R}^3$. As discussed above we already have an explicit smooth parametrization of $G_\text{trans.}$ via \mbox{$(\epsilon_1,\epsilon_2,\epsilon_3)\in\mathbb{R}^3$}, see Eq.~\eqref{Tgeneric}. In addition to this, concretely realizing our translation-matching constraint requires us to fix a smooth map from $G_\text{trans.}$ to $\mathcal{M}$. Combined these two maps give us a global coordinate system $(\epsilon_1,\epsilon_2,\epsilon_3)\in\mathbb{R}^3$ for $\mathcal{M}$. We can then rescale these to give new coordinates $(t,x,y)\in\mathbb{R}^3$ by adding in a length scale $a>0$ with $t=\epsilon_1 a$, $x=\epsilon_2 a$, and $y=\epsilon_3 a$ for some fixed length scale $a$. 

In total this picks out a diffeomorphism \mbox{$d_\text{trans.}:\mathbb{R}^3\to\mathcal{M}$} which assigns global coordinates $(t,x,y)\in\mathbb{R}^3$ to $\mathcal{M}$. For much of the next sections I will work with $\phi$ in these coordinates. Namely I will work with the pull back, $\phi\circ d_\text{trans.}:\mathbb{R}^3\to\mathbb{R}$. One may worry that this choice of coordinates is arbitrary. Indeed, if one changes the smooth map from $G_\text{trans.}$ to $\mathcal{M}$ referenced above we end up with a different coordinate system $d_\text{trans.}$ related to the original by some diffeomorphism. While this is true, no matter how one realizes the translation-matching constraint there will be some coordinate system would result from the above construction. In what follows, nothing depends on how the translation-matching constraint is realized.

A word of warning. The coordinates used above might end up lying on the manifold in a curvilinear way. Crucially, the naive distance and metric structure associated with these coordinates do not automatically have any spaciotemporal significance. At present, our manifold $\mathcal{M}$ still has no metric. When a metric arises later, it will be due to dynamical considerations, not from any choice of coordinates. A coordinate-independent view of these theories will be given in Sec.~\ref{SecFullGenCov}.

With this said, by construction we have the following in correspondences in these coordinate:
\begin{flushleft}\begin{enumerate}
    \item Action by $T_\text{j}^\epsilon$ on $\mathbb{R}^L\cong\mathbb{R}^\mathbb{Z}\otimes\mathbb{R}^\mathbb{Z}\otimes\mathbb{R}^\mathbb{Z}$ acts on the manifold $\mathcal{M}$ in these coordinates as $(t,x,y)\mapsto(t-\epsilon\,a,x,y)$,
    \item Action by $T_\text{n}^\epsilon$ on $\mathbb{R}^L\cong\mathbb{R}^\mathbb{Z}\otimes\mathbb{R}^\mathbb{Z}\otimes\mathbb{R}^\mathbb{Z}$ acts on the manifold $\mathcal{M}$ in these coordinates as $(t,x,y)\mapsto(t,x-\epsilon\,a,y)$,
    \item Action by $T_\text{m}^\epsilon$ on $\mathbb{R}^L\cong\mathbb{R}^\mathbb{Z}\otimes\mathbb{R}^\mathbb{Z}\otimes\mathbb{R}^\mathbb{Z}$ acts on the manifold $\mathcal{M}$ in these coordinates as $(t,x,y)\mapsto(t,x,y-\epsilon\,a)$.
\end{enumerate}\end{flushleft}
In particular, matching up differential translations on $\mathbb{R}^L\cong\mathbb{R}^\mathbb{Z}\otimes\mathbb{R}^\mathbb{Z}\otimes\mathbb{R}^\mathbb{Z}$ with those on $\mathcal{M}$ requires
\begin{align}
E(D_\text{j}\,\bm{\Phi})&=a\,\partial_t E(\bm{\Phi})\\
\nonumber
E(D_\text{n}\bm{\Phi})&=a\,\partial_x E(\bm{\Phi})\\
\nonumber
E(D_\text{m}\bm{\Phi})&=a\,\partial_y E(\bm{\Phi})
\end{align}
in these coordinates for all $\bm{\Phi}\in\mathbb{R}^L$.

Evaluating these conditions in the planewave basis for $\mathbb{R}^L\cong\mathbb{R}^\mathbb{Z}\otimes\mathbb{R}^\mathbb{Z}\otimes\mathbb{R}^\mathbb{Z}$ (namely $\bm{\Phi}(\omega,k_1,k_2)$ for \mbox{$\omega,k_1,k_2\in[-\pi,\pi]$}) we see that $E(\bm{\Phi}(\omega,k_1,k_2))$ must be simultaneously an eigenvector of $\partial_t$, $\partial_x$, and $\partial_y$ with eigenvalues $-\ii \omega/a$, $-\ii k_1/a$, and $-\ii k_2/a$ respectively. This uniquely picks out the continuum planewaves:
\begin{align}\label{PlaneWaveCont}
\text{KG1-KG3:}\quad&\phi(t,x;\omega,k)
\coloneqq e^{-\ii \omega t -\ii k x}\\
\nonumber
\text{KG4-KG7:}\quad&\phi(t,x,y;\omega,k_1,k_2)
\coloneqq e^{-\ii \omega t -\ii k_1 x-\ii k_2 y}.
\end{align}
In particular, we are forced to take
\begin{align}\label{Eftilde}
E:\,&\bm{\Phi}(\omega,k_1,k_2)\\
\nonumber
&\mapsto\tilde{f}(\omega,k_1,k_2)\, \phi(t,x,y;\omega/a,k_1/a,k_2/a)
\end{align}
for some complex function $\tilde{f}(\omega,k_1,k_2)\in\mathbb{C}$. That is, each discrete planewave must map onto something proportional to the corresponding continuous planewaves (rescaled by $a$).%\Dan{Add affine possibility here and in Part 1.}

Note that up to a local rescaling of Fourier space (which recall is a symmetry of the dynamics) our embedding is uniquely fixed by translation-matching. Even without this consideration however, our translation-matching constraint still fixes the vector space $F\subset F(\mathcal{M})$ we can embed \mbox{$\bm{\Phi}\in\mathbb{R}^L$} onto, note the following. We must have \mbox{$\tilde{f}(\omega,k_1,k_2)\neq0$} for all \mbox{$\omega,k_1,k_2\in[-\pi,\pi]$} otherwise $E$ will not be invertible an hence not an isomorphism. Therefore $F$ is fixed as,
\begin{align}
&F=\text{span}(E(\bm{\Phi}(\omega,k_1,k_2))\vert \omega,k_1,k_2\in[-\pi,\pi])\\
\nonumber
&=\!\text{span}(\tilde{f}(\omega,k_1,k_2)\,\phi(t,x,y;\frac{\omega}{a},\frac{k_1}{a},\frac{k_2}{a})\vert \omega,k_1,k_2\in\![-\pi,\pi])\\
\nonumber
&=\text{span}(\phi(t,x,y;\frac{\omega}{a},\frac{k_1}{a},\frac{k_2}{a})\vert \omega,k_1,k_2\in[-\pi,\pi])\\
\nonumber
&=\text{span}(\phi(t,x,y;\omega,k_1,k_2)\vert \omega,k_1,k_2\in[-\pi/a,\pi/a]).
\end{align}
In light of this, let us define
\begin{align}
&\text{KG1-KG3:}\\
\nonumber
&F^K\coloneqq\text{span}(\phi(t,x;\omega,k)\vert \omega,k\in[-K,K])\\
&\text{KG4-KG7:}\\
\nonumber
&F^K\coloneqq\text{span}(\phi(t,x,y;\omega,k_1,k_2)\vert \omega,k_1,k_2\in[-K,K])
\end{align}
where $K=\pi/a$. These are the spaces of bandlimited functions with bandwidth $\omega,k\in[-K,K]$ and \mbox{$\omega,k_1,k_2\in[-K,K]$} respectively. Thus, demanding translation-matching we are forced to map $\bm{\Phi}\in\mathbb{R}^L$ onto some bandlimited $\phi(t,x,y)\in F=F^K$ (at least in the coordinates $d_\text{trans.}$).

The next section will discuss in detail some remarkable properties of bandlimited functions, namely their sampling property. Before that however, let's find an interpretation for the $\tilde{f}(\omega,k_1,k_2)$ function appearing in Eq.~\eqref{Eftilde}. First note that applying $E$ to the basis vector $\bm{e}_0\otimes\bm{e}_0\otimes\bm{e}_0$ we have
\begin{align}\label{Embedf}
E(\bm{e}_0\otimes\bm{e}_0\otimes\bm{e}_0)=f\!\left(\frac{t}{a},\frac{x}{a},\frac{y}{a}\right)
\end{align}
where $f(t,x,y)$ is the inverse Fourier transform of $\tilde{f}(\omega,k_1,k_2)$. Next note that by applying $T_\text{j}^\epsilon$, $T_\text{n}^\epsilon$, and $T_\text{m}^\epsilon$ with integer arguments we can get from $\bm{e}_0\otimes\bm{e}_0\otimes\bm{e}_0$ to any other basis vector, $\bm{e}_j\otimes\bm{e}_n\otimes\bm{e}_m$. In these coordinates this means,
\begin{align}
&E(\bm{e}_j\otimes\bm{e}_n\otimes\bm{e}_m)=f\left(\frac{t}{a}-j,\frac{x}{a}-n,\frac{y}{a}-m\right)
\end{align}
Thus, we can understand a choice of $\tilde{f}$ as picking a profile $f(t,x,y)\in F^\pi$. A translated and rescaled copy of this profile is then associated with each basis vector.

To review: Our translation-matching considerations have greatly constrained our choice of both the spacetime manifold $\mathcal{M}$ and the embedding of $\bm{\Phi}$ onto $F(\mathcal{M})$. In particular, $\mathcal{M}$ was forced to be diffeomorphic to $\mathbb{R}^2$ for KG1-KG3 and to $\mathbb{R}^3$ for KG4-KG7. Moreover, the vector space $F\subset F(\mathcal{M})$ which we map $\bm{\Phi}\in\mathbb{R}^L$ into is forced to be the space of bandlimited functions with some bandwidth (i.e., bandlimited in coordinates $d_\text{trans.}$ with $\omega,k_1,k_2\in[-K,K]$).

Beyond this, our only freedom left is in picking a profile $f$ to associate with each basis vector $\bm{e}_j\otimes\bm{e}_n\otimes\bm{e}_m$. In Sec.~\ref{SecExtPart2} I will motivate a principled choice for $f$. Before that however, let's discuss bandlimited function in some detail.

\section{Brief Review of Bandlimited Functions and Nyquist-Shannon Sampling Theory}\label{SecSamplingTheory}
 The previous section has given us reason to care about bandlimited functions. A bandlimited function is one whose Fourier transform has compact support. The bandwidth of such a function is the extent of its support in Fourier space. As I will now discuss, such functions have a remarkable sampling property: they can be exactly reconstructed knowing only their values at a (sufficiently dense) set of sample points. The study of such functions constitutes Nyquist-Shannon sampling theory. For a selection of introductory texts on sampling theory see ~\cite{GARCIA200263,SamplingTutorial,UnserM2000SyaS}.

To introduce the topic I will at first restrict our attention to the one-dimensional case with uniform sample lattice before generalizing to higher dimensions and non-uniform samplings later on.

\subsection{One Dimension Uniform Sample Lattices}\label{Sec1DUniform}
Consider a generic bandlimited function, $f_\text{B}(x)$, with a bandwidth of $K$. That is, a function $f_\text{B}(x)$ such that its Fourier transform,
\begin{align}
\mathcal{F}[f_\text{B}(x)](k)\coloneqq\int_{-\infty}^\infty f_\text{B}(x) \, e^{-\ii k x} \d x,
\end{align}
has support only for wavenumbers $\vert k\vert< K$.

Suppose that we know the value of $f_\text{B}(x)$ only at the regularly spaced sample points, $x_n=n\,a+b$, with some spacing, \mbox{$0\leq a\leq a^*\coloneqq\pi/K$}, and offset, $b\in\mathbb{R}$. Let \mbox{$f_n=f_\text{B}(x_n)$} be these sample values. Having only the discrete sample data, $\{(x_n,f_n)\}_{n\in\mathbb{Z}}$, how well can we approximate the function? 

The Nyquist-Shannon sampling theorem~\cite{ShannonOriginal} tells us that from this data we can reconstruct $f_\text{B}$ exactly everywhere! That is, from this discrete data, $\{(x_n,f_n)\}_{n\in\mathbb{Z}}$, we can determine everything about the function $f_\text{B}$ everywhere. In particular, the following reconstruction is exact, 
\begin{align}\label{SincRecon}
f_\text{B}(x) 
= S_n\!\left(\frac{x-b}{a}\right) f_n,
\end{align}
where
\begin{align}
S(y)=\frac{\sin(\pi y)}{\pi y}, \quad\text{and}\quad
S_n(y)=S(y-n), 
\end{align}
are the normalized and shifted sinc functions. Note that $S_n(m)=\delta_{nm}$ for integers $n$ and $m$. Moreover, note that each $S_n(x)$ is both $L_1$ and $L_2$ normalized and that taken together the set $\{S_n(x)\}_{n\in\mathbb{Z}}$ forms an orthonormal basis with respects to the $L_2$ inner product. The fact that any bandlimited function can be reconstructed in this way is equivalent to the fact that this orthonormal basis spans the space of bandlimited functions with bandwidth of $K=\pi$.

\begin{figure}
\includegraphics[width=0.4\textwidth]{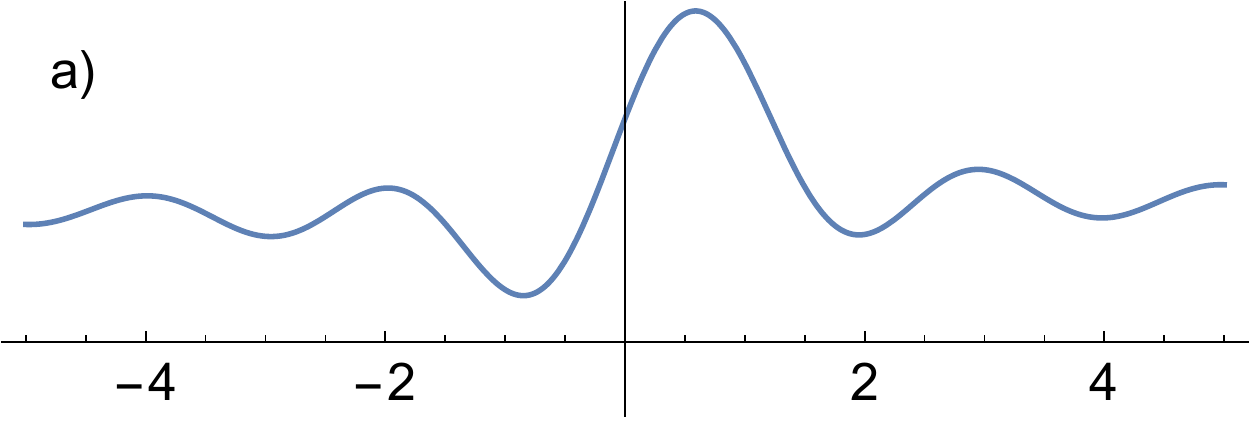}
\includegraphics[width=0.4\textwidth]{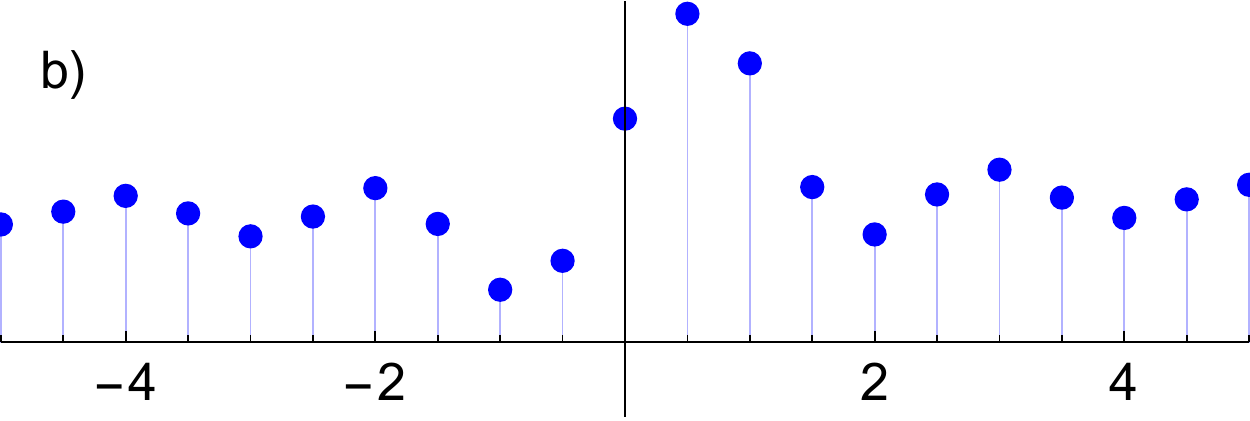}
\includegraphics[width=0.4\textwidth]{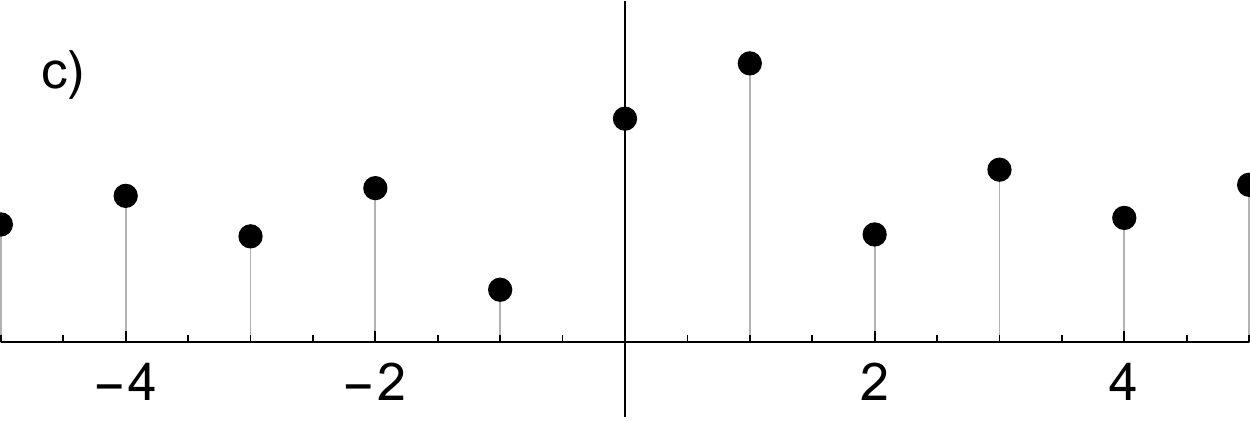}
\includegraphics[width=0.4\textwidth]{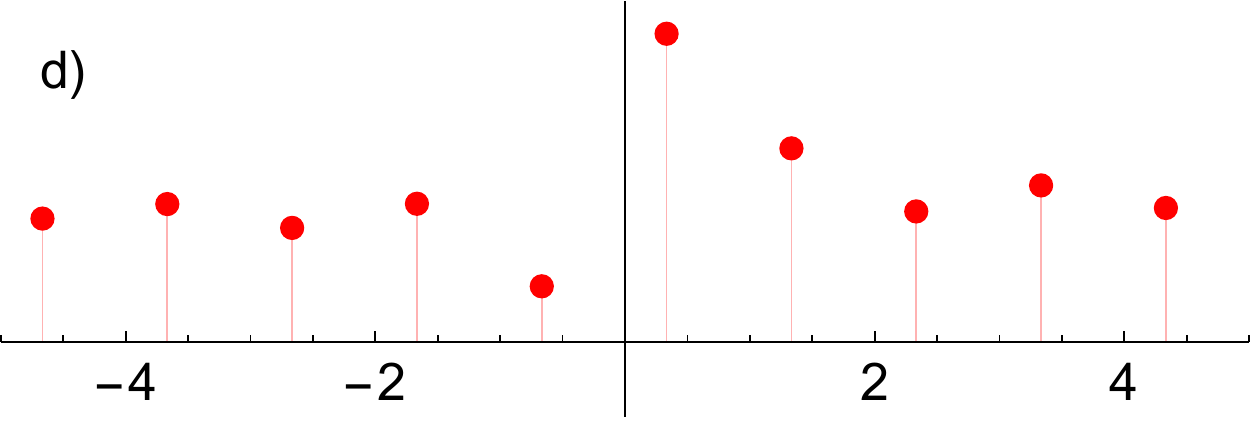}
\includegraphics[width=0.4\textwidth]{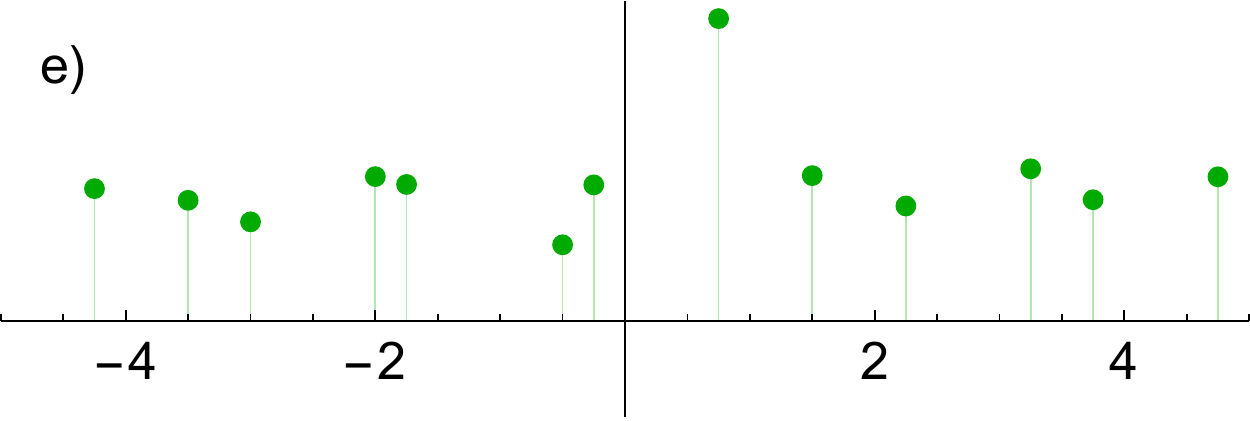}
\caption{Several different (but completely equivalent) graphical representations of the bandlimited function \mbox{$f_\text{B}(x)=1+S(x-1/2)+x\,S(x/2)^2$} with bandwidth of $K=\pi$ and consequently a critical spacing of $a^*=\pi/K=1$. Subfigure a) shows the function values for all $x$. b) shows the values of $f_\text{B}$ at $x_n=n/2$. Since $1/2<a^*=1$ this is an instance of oversampling. c) shows the values of $f_\text{B}$ at $x_n=n$. This is an instance of critical sampling. d) shows the values of $f_\text{B}$ at $x_n=n+1/3$. This too is an instance of critical sampling. e) shows a non-uniform sampling of $f_\text{B}$. From any of these samplings we can recover the function $f_\text{B}$ exactly everywhere. [\textit{Reproduced with permission from} \cite{DiscreteGenCovPart1}.]}\label{Fig1DSamples}
\end{figure}

As a concrete example, let us consider the function $f_\text{B}(x)=1+S(x-1/2)+x\,S(x/2)^2$, shown in Fig.~\ref{Fig1DSamples}a). This function has a bandwidth of $K=\pi$ and so has a critical sample spacing of $a^*=\pi/K=1$. Thus, we can fully reconstruct $f_\text{B}(x)$ knowing only its values at $x_n=n\,a+b$ for any spacing $a\leq a*=1$. In particular the sample values at $x_n=n/2$ are sufficient to exactly reconstruct the function, see Fig.~\ref{Fig1DSamples}b). So too are the sample values at the integers $x_n=n$ and at $x_n=n+1/3$, see Fig.~\ref{Fig1DSamples}c) and Fig.~\ref{Fig1DSamples}d). In each of these cases the reconstruction is given by Eq.~\eqref{SincRecon}.

Everything about this function can be reconstructed from any uniform sample lattice with $a\leq a^*=1$. In particular, the value of $f_\text{B}$ two third's of the way between sample point, $f_\text{B}(2/3)$, is fixed by $\{(n,f_\text{B}(n))\}_{n\in\mathbb{Z}}$ even though we have no sample at or even near $x=2/3$. The derivative of $f_\text{B}$ at zero, $f_\text{B}'(0)$, is fixed by $\{(n,f_\text{B}(n))\}_{n\in\mathbb{Z}}$ even though the only sample point we have in this neighborhood is $f_\text{B}(0)$. Moreover, the derivative at $x=2/3$, namely $f_\text{B}'(2/3)$, is fixed by $\{(n,f_\text{B}(n))\}_{n\in\mathbb{Z}}$ even we have no sample points in the neighborhood.

On first exposure this may be shocking: how can a function's behavior everywhere be fixed by its value at a discrete set of points? When $f_\text{B}$ is represented discretely, where has all of the information gone? Where is the information about the derivative at $x=2/3$ stored in the discrete representation? 

To see this, it is convenient (but not necessary) to organize our sample values $f_n=f(x_n)$ into a vector as,
\begin{align}
\bm{f}&=(\dots,f_{-1},f_0,f_1,f_2,\dots)^\intercal
= \sum_{n\in\mathbb{Z}} f_{n} \, \bm{w}_n
\end{align}
where $\bm{w}_n = (\dots,0,0,1,0,0,\dots)^\intercal\in\mathbb{R}^\mathbb{Z}$ with the 1 in the $n^\text{th}$ position. 

The values that $f_\text{B}$ takes at the sample points $x_n$ can be recovered from $\bm{f}$ as
\begin{align}
\nonumber
f_\text{B}(x_n) 
&= f_n
= \bm{w}_n^\intercal\bm{f},
\end{align}
Next notice that translating a bandlimited function preserves its bandwidth. As such both $f_\text{B}(x)$ and $f_\text{B}(x+\epsilon a)$ can be represented as Eq.~\eqref{SincRecon}. Using this fact, we can recover the values that $f_\text{B}$ takes away from the sample points (i.e., at \mbox{$x=x_n+\epsilon\,a$} for $\epsilon\in\mathbb{R}$) as
\begin{align}\label{TBdef}
f_\text{B}(x_n+\epsilon\,a) 
&= \sum_{m=-\infty}^\infty S_m(n+\epsilon) \ f_m
=\bm{w}_n^\intercal \, T_\text{B}^\epsilon \, \bm{f}
\end{align}
where the entries of the matrix $T^\epsilon_\text{B}$ are $[T^\epsilon_\text{B}]_{i,j}=S_i(j+\epsilon)$. Note $T_\text{B}^\epsilon$ acts as the translation operator for this representation of bandlimited functions. If $\bm{f}$ represents $f_\text{B}(x)$ then $T^\epsilon_\text{B}\bm{f}$ represents $f_\text{B}(x+\epsilon \,a)$.

From this translation operator we can identify the derivative operator for bandlimited functions, $D_\text{B}$, as 
\begin{align}
\nonumber
D_\text{B}\coloneqq\lim_{\epsilon\to0} \frac{T^\epsilon_\text{B}-\openone}{\epsilon}. \end{align}
It should be noted that $D_\text{B}$ and $T_\text{B}^\epsilon$ commute and moreover we have the usual relationship between derivatives and translations, $T_\text{B}^\epsilon=\exp(\epsilon\, D_\text{B})$.

From the above definition of $D_\text{B}$ one can easily work out its matrix entries as \mbox{$[D_\text{B}]_{i,j}=(-1)^{i-j}/(i-j)$} when $i\neq j$ and 0 when $i=j$. That is, 
\begin{align}\label{DBMatrix}
D_\text{B}&=\text{Toeplitz}(\dots,\!\frac{1}{4},\!\frac{-1}{3},\!\frac{1}{2},\!-1,\!0,\!1,\!\frac{-1}{2},\!\frac{1}{3},\!\frac{-1}{4},\!\dots)
\end{align}
Note that $D_\text{B}$ acts as the derivative operator for this representation of bandlimited functions. If $\bm{f}$ represents $f_\text{B}(x)$ then $\frac{1}{a}D_\text{B}\bm{f}$ represents $f_\text{B}'(x)$.

Comparing this with the $D$ operator introduced in Eq.~\eqref{BigToeplitz} we see that they are numerically identical. Indeed, $D_\text{B}=D$ and moreover $T_\text{B}^\epsilon=T^\epsilon$. If we were to extend our discussion to two-dimensional functions we could find a discrete representation of the rotation operator for bandlimited functions, $R_\text{B}^\theta$. This would come out numerically equal to the $R^\theta$ operator introduced earlier in Eq.~\eqref{RthetaDef}, namely $R_\text{B}^\theta=R^\theta$. See Appendix~\ref{AppA} for further discussion. Thus, the discrete notions of derivative, translation, and rotation that we have been using up until now are intimately connected with bandlimited functions.

It should be noted that $D=D_\text{B}$ gives us the following remarkable derivative approximation (which is exact for bandlimited functions):
\begin{align}\label{ExactDerivative}
\partial_x f(x)
&\approx2\sum_{m=1}^\infty (-1)^{m+1} \frac{f(x+m\,a)-f(x-m\,a)}{2\,m\,a}.
\end{align}
Relatedly, we have the second derivative approximation (which is exact for bandlimited functions):
\begin{align}\label{ExactDerivative2}
\partial_x^2 f(x)
&\approx2\sum_{m=1}^\infty (-1)^{m+1} \frac{f(x+m\,a)+f(x-m\,a)}{m^2\,a^2}\\
\nonumber
&-\frac{\pi^2}{3}f(x).
\end{align}
Namely, when $f$ is bandlimited with bandwidth of $K$ and $a\leq\pi/K$ then these formulas are exact. Moreover, if the Fourier transform of $f$ is mostly supported in $[-K,K]$ with thin tails (e.g, Gaussian tails) outside this region, then these are very good derivative approximations. 

Ultimately we can compute any derivative of $f_\text{B}$ anywhere from our sample data as,
\begin{align}
\nonumber
\partial_x^r\,f_\text{B}(x_n+\epsilon\,a) 
=\frac{1}{a^r}\,\bm{w}_n^\intercal \, D_\text{B}^r T_\text{B}^\epsilon \, \, \bm{f}.
\end{align}
Thus, we can recover any value or derivative of $f_\text{B}$ from its values on any sufficiently dense uniform sample lattice.

Note that each $f_\text{B}$ is representable in this way on a wide number of sample lattices with differing spacings $a<\pi/K$ and differing offsets $b$. Translating between these different descriptions of $f_\text{B}$ is equivalent to a change of basis on the vector of sample values $\bm{f}$.

\subsection{Non-Uniform Sample Lattices}\label{Sec2DSampling}
The previous subsection showed how any value or derivative of $f_\text{B}$ can be recovered from its values on any sufficiently dense uniform sample lattice. Moreover, it showed how changing between representing $f_\text{B}$ with different uniform sample lattices is ultimately just a change of basis on $\bm{f}$. 

As I will now discuss, we can also represent $\bm{f}$ on any sufficiently dense non-uniform lattice. To motivate this, consider first an oversampling of $f_\text{B}$. For example, figure Fig.~\ref{Fig1DSamples}b) shows $f_\text{B}$ sampled at twice the necessary frequency. This is a representation of $\bm{f}$ in an overcomplete basis. Imagine oversampling by a factor of ten with a spacing of $a=a^*/10$. Intuitively, this sample lattice has ten times the information needed to recover the function exactly. If we were to delete all but every tenth data point we would still be able to recover the function. But what if we just half of the sample points, but did so randomly? This would result in a non-uniform sample lattice. See for instance Fig.~\ref{Fig1DSamples}e). Hopefully, the reader has some intuition that at least some non-uniform sample lattices are sufficient to exactly reconstruct $f_\text{B}$.

The scope of such non-uniform samplings is established by various non-uniform sampling theorems~\cite{GARCIA200263,SamplingTutorial}. The details of these theorems are not important here; They can all be summarized as saying that reconstruction is possible when our non-uniform sample points are ``sufficiently dense'' in some technical sense. The sampling shown in Fig. \ref{Fig1DSamples}e) is sufficiently dense. The reconstruction in the non-uniform case is significantly more complicated than it is in the uniform case. Rather than Eq.~\eqref{SincRecon}, in the non-uniform case our reconstruction is of the form,
\begin{align}\label{GenRecon}
f_\text{B}(x)=\sum_{m=-\infty}^\infty G_m(x;\{z_n\}_{n\in\mathbb{Z}}) \, f_\text{B}(z_m) 
\end{align}
for some reconstruction functions, $G_m$, which depend in a complicated way on the location of all of the other sample points, $\{z_n\}_{n\in\mathbb{Z}}$.

%If we do not know the values of either $z_n$ or $f_\text{B}(z_n)$ exactly (for instance, if they are discovered through experiment) then there are stability concerns~\cite{ReconstructionStability} about these reconstruction from a non-uniform sample lattice. Small errors in $z_n$ or $f_\text{B}(z_n)$ can lead to large reconstruction errors. In some senses taking a uniform sample lattice minimizes these stability issues. However, if the values of $z_n$ and $f_\text{B}(z_n)$ are known exactly then these issues do not arise and the above reconstruction is exact.

\subsection{Higher Dimensional Sampling}
\begin{figure*}[t!]
\centering
\includegraphics[width=0.95\textwidth]{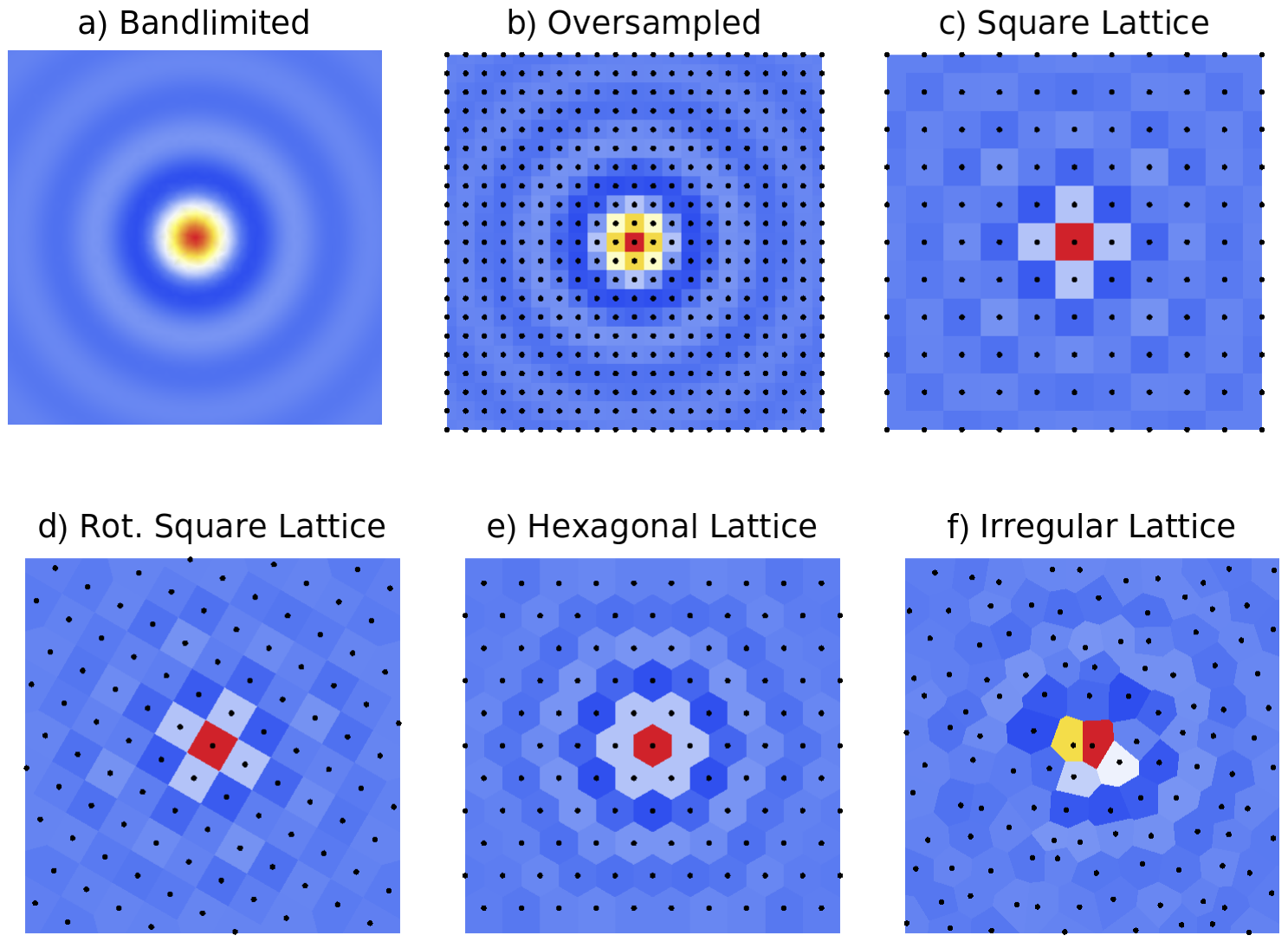}
\caption{Several different (but completely equivalent) graphical representations of the bandlimited function given by Eq.~\eqref{J1}. This function has a bandwidth of $\sqrt{k_x^2+k_y^2}<K=\pi$ and so has a critical spacing of $a^*=\pi/K=1$ in every direction. The scale of each subfigure is 5x5. In each subfigure, the colored regions are the Voronoi cells around the sample points (black). Subfigure a) shows the function values for all $x$. b) shows $f_\text{B}$ sampled on a square lattice with $z_{n,m}=(n/2,m/2)$. Since $1/2<a^*=1$ this is an instance of oversampling. c) shows $f_\text{B}$ sampled on a square lattice with $z_{n,m}=(n,m)$. This is an instance of critical sampling since $a=a^*=1$. d) shows $f_\text{B}$ sampled on a square lattice with $z_{n,m}=(n+m,n-m)/\sqrt{2}$. e) $f_\text{B}$ sampled on a hexagonal lattice of with \mbox{$z_{n,m}=(n+m/2,\sqrt{3}m/2)\in\mathbb{R}^2$}. f) shows $f_\text{B}$ sampled on an irregular lattice. [\textit{Reproduced with permission from} \cite{DiscreteGenCovPart1}.]}
\label{Fig2DSamples}
\end{figure*}

The same story about bandlimited functions is largely true in higher dimensions as well. A two-dimensional function $f_\text{B}(x,y)$ is bandlimited if is Fourier transform $\mathcal{F}[f_\text{B}(x,y)](k_x,k_y)$ is compactly supported in the \mbox{($k_x$, $k_y$)-plane}. Specifying the value of the bandwidth is less straightforward in the high dimensional case as the Fourier transform's support may have different extents in different directions. However, any compact region can be bounded in a square. We can thus always imagine $f_\text{B}(x,y)$ as being bandlimited with \mbox{$k_x,k_y\in[-K,K]$} for some $K>0$. As such, we can represent $f_\text{B}(x,y)$ with a (sufficiently dense) uniform sample lattice in both the $x$ and $y$ directions. That is, we can represent any bandlimited $f_\text{B}(x,y)$ in terms of its sample values on a sufficiently dense square lattice. Once we have such a uniform sampling, the reasoning carried out above applies unchanged. We can represent $f_\text{B}(x,y)$ on any sufficiently dense non-uniform lattice.

For a concrete example consider the bandlimited function shown shown in Fig.~\ref{Fig2DSamples}a), namely,
\begin{align}\label{J1}
f_\text{B}(x,y)=J_1(\pi\,r)/(\pi\,r) 
\end{align}
where $J_1$ is the first Bessel function and $r=\sqrt{x^2+y^2}$. This function is bandlimited with $\sqrt{k_x^2+k_y^2}<K=\pi$ and hence critical spacing $a^*=\pi/K=1$. Moreover, note that this function is rotation invariant.

Given this function's bandwidth of $K=\pi$, we can represent it via its sample values taken on a square lattice with spacing $a=1/2\leq a^*=1$, see Fig. \ref{Fig2DSamples}b). We can also use a coarser square lattice with a spacing of $a=a^*=1$, see Fig. \ref{Fig2DSamples}c). We could also use a rotated square lattice, see Fig. \ref{Fig2DSamples}d). Sampling the function on a hexagonal lattice also works, see Fig. \ref{Fig2DSamples}e). Finally we can use a non-uniform lattice of sample points, see Fig. \ref{Fig2DSamples}f). From each of these discrete representations, we could recover the original bandlimited function everywhere exactly via some generalization of Eq.~\eqref{SincRecon} in the uniform cases and Eq.~\eqref{GenRecon} in the irregular case.

Thus, there is no conceptual barrier to representing a rotationally invariant bandlimited function on a square lattice. Indeed, there is no issue with representing such a function on any sufficiently dense lattice. In light of the analogy proposed in this paper, we can see this as analogous to the unsurprising fact that there is no conceptual barrier to representing rotationally invariant functions in Cartesian coordinates. There is no requirement that our representation (be it a choice of coordinates or a choice of sample points) latches onto the symmetries of what is being represented. 

Thus we have a non-uniform sampling theory for higher dimensions. But what about a sampling theory on curved spaces? While such things are not relevant for the aims of this paper, recently notable progress has been made on developing a sampling theory for curved manifolds~\cite{CurvedSampling,Martin2008}.

\section{Externalizing these theories - Part 2}\label{SecExtPart2}
Having reviewed the sampling property of bandlimited functions, let's return to the task of embedding $\bm{\Phi}$ into our spacetime manifold $\mathcal{M}$. As discussed in Sec.~\ref{SecExtPart1}, for KG4-KG7 our translation-matching constraint forces the new field \mbox{$\phi:\mathcal{M}\to\mathbb{R}$} to be bandlimited in the coordinate system $d_\text{trans.}$ with bandwidth $\omega,k_1,k_2\in[-K,K]$ where $K=\pi/a$. Similarly for KG1-KG3. In either case, let's note this by giving the field a subscript $\text{B}$ as \mbox{$\phi_\text{B}:\mathcal{M}\to\mathbb{R}$}. 

Before continuing on, it should be noted that, the above discussed translation-matching constraint guarantees that $\phi_\text{B}\circ d_\text{trans.}$ will have the sampling property discussed in the previous section.

\subsection{A Principled Choice of Profile}
What freedom remains for our choice of embedding? Focusing on KG4-KG7, in coordinate $d_\text{trans.}$ our remaining freedom is a choice of profile $f(t,x,y)\in F^\pi$ appearing in Eq.~\eqref{Embedf}. A translated and rescaled copy of this profile is associated with each basis vector $\bm{e}_j\otimes\bm{e}_n\otimes\bm{e}_m$. As noted following Eq.~\eqref{Eftilde}, up to a local recaling of Fourier space (which is a symmetry of the dynamics) our embedding is fixed. However, in order to completely fix this embedding, we need to pick a profile $f(t,x,y)$.

The discussion in the previous section offers us a promising candidate for this profile namely \mbox{$f(t,x,y)=S_0(t)S_0(x)S_0(y)$}. This profile is the unique one in $F^\pi$ with the following property\footnote{To give the technical details, this $f$ is the unique function in $F^\pi$ which evaluates to zero at all integer arguments except $(0,0,0)$ where it returns $1$. That is, \mbox{$f(j,n,m)=\delta_{j0}\delta_{n0}\delta_{m0}$} for $j,n,m\in\mathbb{Z}$. This uniqueness is a direct consequence of the Nyquist-Shannon sampling theorem.}: it makes $\phi_\text{B}(t,x,y)$ evaluated at $z_{j,n,m}=(j\,a,n\,a,m\,a)$ take the value of $\phi_{j,n,m}$. That is, it makes our original discrete variables $\phi_{j,n,m}$ sample values of $\phi_\text{B}$ at the sample points $z_{j,n,m}$. Let us call this choice of profiles the lattice-as-sample-points constraint.

To be clear, what hinges on our choice of profile $f$ is not whether or not $\phi_\text{B}$ has the sampling property discussed in the previous section; This property is guaranteed independent of our choice of $f$. Rather, what is at stake is just whether our original discrete variables $\phi_{j,n,m}$ \mbox{\textit{themselves}} are sample values.

It follows from the lattice-as-sample-points constraint that $E$ acts on the planewave basis as
\begin{align}\label{Embed0}
E:\bm{\Phi}(\omega,k_1,k_2)\mapsto\phi(t,x,y;\omega/a,k_1/a,k_2/a)
\end{align}
That is, discrete planewaves are mapped onto continuous planewaves (rescaled by $a$). Moreover, this constraint forces $E$ to act on the basis $\bm{e}_j\otimes\bm{e}_n\otimes\bm{e}_m$ as
\begin{align}\label{Embed}
E: \bm{e}_j\otimes\bm{e}_n\otimes\bm{e}_m\mapsto
S_{j}(t/a) \, S_{n}(x/a) \, S_{m}(y/a).
\end{align}
In particular, applying this to  Eq.~\eqref{PhiVec2} gives us
\begin{align}\label{PhiSincRecon1}
&\text{KG1-KG3:}\\
\nonumber
&\quad\phi_\text{B}(t,x)
=\sum_{j,n\in\mathbb{Z}} S_{j}(t/a)\,S_{n}(x/a) \ \phi_{j,n}\\
\label{PhiSincRecon2}
&\text{KG4-KG7:}\\
\nonumber
&\quad\phi_\text{B}(t,x,y)
=\sum_{j,n,m\in\mathbb{Z}} S_{j}(t/a) \, S_{n}(x/a) \, S_{m}(x/a) \, \phi_{j,n,m}.
\end{align}
where analogous reasoning applies for KG1-KG3.

Thus from these translation-matching and lattice-as-sample-points constraints it follows that our new field must be the result of a certain bandlimited reconstruction. Namely, for KG4-KG7 it must come from a reconstruction using sample values $\phi_{j,n,m}$ at sample point $z_{j,n,m}$. Similarly for KG1-KG3 it must come from using sample values $\phi_{j,n}$ at sample points $z_{j,n}=(j\,a,n\,a)$.

\begin{figure*}[t]
\includegraphics[width=0.95\textwidth]{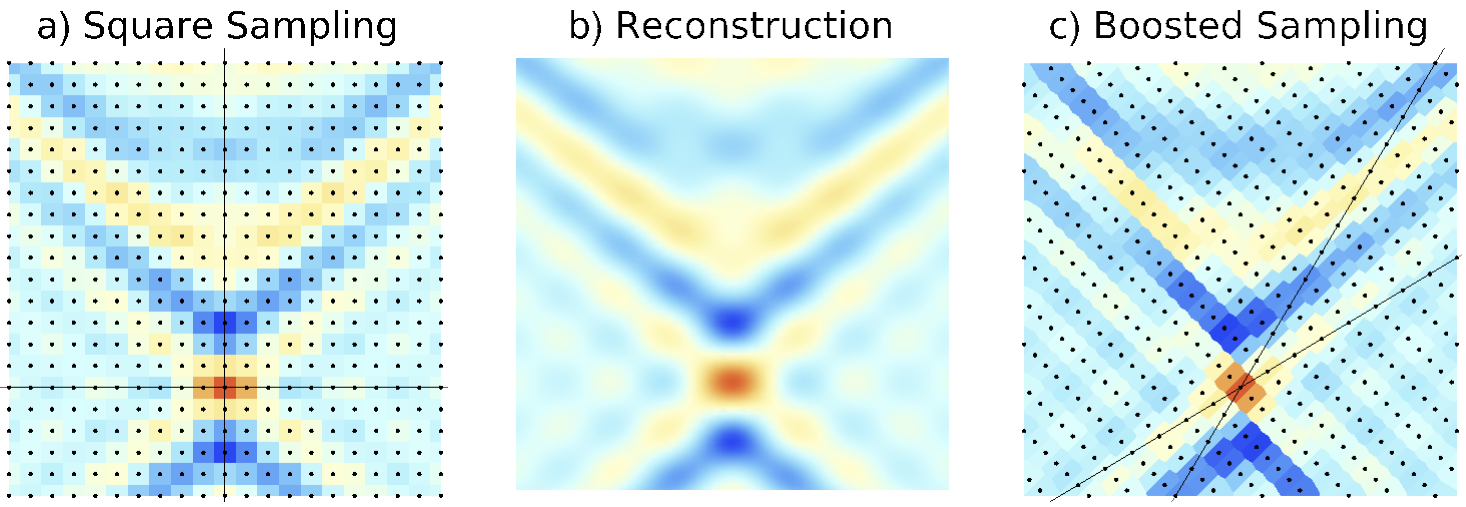}
\includegraphics[width=0.95\textwidth]{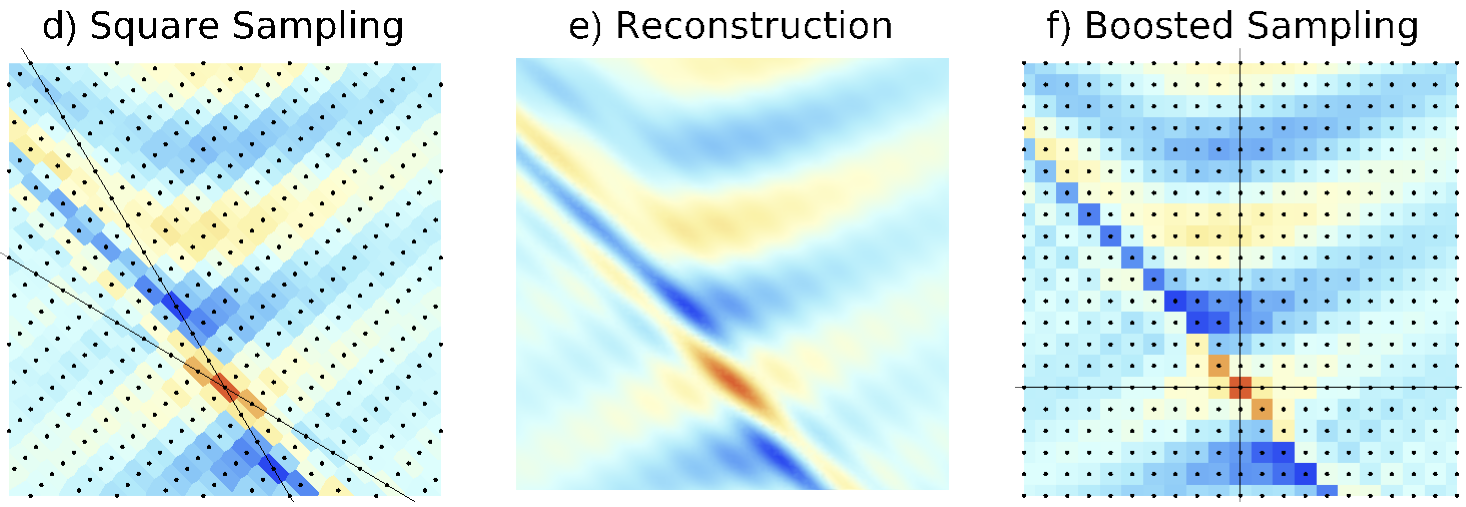}
\caption{Subfigure b) shows a solution to KG3 formulated as a bandlimited function in the coordinates given by $d_\text{trans.}$.
This field in these coordinates obeys Eq.~\eqref{DKG3bandlimited}. Subfigure a) shows the values that this function takes on a square lattice in spacetime.  These discrete field values obey Eq.~\eqref{DKG3}. From this embedding we can uniquely reconstruct the bandlimited field $\phi_\text{B}$. However, these sample points play no special role in the theory. We are free to redescribe the state and its dynamics by resampling it on any sufficiently dense collection of sample points. For instance, we might describe $\phi_\text{B}$ using the boosted sample points shown in subfigure c). Not all states will be describable on both of these lattices, but some are. If we boost both the sample values and the field as shown in subfigure d) e) and f) nothing changes. Both sets of sample values still obey Eq.~\eqref{DKG3}. Moreover, in the bandlimited field still obeys Eq.~\eqref{DKG3bandlimited}.}\label{FigKleinSpaceTime}
\end{figure*}

Fig.~\ref{FigKleinSpaceTime}a) shows for KG1-KG3 these sample points \mbox{$z_{j,n}=(j\,a,n\,a)$} and sample values as they lie on the spacetime manifold $\mathcal{M}\cong\mathbb{R}^2$ in the coordinates $d_\text{trans.}$ Fig.~\ref{FigKleinSpaceTime}b) shows for KG3 what the reconstructed bandlimited function $\phi_\text{B}:\mathcal{M}\to\mathbb{R}$ might look like in these coordinates. One can imagine an analogous figure for KG4-KG7 with the sample points forming a cubic lattice in the coordinates $d_\text{trans.}$.

One may worry that we are here taking a cubic 3D lattice for each of KG4-KG7 whereas for KG5 and KG7 we naturally ought to embed onto a hexagonal 3D lattice. This was after all, the lattice structure picked out by our symmetry analysis in the first interpretation. However, one need not worry for two reasons: Firstly, there is no real sense in which the sample points $z_{j,n}$ and $z_{j,n,m}$ form a square and cubic lattice respectively. These sample lattices only appear square and cubic in this particular coordinate system; in other coordinate systems they won't, see Fig.~\ref{FigKleinSpaceTime}d). Indeed, the coordinates we are using here may lie on the spacetime manifold $\mathcal{M}$ in a curvilinear way. As such one ought not to think of these sample points as being arranged in a square or cubic lattice on the spacetime manifold. Indeed, at this point the manifold still does not have a metric.

Secondly, after we have used the sample points $z_{j,n}$ and $z_{j,n,m}$ to build $\phi_\text{B}$ they no longer play any special role in the theory. As discussed in the previous section, we are always free to resample a bandlimited function on any sufficiently dense sample lattice. Recall Fig.~\ref{Fig2DSamples}. Thus even staying in this coordinate system (where our initial sample points do form a cubic 3D lattice) we are free to resample $\phi_\text{B}$ on a hexagonal 3D lattice. Moreover, we also might be able to resample $\phi_\text{B}$ on a boosted lattice as shown in Fig.~\ref{FigKleinSpaceTime}c).

Thus, in a sense the lattice sites $L$ from our first interpretation still exist here they are embedded onto our new spacetime manifold, $\ell\in L\mapsto z_\ell\in\mathcal{M}$. However, once embedded, these lattice sites no longer play any special role in the theory; We can always resample.

\vspace{0.25cm}

To review: In Sec.~\ref{SecExtPart1} we saw how the translation-matching constraint forced us to embed $\bm{\Phi}$ onto $\mathcal{M}$ as a some bandlimited function, $\phi_\text{B}:\mathcal{M}\to\mathbb{R}$ (i.e., bandlimited in coordinates $d_\text{trans.}$ with $\omega,k_1,k_2\in[-K,K]$). In this section, we further demanded the lattice-as-sample-points constraint. This allowed us to think of our initial discrete variables $\phi_\ell$ as being the values that $\phi_\text{B}$ takes at some sample points $z_\ell\in\mathcal{M}$. Taken together these two constraints completely fix how $\phi_\text{B}:\mathcal{M}\to\mathbb{R}$ is built from $\bm{\Phi}$ (or equivalently $\phi_\ell$). In particular, in the global coordinate system given by $d_\text{trans.}$ we must have Eq.~\eqref{PhiSincRecon1} and Eq.~\eqref{PhiSincRecon2}.

In the next subsection, I will rewrite the dynamics of each of KG1-KG7 in terms of this new field $\phi_\text{B}$ in the coordinates $d_\text{trans.}$. Following this, in Secs.~\ref{SecKlein3} and \ref{SecKlein3Extra}, I will begin interpreting this theory independently of how they were constructed. In Sec.~\ref{SecFullGenCov} I will give a coordinate-independent formulation of these theories.

\subsection{Bandlimited Dynamics}\label{SecKlein3B}
The previous subsections has given us a principled way to embed $\phi_\ell$ onto  our new spacetime manifold $\mathcal{M}$ as a bandlimited function. Namely, Eq.~\eqref{PhiSincRecon1} and Eq.~\eqref{PhiSincRecon2}. Equivalently we can think of embedding $\bm{\Phi}$ onto our new spacetime manifold $\mathcal{M}$ as a bandlimited function. Namely, Eq.~\eqref{Embed}. Either of these perspectives allow us to translate the kinematics of KG1-KG7 into this new continuous setting. This subsection will additionally translate over the dynamics into this new continuous setting.

Let's begin with KG1. This translation is aided by the fact that the derivative is the generator of translations, i.e.,  $h(x+a)=\text{exp}(a\, \partial_x) h(x)$. Mapping the left hand side of the dynamics of KG1 (i.e., Eq.~\eqref{KG1Long}) onto a bandlimited function we have
\begin{align}
&\sum_{j,n\in\mathbb{Z}} S_{j}(t/a)\, S_{n}(x/a)\, \left(\text{L.H.S. of Eq.~\eqref{KG1Long}}\right)\\
\nonumber
&=\sum_{j,n\in\mathbb{Z}} S_{j}(t/a)\,S_{n}(x/a) \big[\phi_{j+1,n}-2\phi_{j,n}+\phi_{j-1,n}\big]\\
\nonumber
&=\phi_\text{B}(t - a,x) - 2\phi_\text{B}(t,x) + \phi_\text{B}(t + a,x)\\
\nonumber
&=[\exp(-a\,\partial_t)-2+\exp(a\,\partial_t)] \phi_\text{B}(t,x)\\
\nonumber
&=[2\cosh(a\,\partial_t)-2] \ \phi_\text{B}(t,x).
\end{align}
Similarly, for the right hand side of Eq.~\eqref{KG1Long} we have
\begin{align}
&\sum_{j,n\in\mathbb{Z}} S_{j}(t/a)\, S_{n}(x/a)\, \left(\text{R.H.S. of Eq.~\eqref{KG1Long}}\right)\\
\nonumber
&=[\mu^2+2\cosh(a\,\partial_t)-2] \ \phi_\text{B}(t,x).
\end{align}
Thus, when the sample values $\phi_{j,n}=\phi_\text{B}(z_{j,n})$ obey Eq.~\eqref{KG1Long}, the bandlimited field obeys,
\begin{align}\label{DKG1bandlimited}
\text{KG1:}\,[2\cosh(a\partial_t)\!-\!2] \phi_\text{B}\!=\![\mu^2+2\cosh(a\partial_t)\!-\!2] \phi_\text{B}
\end{align}
Similarly for the other six theories we have:
\begin{align}\label{DKG2bandlimited}
\text{KG2:}&\ \frac{1}{6}[-\cosh(2a\partial_t)\!+\!16\cosh(a\partial_t)\!-\!15]\phi_\text{B}\\
\nonumber
&=\left(\mu^2+\frac{1}{6} [-\cosh(2a\partial_x)\!+\!16\cosh(a\partial_x)\!-\!15]\right)\phi_\text{B}\\
\label{DKG3bandlimited}
\text{KG3:}&\ \partial_t^2\phi_\text{B}
=(\mu_0^2+\partial_x^2) \, \phi_\text{B}\\
\label{DKG4bandlimited}
\text{KG4:}&\  [2\cosh(a\partial_t)\!-\!2] \phi_\text{B}\\
\nonumber
&=[\mu^2+2\cosh(a\,\partial_x)+2\cosh(a\,\partial_y)\!-\!4]\phi_\text{B}\\
\label{DKG5bandlimited}
\text{KG5:}&\ [2\cosh(a\,\partial_t)\!-\!2] \phi_\text{B}=\Big[\mu^2+\frac{4}{3}\cosh(a\,\partial_x)\\
\nonumber
&\qquad\quad+\frac{4}{3}\cosh(a\,\partial_y) +\frac{4}{3}\cosh(a\,(\partial_y-\partial_x))\!-\!4\Big]
\phi_\text{B},\\
\label{DKG6bandlimited}
\text{KG6:}&\  
\partial_t^2\phi_\text{B}
=(\mu_0^2+\partial_x^2+\partial_y^2) \, \phi_\text{B}\\
\label{DKG7bandlimited}
\text{KG7:}&\  
\partial_t^2\phi_\text{B}
\!=\!\Big(\mu_0^2\!+\!\frac{2}{3}\big[\partial_x^2\!+\!\partial_y^2\!+\!(\partial_x\!-\!\partial_y)^2\big]\!\Big) \phi_\text{B}
\end{align}
where $\mu_0\coloneqq\mu/a$. 

\begin{figure}[t]
\includegraphics[width=0.4\textwidth]{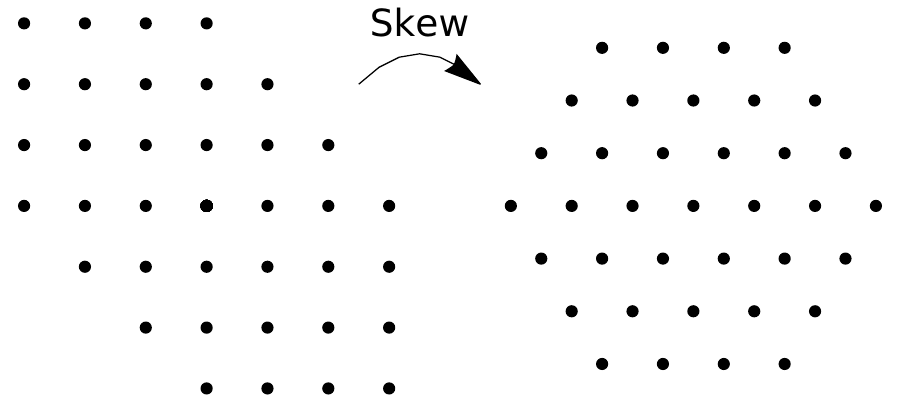}
\caption{As this figure shows, a certain linear transformation of coordinates (namely Eq.~\eqref{CoorKG7KG6}) maps a square lattice to a hexagonal one. [\textit{Reproduced with permission from} \cite{DiscreteGenCovPart1}.]}\label{FigSkew}
\end{figure}

Note that the dynamics of KG3 and KG6 are exactly the same as those of the continuum Klein Gordon equations KG00 and KG0 with a field mass \mbox{$M=\mu_0$}. Moreover, after a coordinate transformation, 
\begin{align}\label{CoorKG7KG6}
t\mapsto t,\quad
x\mapsto x + \frac{1}{2} y,\quad
y\mapsto \frac{\sqrt{3}}{2}y
\end{align}
the dynamics of KG7 exactly maps onto the dynamics of KG6 (and consequently, KG0). Note that this change of coordinates is equivalent to Eq.~\eqref{SkewKG7KG6} applied in continuum Fourier space. Moreover, note that Eq.~\eqref{CoorKG7KG6} maps a square lattice to a hexagonal one, see Fig.~\ref{FigSkew}. As I will discuss later, this means that KG6 and KG7 can be seen as describing the same bandlimited theory, just using different sample points.

Applying this coordinate transformation to KG5's bandlimited dynamics gives us:
\begin{align}\label{DKG5bandlimitedSkew}
\text{KG5:}\ &[2\cosh(a\partial_t)\!-\!2] \phi_\text{B}=\Big[\mu^2-4+\frac{4}{3}\cosh(a\,\partial_x)\\
\nonumber
&\qquad\qquad\qquad+\frac{4}{3}\cosh(a(\sqrt{3} \partial_y+\partial_x)/2)\\
\nonumber
&\qquad\qquad\qquad+\frac{4}{3}\cosh(a(\sqrt{3} \partial_y-\partial_x)/2)\Big]
\phi_\text{B}.
\end{align}
Note that this dynamics now manifestly has a one-sixth rotational symmetry. Sampling the above dynamics on a hexagonal lattice gives us our original dynamics for KG5. Sampling the above dynamics on a square lattice gives us the following. Given Eq.~\eqref{ExactDerivative} sampling on a square lattice effectively means taking $a\,\partial_t\to D_\text{j}$, $a\,\partial_x\to D_\text{n}$ and $a\,\partial_y\to D_\text{m}$ and replacing $\phi_\text{B}$ with $\bm{\Phi}$. For the above dynamics of KG5 this gives:
\begin{align}\label{DKG5Skew}
\text{KG5:}\ \Delta_{(1),\text{j}}^2\,\bm{\Phi}&=\Big[\mu^2-4+\frac{4}{3}\cosh(D_\text{n})\\
\nonumber
&+\frac{4}{3}\cosh((\sqrt{3} D_\text{m}+D_\text{n})/2)\\
\nonumber
&+\frac{4}{3}\cosh((\sqrt{3} D_\text{m}-D_\text{n})/2)\Big]
\phi_\text{B}.
\end{align}
Note that before this resampling KG5 was local in the intuitive sense nearest-neighbor couplings only, see Eq.~\eqref{KG5Long}. After resampling it is infinite range. Thus, the intuitive notion of locality in terms of nearest neighbors is unstable under resampling. For this and other reasons (see \cite{DiscreteGenCovPart1}) the notion of locality arising in the bandlimited formulation is preferable. Sample values can obey what appears to be non-local dynamics, because the sample values themselves are non-local objects: they ought to be associated with sinc profiles not spacetime points, see \cite{DiscreteGenCovPart1}.

We can easily solve each of these dynamical equations. Just as in Sec. \ref{SecSevenKG} these dynamics admit a complete basis of planewave solutions, given by Eq.~\eqref{PlaneWaveCont}. A few key differences should be noted between these continuum planewaves and the discrete planewaves considered earlier, namely $\bm{\Phi}(\omega,k)$ and $\bm{\Phi}(\omega,k_1,k_2)$. Firstly, there is a difference of scale: note how our embedding Eq.~\eqref{Embed0} rescales the planewaves by a factor of $1/a$. Secondly, the discrete planewaves $\bm{\Phi}(\omega,k)$ and $\bm{\Phi}(\omega,k_1,k_2)$ repeated themselves cyclically with period $2\pi$ outside of the regions \mbox{$\omega,k\in[-\pi,\pi]$} and \mbox{$\omega,k_1,k_2\in[-\pi,\pi]$}. The continuum planewaves do not do this. Now each planewave with $\omega,k_1,k_2\in\mathbb{R}$ is distinct. Here, however, by its construction $\phi_\text{B}$ only has support over planewaves with \mbox{$\omega,k_1,k_2\in[-K,K]$} with $K=\pi/a$.

Each of these planewaves are solutions only when they obey the following dispersion relations: 
\begin{align}
\text{KG1:}& \quad \!\! 2-2\,\cos(\omega\,a)= \mu^2+2-2\,\cos(k\,a)\\
\text{KG2:}& \quad \!\! \frac{1}{6}\,(\cos(2\,\omega\,a)-16\,\cos(\omega\,a)+15)\\
\nonumber
&= \mu^2+\frac{1}{6}\,(\cos(2\,k\,a)-16\,\cos(k\,a)+15)\\
\text{KG3:}& \quad \!\! \omega^2= \mu_0^2+k^2\\
\text{KG4:}& \ 2\!-\!2\cos(\omega a)\!=\! \mu^2\!+\!4\!-\!2\cos(k_1 a)\!-\!2\cos(k_2 a)\\
\text{KG5:}& \  2-2\,\cos(\omega\,a)= \mu^2 \\
\nonumber
&+\frac{4}{3}\big[3-\cos(k_1\,a)-\cos(k_1\,a)-\cos(k_2\,a-k_1\,a)\big]\\
\text{KG6:}& \ \omega^2 = \mu_0^2 +k_1^2 +k_2^2\\
\text{KG7:}& \ \omega^2= \mu_0^2 +\frac{2}{3}\big[k_1^2 +k_2^2 +(k_2-k_1)^2\big].
\end{align}
Note that the dispersion relation of KG3 and KG6 are exactly the same as those of the continuum Klein Gordon equations KG00 and KG0 with a field mass $M=\mu_0$. Moreover, after the coordinate transformation, Eq.~\eqref{SkewKG6KG7}, 
the dispersion relation of KG7 exactly maps onto that of KG6 (and consequently, KG0). More will be said about this later.

We are now ready to make a third attempt at interpreting these theories. 

\section{A Third Attempt at Interpreting KG1-KG7 - Part 1}\label{SecKlein3}
The previous three sections have (after much effort) given us a new formulation of KG1-KG7 to interpret. In particular, this new formulation has aimed to externalize many of the internal symmetries which we discovered in Sec.~\ref{SecKlein2}. 

On this third interpretation I will be taking the formulations of KG1-KG7 in terms of $\phi_\text{B}$ seriously: namely Eq.~\eqref{DKG1bandlimited}-\eqref{DKG7bandlimited}. Taken literally as written, what are these theories about? 
Intuitively these theories are about a field $\phi_\text{B}$ which maps points on a manifold ($p\in\mathcal{M}$) into field amplitudes ($\phi_\text{B}(p)\in\mathbb{R}$). That is a field $\phi_\text{B}:\mathcal{M}\to \mathcal{V}$ with a manifold $\mathcal{M}$ and value space $\mathcal{V}=\mathbb{R}$. Thus, taking $\phi_\text{B}:\mathcal{M}\to \mathcal{V}$ seriously as a fundamental field leads us to thinking of $\mathcal{M}$ as the theory's underlying manifold and $\mathcal{V}=\mathbb{R}$ as the theory's value space. Indeed, on this interpretation
KG1-KG7 are continuum spacetime theories of the sort we
are used to interpreting (albeit ones which consider only bandlimited fields). 

Note that here we have $\mathcal{M}\cong\mathbb{R}^d$ with $d=2$ for KG1-KG3 and $d=3$ for KG4-KG7. As such, in either case we have access to global coordinate systems. The fields $\phi_\text{B}$ considered by this interpretation are bandlimited in the following sense. There exists a fixed special diffeomorphism $d_\text{coor.}:\mathbb{R}^d\to\mathcal{M}$ which giving us a fixed special global coordinate system for $\mathcal{M}$. In these special coordinates the field is bandlimited with $\omega,k_1,k_2\in[-K,K]$ where $K=\pi/a$. That is $\phi_\text{B}\circ d_\text{trans.}\in F^K$. It is in these special coordinates that the fields $\phi_\text{B}$ obey the dynamical equations Eq.~\eqref{DKG1bandlimited}-Eq.~\eqref{DKG7bandlimited}.

In our construction of $\phi_\text{B}$, this special coordinate system is the one which is associated to the translation operations $T^\epsilon$ by our translation-matching condition, namely $d_\text{coor.}=d_\text{trans.}$. Here however, we are trying to interpret KG1-KG7 as formulated above independent of how we arrived here. Thus, presently, this special coordinate system $d_\text{coor.}$ is just something promised to us by God. Of course, we know from the lessons of general covariance to be skeptical of ``special coordinate systems''. Coordinate systems are representational artifacts, reflecting no physics. We can and should reformulate our theories in a generally covariant way so as to avoid them. A coordinate independent view of them will be given in Sec.~\ref{SecFullGenCov}.

As I will discuss, taking $\mathcal{M}$ to be these theories' underlying manifold has substantial consequences for these theories' locality properties and symmetries. To preview: this third interpretation either fixes all of our issues with the first and second interpretations. 

In our first interpretation there was some tension between our theories' differences in locality and their differences in convergence rate in the continuum limit. The second interpretation addressed this tension in a hamfistedly way: denying that there are differences in locality in the first place. This move also had the unfortunate consequence of undercutting our ability to use the lattice sites to reason about locality. This interpretation improves on this by instead bringing the locality of these theories into harmony with their convergence rates in the continuum limit. Not much changes between the discrete heat equations considered in \cite{DiscreteGenCovPart1} and the discrete Klein Gordon equations considered here. As such, I will leave discussion of this issue to \cite{DiscreteGenCovPart1} and direct the interested reader there.

In our first interpretation, KG6 and KG7 were seen as distinct theories with radically different symmetries despite there being a nice one-to-one correspondence between their solutions. In our second interpretation, KG6 and KG7 were more satisfyingly seen to be completely equivalent, differing only by a change of basis in the value space. Here too KG6 and KG7 will turn out to be completely equivalent differing only by a change of coordinates. Moreover, as I will discuss, we can here see KG6 and KG7 as parts of a larger unified theory viewed in two ways with differently limited representational capacities.

Our first interpretation outright denied the possibility that KG1-KG7 could have continuous symmetries (e.g., translations, rotations, Lorentz boosts). Our second interpretation then fixed this oversight by revealing KG1-KG7's hidden continuous translation and rotation symmetries and even their (limited) Lorentzian boost symmetries. However, these were unintuitively categorized as internal symmetries. This interpretation maintains all of these hidden symmetries, but more satisfyingly categorizes them as external symmetries. Moreover, seeing KG6 and KG7 as parts of a larger unified theory will guide us towards a perfectly Lorentzian lattice theory.

As I discussed in Sec.~\ref{SecKlein2}, the improvements that the second interpretation made over the first one all centered around the following realization. The lattice structures appearing in KG1-KG7 are merely representational artifacts and as such should not be taken seriously as a substantive part of the theory. As I will discuss, this realization is deepened and clarified on the third interpretation.

\subsection{Bandlimited Symmetries}
How does this externalization move affect our theory's capacity for symmetry? Now that we have a continuous spacetime manifold $\mathcal{M}$ underlying these theories, we have a greatly improved capacity for external symmetries, namely $\text{Auto}(\mathcal{M})=\text{Diff}(\mathcal{M})$. By construction all of the symmetries which we wished to externalize $G^\text{dym}_\text{to-be-ext}$ fit inside of $\text{Diff}(\mathcal{M})$. 

In addition to these external symmetries, we also have some capacity for internal symmetries, namely $\text{Auto}(\mathcal{V})$. Note that the space of (potentially off-shell) bandlimited fields are themselves vectors $\phi_\text{B}(t,x,y)\in F^K$. Namely, they are closed under addition and scalar multiplication and hence a vector space. This addition and scalar multiplication is carried out in parallel at each spacetime point. Thus, the field's value space $\mathcal{V}=\mathbb{R}$ is also structured like a vector space. Therefore, $\text{Auto}(\mathcal{V})=\text{Aff}(\mathbb{R})$ such that our internal symmetries are linear-affine rescalings of $\phi_\text{B}$. We might also find gauge symmetries by allowing these to vary smoothly across the manifold. Thus, in total the possibly symmetries for our theories under this interpretation are,
\begin{align}\label{BandlimitedSymmetries}
s:\quad\phi_\text{B}\mapsto C \, \phi_\text{B}\circ d +c
\end{align}
where $C:\mathcal{M}\to\mathbb{R}$ and $c:\mathcal{M}\to\mathbb{R}$ are some real scalar functions and $d:\mathcal{M}\to\mathcal{M}$ is some  diffeomorphism \mbox{$d\in\text{Diff}(\mathcal{M})$}. Note that 

Are these more or less possible symmetries than we had on our second interpretation? $\text{Diff}(\mathcal{M})$ is a tremendously large group and is in fact bigger than $\text{Aff}(\mathbb{R}^L)$. One might expect that we have strictly more possible symmetries here. However, looks can be deceiving. 

We can translate our old potential symmetries into the new continuum setting as follows. Note that because $\mathbb{R}^L \cong F^K$ we have $\text{Aff}(\mathbb{R}^L)\cong \text{Aff}(F^K)$. Using $d_\text{coor.}$ we can think of $\text{Aff}(F^K)$ as a subgroup of $\text{Aff}(F(\mathcal{M}))$. Note that we can also see the above discussed transformations (i.e., \mbox{$\text{Aff}(\mathbb{R})$} varying smoothly over \mbox{$\text{Diff}(\mathcal{M})$}) as a subgroup of $\text{Aff}(F(\mathcal{M}))$. 

Viewed this way, neither is strictly larger. Indeed, $\text{Aff}(F^K)$ includes transformations in Fourier space which are inaccessible from $\text{Aff}(\mathbb{R})$ and $\text{Diff}(\mathcal{M})$. In particular, the local Fourier rescaling symmetry from Sec.~\ref{SecKlein2} is no longer available to us. On the other hand, our new symmetries contain transformations which are not closed over $F^K$. However, this comparison seems unfair, what if we restrict our attention transformations closed over $F^K$? With this restriction, our new range of possible symmetries is a strict subset of $\text{Aff}(F^K)$. 

Thus, in moving from the second to the third interpretation, we have actually lost some capacity for symmetry. The source of this reduction in capacity is us imposing structure on the theory by our choice of manifold and value space.

As I will now discuss, besides the local Fourier rescaling symmetry, all of our previous symmetries are maintained on this third interpretation. Moreover, these remaining symmetries are reformulated and recategorized. As I will now show, on this interpretation the continuous translation, rotation and (limited) Lorentz boost symmetries identified earlier are now honest-to-goodness manifold symmetries, represented by diffeomorphisms $d\in\text{Diff}(\mathcal{M})$.

\subsubsection*{Symmetries of KG1-KG7: Third Attempt}
Which transformations of the form Eq.~\eqref{BandlimitedSymmetries} are symmetries for KG1-KG7? A technical investigation of the symmetries of KG1-KG7 on this interpretation is carried out in Appendix~\ref{AppA}, but the results are the following. For KG1 and KG2 the dynamical symmetries of the form Eq.~\eqref{BandlimitedSymmetries} are:
\begin{flushleft}
\begin{enumerate}
    \item[1)]  continuous translation which maps \mbox{$\phi_\text{B}(t,x)\mapsto \phi_\text{B}(t-t_0,x-x_0)$} for $t_0,x_0\in\mathbb{R}$,
    \item[2)] two negation symmetries which map \mbox{$\phi_\text{B}(t,x)\mapsto \phi_\text{B}(-t,x)$} and \mbox{$\phi_\text{B}(t,x)\mapsto \phi_\text{B}(t,-x)$} respectively,
    \item[3)] global linear rescaling which maps \mbox{$\phi_\text{B}(t,x)\mapsto c_1\,\phi_\text{B}(t,x)$} for some $c_1\in\mathbb{R}$,
    \item[4)] local affine rescaling which maps \mbox{$\phi_\text{B}(t,x)\mapsto c_1\,\phi_\text{B}(t,x)+c_2(t,x)$} for some $c_2(t,x)$ which is also a solution of the dynamics.
\end{enumerate}
\end{flushleft}
These are exactly the same symmetries that we found on the previous interpretation (sans local Fourier rescaling) just recategorized: the translation and reflection symmetries are here external symmetries. Translation here is generated by \mbox{$\exp(-t_0\partial_t-x_0\partial_x)$} for some $t_0,x_0\in\mathbb{R}$. Notice that translating a bandlimited function by any amount preserves its bandwidth. Thus, $\phi_\text{B}(t,x)$ remains in $F^K$ upon translation.

Thus, the first big lesson from Sec.~\ref{SecKlein2} is repeated here: despite the fact that our discrete theories KG1 and KG2 can be represented on a lattice, they nonetheless have a continuous translation symmetry. This continuous translation symmetry was hidden on our first interpretation because we there took the lattice to be a fundamental part of the manifold. Here, we do not take the lattice structure so seriously. We have embedded it onto the manifold where it then disappears from view as just one of many possible sample lattices.

Allow me to skip over KG3 temporarily. For KG4 the dynamical symmetries of the form Eq.~\eqref{BandlimitedSymmetries} are:
\begin{flushleft}\begin{enumerate}
    \item[1)] continuous translation which maps \mbox{$\phi_\text{B}(t,x,y)\mapsto \phi_\text{B}(t-t_0,x-x_0,y-y_0)$} for some $t_0,x_0,y_0\in\mathbb{R}$,
    \item[2)] three negation symmetries which map
    \mbox{$\phi_\text{B}(t,x,y)\mapsto \phi_\text{B}(-t,x,y)$}, and \mbox{$\phi_\text{B}(t,x,y)\mapsto \phi_\text{B}(t,-x,y)$}, and \mbox{$\phi_\text{B}(t,x,y)\mapsto \phi_\text{B}(t,x,-y)$} respectively,
    \item[3)] a 4-fold symmetry which maps \mbox{$\phi_\text{B}(t,x,y)\mapsto \phi_\text{B}(t,y,-x)$},
    \item[4)] global linear rescaling which maps \mbox{$\phi_\text{B}(t,x)\mapsto c_1\,\phi_\text{B}(t,x)$} for some $c_1\in\mathbb{R}$,
    \item[5)] local affine rescaling which maps \mbox{$\phi_\text{B}(t,x)\mapsto c_1\,\phi_\text{B}(t,x)+c_2(t,x)$} for some $c_2(t,x)$ which is also a solution of the dynamics.
\end{enumerate}\end{flushleft}
These are exactly the same symmetries that we found on the previous interpretation (sans local Fourier rescaling) just recategorized: the translation, reflection, and 4-fold symmetries are here external symmetries. Thus, KG4 has three continuous translation symmetries despite initially being represented on a lattice, Eq.~\eqref{KG4Long}.

Let's next consider KG5. For KG5 the dynamical symmetries of the form Eq.~\eqref{BandlimitedSymmetries} are:
\begin{flushleft}\begin{enumerate}
    \item[1)] continuous translation which maps \mbox{$\phi_\text{B}(t,x,y)\mapsto \phi_\text{B}(t-t_0,x-x_0,y-y_0)$} for some $t_0,x_0,y_0\in\mathbb{R}$,
    \item[2)] a negation symmetry which maps
    \mbox{$\phi_\text{B}(t,x,y)\mapsto \phi_\text{B}(-t,x,y)$}, and an exchange symmetry which maps \mbox{$\phi_\text{B}(t,x,y)\mapsto \phi_\text{B}(t,y,x)$},
    \item[3)] a 6-fold symmetry which maps \mbox{$\phi_\text{B}(t,x,y)\mapsto \phi_\text{B}(t,-y,x+y)$}. (Roughly, this permutes the three terms in Eq.~\eqref{DKG5bandlimited}),
    \item[4)] global linear rescaling which maps \mbox{$\phi_\text{B}(t,x)\mapsto c_1\,\phi_\text{B}(t,x)$} for some $c_1\in\mathbb{R}$,
    \item[5)] local affine rescaling which maps \mbox{$\phi_\text{B}(t,x)\mapsto c_1\,\phi_\text{B}(t,x)+c_2(t,x)$} for some $c_2(t,x)$ which is also a solution of the dynamics.
\end{enumerate}\end{flushleft}
These are exactly the same symmetries that we found on the previous interpretation (sans local Fourier rescaling) just recategorized: the translation, reflection, and 6-fold symmetries are here external symmetries. Thus, KG5 has three continuous translation symmetries despite initially being represented on a lattice, Eq.~\eqref{KG5Long}. Note that after a coordinate transformation Eq.~\eqref{CoorKG7KG6} the above noted 6-fold symmetry is realized as one-sixth rotations, see Eq.~\eqref{DKG5Skew}.

Before moving on to analyze the symmetries of KG6 and KG7, let's first see what this interpretation has to say about them being equivalent to one another. As discussed in Sec.~\ref{SecKlein2} KG6 and KG7 have a solution-preserving vector space isomorphism between a substantial portion of their models, including all of their solutions. Since our second and third interpretations are related by a solution-preserving vector space isomorphism, \mbox{$E:\mathbb{R}^L\to F^K$}, the same is broadly true here. However, some of the details change.

As I noted in Sec.~\ref{SecKlein2} each of KG4-KG7 are equivalent to each other in a weak sense: approximately in the continuum limit regime, $\vert \omega\vert,\vert k_1\vert,\vert k_2\vert\ll \pi$. Here KG4-KG7 are approximately equivalent in the $\vert \omega\vert,\vert k_1\vert,\vert k_2\vert\ll K=\pi/a$ regime. Note however that this is not the continuum limit regime, we are already in the continuum. Rather this is the regime which is significantly below the bandwidth $K$. While each of KG4-KG7 are equivalent in this weak sense, KG6 and KG7 are equivalent in a much stronger sense: KG6 and KG7 is in \textit{exact} one-to-one correspondence over \textit{the whole of} $\sqrt{k_1^2+k_2^2} < K$ and indeed more. This includes all of their solutions but not all of $\omega,k_1,k_2\in[-\pi,\pi]$.

This exact correspondence is mediated by Eq.~\eqref{SkewKG7KG6} and Eq.~\eqref{SkewKG6KG7} in Fourier space or equivalently the coordinate transformations Eq.~\eqref{CoorKG7KG6} and its inverse. On our second interpretation, we ran into technical trouble here due to the $2\pi$ periodic identification of the discrete planewaves $\bm{\Phi}(\omega,k_1,k_2)$ stemming from Euler's identity. This caused our maps between KG6 and KG7 not to be each other's inverses over the whole of $\mathbb{R}^L\cong\mathbb{R}^\mathbb{Z}\otimes\mathbb{R}^\mathbb{Z}\otimes\mathbb{R}^\mathbb{Z}$. We were led to consider instead the subspaces $\mathbb{R}^L_\text{KG6}$ and $\mathbb{R}^L_\text{KG7}$ where these transformations were each other's inverse.

In the present interpretation, something similar happens but with some key differences. Unlike before, here our maps between KG6 and KG7 given wide scope are invertible and indeed each other's inverses. Wide scope here meaning seen as acting on $F(\mathbb{R}^3)$ the set of all functions $f(t,x,y)$. However, as before, seen as acting on the relevant vector space $F^K$ we have issues. Namely, Eq.~\eqref{SkewKG7KG6} and Eq.~\eqref{SkewKG6KG7}, understood as acting on $F^K$ are not closed: a function supported over $\omega,k_1,k_2\in[-K,K]$ may have support outside of here after these transformations.

In Sec.~\ref{SecKlein2} we fixed the non-invertible issue by focusing on the subspaces $\mathbb{R}^L_\text{KG6}$ and $\mathbb{R}^L_\text{KG7}$ where these transformations are each other's inverses. We might overcome our current not-closed-over-$F^K$ issue here in the same way defining 
\begin{align}
F^K_\text{KG7}\coloneqq&\text{span}(\phi(t,x,y;\omega,k_1,k_2)\vert\,\text{in }F^K\\
\nonumber
&\text{ before and after applying Eq.~\eqref{SkewKG7KG6}})\\
F^K_\text{KG6}\coloneqq&\text{span}(\phi(t,x,y;\omega,k_1,k_2)\vert\,\text{in }F^K\\
\nonumber
&\text{ before and after applying Eq.~\eqref{SkewKG6KG7}})\\
\label{FKrotinv}
F^K_\text{rot.inv.}\!\!\coloneqq&\text{span}(\phi(t,x,y;\omega,k_1,k_2)\vert k_1^2+k_2^2 < K^2).
\end{align}
Note that under our embedding $E$ these subspaces are isomorphic to $\mathbb{R}^L_\text{KG6}$, $\mathbb{R}^L_\text{KG7}$ and $\mathbb{R}^L_\text{rot.inv.}$.

Restricted to $F^K_\text{KG6}$ and $F^K_\text{KG7}$ these transformations are invertible and indeed are each other's inverses. All of KG6's solutions are in $F^K_\text{KG6}$ (and moreover they are in $F^K_\text{rot.inv}$ as well). Similarly, all of KG7's solutions are in $F^K_\text{KG7}$. Thus, Eq.~\eqref{CoorKG7KG6} maps generic solutions to KG7 onto generic solutions for KG6 in an invertible way. Therefore, Eq.~\eqref{CoorKG7KG6} and its inverse give us not only a one-to-one correspondence between the solutions to KG6 and KG7 but a solution-preserving vector-space isomorphism between KG6 and KG7. One can gloss this situation saying: the KPMs of KG6 and KG7 are not isomorphic, but their DPMs are.

The fact that these transformations (namely, Eq.~\eqref{CoorKG7KG6} and its inverse) are of the form Eq.~\eqref{BandlimitedSymmetries} means that for any symmetry transformation for KG6 there is a corresponding symmetry transformation for KG7 and vice versa. In the present interpretation KG6 and KG7 are thoroughly equivalent: We have a solution-preserving vector-space isomorphism between them. Indeed, on this interpretation the only difference between KG6 and KG7 is a change of coordinates. 

Thus our second big lesson from Sec.~\ref{SecKlein2} is repeated here: despite KG6 and KG7 being initially presented to us with very different lattice structures (i.e., a square lattice in space versus a hexagonal lattice in space) they have nonetheless turned out to be completely equivalent to one another. This equivalence was hidden from us on our first interpretation because we there took the lattice too seriously. As I discuss in Sec.~\ref{SecKlein2}, this reduced their continuous rotation symmetries down to quarter rotations and one-sixth rotations respectively and thereby made them inequivalent. Here, we do not take the lattice structure so seriously. We have embedded it onto the manifold where it then disappears from view as just one of many possible sample lattices.

One substantial difference from our second interpretation should be noted here. We are not forced to shrink $F^K$ down to $F^K_\text{KG6}$ and $F^K_\text{KG7}$ to see KG6 and KG7 as equivalent; we have another option. Namely, rather than shrinking $F^K$ we could also expand it. Consider a theory just like KG6 on this interpretation, but with a bandwidth of $2\,K$ instead of $K$. Clearly, KG6 is a subtheory of this expanded theory. Note that the coordinate transformation which maps KG7 onto KG6, namely Eq.~\eqref{SkewKG7KG6}, skews its support in Fourier space. However, its support remains inside of $k_1,k_2\in[-2K,2K]$. Thus, KG7 is also a subtheory of our expanded theory. We can here see KG6 and KG7 as parts of a larger theory. As I will discuss in Sec.~\ref{PerfectLorentz}, KG6 and KG7 are the parts of this extended theory which is visible to us when we restrict ourselves to only certain sets of representational tools. This line of thought will lead us to a perfectly Lorentzian lattice theory.

For the rest of this subsection I will only discuss the symmetries of KG6; analogous conclusions are true for KG7 after applying Eq.~\eqref{CoorKG7KG6}. For KG6 the dynamical symmetries of the form Eq.~\eqref{BandlimitedSymmetries} are:
\begin{flushleft}\begin{enumerate}
    \item[1)] continuous translation which maps \mbox{$\phi_\text{B}(t,x,y)\mapsto \phi_\text{B}(t-t_0,x-x_0,y-y_0)$} for some $t_0,x_0,y_0\in\mathbb{R}$,
    \item[2)] three negation symmetries which map
    \mbox{$\phi_\text{B}(t,x,y)\mapsto \phi_\text{B}(-t,x,y)$}, and \mbox{$\phi_\text{B}(t,x,y)\mapsto \phi_\text{B}(t,-x,y)$}, and \mbox{$\phi_\text{B}(t,x,y)\mapsto \phi_\text{B}(t,x,-y)$} respectively,
    \item[3)] continuous rotation which maps \mbox{$\phi_\text{B}(t,x,y)$} to \mbox{$\phi_\text{B}(t,x\cos(\theta)-y\sin(\theta),x\sin{\theta}+y\cos(\theta))$} for some $\theta\in\mathbb{R}$. (This being a symmetry requires some qualification as I will discuss below.),
    \item[4)] two Lorentz boosts which map \mbox{$\phi_\text{B}(t,x,y)$} to \mbox{$\phi_\text{B}(t\cosh(w)+x\sinh(w),t\sinh(w)+x\cosh(w),y)$} and map \mbox{$\phi_\text{B}(t,x,y)$} to \mbox{$\phi_\text{B}(t\cosh(w)+y\sinh(w),x,t\sinh(w)+y\cosh(w))$} respectively for some $w\in\mathbb{R}$. (This being a symmetry requires some qualification as I will discuss below.),
    \item[5)] global linear rescaling which maps \mbox{$\phi_\text{B}(t,x)\mapsto c_1\,\phi_\text{B}(t,x)$} for some $c_1\in\mathbb{R}$,
    \item[6)] local affine rescaling which maps \mbox{$\phi_\text{B}(t,x)\mapsto c_1\,\phi_\text{B}(t,x)+c_2(t,x)$} for some $c_2(t,x)$ which is also a solution of the dynamics.
\end{enumerate}\end{flushleft}
These are exactly the same symmetries that we found on the previous interpretation (sans local Fourier rescaling) just recategorized: the translation, reflection, and rotation, and Lorentz boost symmetries are here external symmetries. Thus, KG6 (and KG7) each have three continuous translation symmetries despite initially being represented on a lattice.

In addition to this, KG6 has a (qualified) continuous rotation symmetry. But in what sense is this only a qualified symmetry? Much of the discussion from the previous section following Eq.~\eqref{RthetaDef} applies here as well with some key differences.

As I noted in Sec.~\ref{SecKlein2}, each of KG4-KG7 are rotation invariant in a weak sense: approximately in the continuum limit regime, $\vert \omega\vert,\vert k_1\vert,\vert k_2\vert\ll \pi$. Here KG4-KG7 are approximately rotation invariant in the $\vert \omega\vert,\vert k_1\vert,\vert k_2\vert\ll K=\pi/a$ regime. Note however that this is not the continuum limit regime, we are already in the continuum. Rather, this is the regime which is significantly below the bandwidth $K$. While each of KG4-KG7 are rotation invariant in this weak sense, KG6 is rotation invariant in a much stronger sense: KG6 is \textit{exactly} rotation over \textit{the whole of} $\sqrt{k_1^2+k_2^2} < K$.

Rotation is here generated by \mbox{$\exp(-\theta (x \partial_y - y \partial_x))$} for some $\theta\in\mathbb{R}$. Acting on the continuum planewave basis this transformation merely rotates their wavenumbers in Fourier space. On our second interpretation, we ran into technical trouble here due to the $2\pi$ periodic identification of the discrete planewaves $\bm{\Phi}(\omega,k_1,k_2)$ stemming from Euler's identity. This caused $R^\theta$ not to be invertible over the whole of $\mathbb{R}^L\cong\mathbb{R}^\mathbb{Z}\otimes\mathbb{R}^\mathbb{Z}\otimes\mathbb{R}^\mathbb{Z}$. We were led to consider only the rotation invariant subspace $\mathbb{R}^L_\text{rot.inv.}$.

In the present interpretation, something similar happens but with some key differences. Unlike before, here given wide scope rotations are always invertible. Wide scope here meaning seen as acting on $F(\mathbb{R}^3)$ the set of all functions $f(t,x,y)$. However, as before, seen as acting on the relevant vector space $F^K$ we have issues. Namely, understood as acting on $F^K$ rotation is not closed: some $f$ supported over $\omega,k_1,k_2\in[-K,K]$ may have support outside of here after being rotated.

In Sec.~\ref{SecKlein2} we fixed the non-invertible issue by focusing on the rotation invariant subspace $\mathbb{R}^L_\text{rot.inv.}\subset \mathbb{R}^L$. We might overcome our current not-closed-over-$F^K$ issue here in the same way. While rotation is not closed over $F^K$ it is closed over $F^K_\text{rot.inv.}$ defined in Eq.~\eqref{FKrotinv}. Note that under our embedding $E$ this subspace is isomorphic to the vector space $\mathbb{R}^L_\text{rot.inv.}$ defined in Eq.~\eqref{RLrotinv}.

Note that all of KG6's solutions are within $F^K_\text{rot.inv.}$. Within $F^K_\text{rot.inv.}$ the above discussed rotation transformation is of the form Eq.~\eqref{BandlimitedSymmetries} and maps solutions to KG6 onto solution to KG6 in an invertible way, and is hence a symmetry. One can gloss this situation saying: rotation does not map the KPMs of KG6 onto themselves in an invertible way, but it does for the DPMs of KG6. If we cut the KPMs of KG6 down to $F^K_\text{rot.inv}$ this minor issue is fixed.

This adds to our first big lesson from Sec.~\ref{SecKlein2}: despite the fact that KG6 and KG7 can be represented on a cubic 3D lattice and a hexagonal 3D lattice respectively, they nonetheless both have a continuous rotation symmetry. This, in addition to their continuous translation symmetries. These continuous translation and rotations symmetries were hidden from us on our first interpretation because we there took the lattice representations too seriously. Here, we do not take the lattice structure so seriously. We have embedded it onto the manifold where it then disappears from view as just one of many possible sample lattices.

One substantial difference from our second interpretation should be noted here. We are not forced to shrink $F^K$ down to $F^K_\text{rot.inv.}$ to make KG6 rotation invariant; we have another option. Rather than shrinking $F^K$ down to its rotation invariant core, we could also expand it by adding in every state reachable from $F^K$ by rotation yielding $F^{\sqrt{2}K}_\text{rot.inv.}$. Consider a theory just like KG6 on this interpretation, but defined over $F^{\sqrt{2}K}_\text{rot.inv.}$ instead of $F^K$. Clearly, KG6 is a subtheory of this expanded theory. As I will discuss in Sec.~\ref{PerfectLorentz}, KG6 is the part of this extended theory which is visible to us when we restrict ourselves to only a certain set of representational tools. This line of thought will lead us to a perfectly Lorentzian lattice theory.

Let's next discuss how KG6's (limited) Lorentzian boost symmetry is realized on this interpretation. The Lorentz boosts of KG6 are here generated by \mbox{$\exp(-w (x \partial_t + t \partial_x))$} and \mbox{$\exp(-w (y \partial_t + t \partial_y))$} for some boost parameter $w\in\mathbb{R}$. Notice that boosting a bandlimited function also boosts its support in Fourier space. On our second interpretation, we ran into technical trouble here due to the $2\pi$ periodic identification of the discrete planewaves $\bm{\Phi}(\omega,k_1,k_2)$ stemming from Euler's identity. This caused $R^\theta$ not to be invertible over the whole of $\mathbb{R}^L\cong\mathbb{R}^\mathbb{Z}\otimes\mathbb{R}^\mathbb{Z}\otimes\mathbb{R}^\mathbb{Z}$. We were led to consider only the rotation invariant subspace $\mathbb{R}^L_\text{rot.inv.}$.

In the present interpretation, something similar happens but with some key differences. Unlike before, here given wide scope Lorentz boosts are always invertible. Wide scope here meaning seen as acting on $F(\mathbb{R}^3)$ the set of all functions $f(t,x,y)$. However, as before, seen as acting on the relevant vector space $F^K$ we have issues. Namely, understood as acting on $F^K$ Lorentz boosts are not invertible nor are they even functions: some $f$ supported over $\omega,k_1,k_2\in[-K,K]$ may have support outside of here after being boosted. 

In Sec.~\ref{SecKlein2} we fixed the non-invertible issue by focusing on the finite-sized region around $\omega,k_1,k_2=0$ and boost parameter $w=0$ where $\Lambda^w_\text{j,n}$ and $\Lambda^w_\text{j,m}$ are invertible. Moreover, we were even able to find a representation of the Poincar\'e algebra. We might overcome our current not-closed-over-$F^K$ issue here in the same way. There is some finite-sized region around $\omega,k_1,k_2=0$ and boost parameter $w=0$ where Lorentz boosts keep us inside $F^K$. Moreover, as above, we could find a representation of the Poincar\'e algebra over $F^K$ as well as a finite portion of the Poincar\'e group.

So far the situation here is very much like it was in Sec.~\ref{SecKlein2}. Our first big lesson is here repeated: despite the fact that KG6 and KG7 can be represented on a cubic 3D lattice and a hexagonal 3D lattice respectively, they nonetheless both have a Lorentzian boost symmetries (in a finite but limited regime).

However, there is one substantial difference here which ultimately points the way towards a perfectly Lorentzian lattice theory. As I will discuss in Sec.~\ref{PerfectLorentz}, we can see KG6 as the part of some extended theory which is visible to us when we restrict ourselves to only a certain set of representational tools. This extended theory is perfectly Lorentz invariant. Viewed this way, KG6's failure to be Poincar\'e invariant in an unqualified way ought to be seen as a problem of representational capacity and not of physics.

Finally, let's consider KG3. For KG3 the dynamical symmetries of the form Eq.~\eqref{BandlimitedSymmetries} are:
\begin{flushleft}
\begin{enumerate}
    \item[1)]  continuous translation which maps \mbox{$\phi_\text{B}(t,x)\mapsto \phi_\text{B}(t-t_0,x-x_0)$} for $t_0,x_0\in\mathbb{R}$,
    \item[2)] two negation symmetries which map \mbox{$\phi_\text{B}(t,x)\mapsto \phi_\text{B}(-t,x)$} and \mbox{$\phi_\text{B}(t,x)\mapsto \phi_\text{B}(t,-x)$} respectively,
    \item[3)] a Lorentz boosts which maps \mbox{$\phi_\text{B}(t,x)$} to \mbox{$\phi_\text{B}(t\cosh(w)+x\sinh(w),t\sinh(w)+x\cosh(w))$},
    \item[4)] global linear rescaling which maps \mbox{$\phi_\text{B}(t,x)\mapsto c_1\,\phi_\text{B}(t,x)$} for some $c_1\in\mathbb{R}$,
    \item[5)] local affine rescaling which maps \mbox{$\phi_\text{B}(t,x)\mapsto c_1\,\phi_\text{B}(t,x)+c_2(t,x)$} for some $c_2(t,x)$ which is also a solution of the dynamics.
\end{enumerate}
\end{flushleft}
These are exactly the same symmetries that we found on the previous interpretation (sans local Fourier rescaling) just recategorized: the translation, reflection, and (limited) Lorentz boost symmetries are here external symmetries. 

\vspace{0.25cm}

To summarize: this third attempt at interpreting KG1-KG7 has fixed all of the issues with our first and second interpretations. Firstly, the tension between our theories' differences in locality and their differences in convergence rate in the continuum limit has been harmoniously resolved. This improves upon the hamfisted dissolution of tension given by our second interpretation, see \cite{DiscreteGenCovPart1}. Secondly, like on the second interpretation, here KG6 and KG7 are seen to be completely equivalent. Here however, we can moreover see KG6 and KG7 in their entirety as parts of a larger unified theory. Finally, like on the second interpretation, we have here exposed KG1-KG7's hidden continuous translation and rotation symmetries and even limited Lorentz symmetries. Here however, these are more satisfyingly categorized as external symmetries.

Like the previous interpretation, this interpretation by and large invalidates all of the first intuitions laid out in Sec.~\ref{SecIntro}. Indeed here the lattice seems to play a merely representational role in the theory: it does not restrict our symmetries. Moreover, theories initially appearing with different lattices may nonetheless turn out to be substantially equivalent. The process for switching between lattice structures is here done by Nyquist-Shannon resampling. Indeed, the third big lesson from Sec.~\ref{SecKlein2} is repeated here: there is no sense in which these lattice structures are essentially ``baked-into'' these theories; our bandlimited theories make no reference to any lattice structure. 

As I will discuss in Sec.~\ref{SecDisGenCov}, these three lessons lay the foundation for a strong analogy between the lattice structures which appear in our discrete spacetime theories and the coordinate systems which appear in our continuum spacetime theories. This ultimately gives rise to a discrete analog of general covariance.

Before that however, a bit more must be said about this third interpretation.

\section{A Third Attempt at Interpreting KG1-KG7 - Part 2}\label{SecKlein3Extra}
The previous section presented a third interpretation of our seven discrete theories KG1-KG7. This section will flesh out this interpretation in two ways. Firstly, I will provide some more explicit demonstrations of these theories' symmetries. In particular, I will explicitly demonstrate the fact that the symmetries of these theories are independent of which sample lattice we use to represent them. Secondly, I will write our above formulations of KG1-KG7 in terms of $\phi_\text{B}$ in the coordinate-independent language of differential geometry. This will help clarify their symmetries and assumed background structures.

\subsection{Demonstration of Bandlimited Symmetries}
\begin{figure*}[t]
\centering
\includegraphics[width=0.9\textwidth]{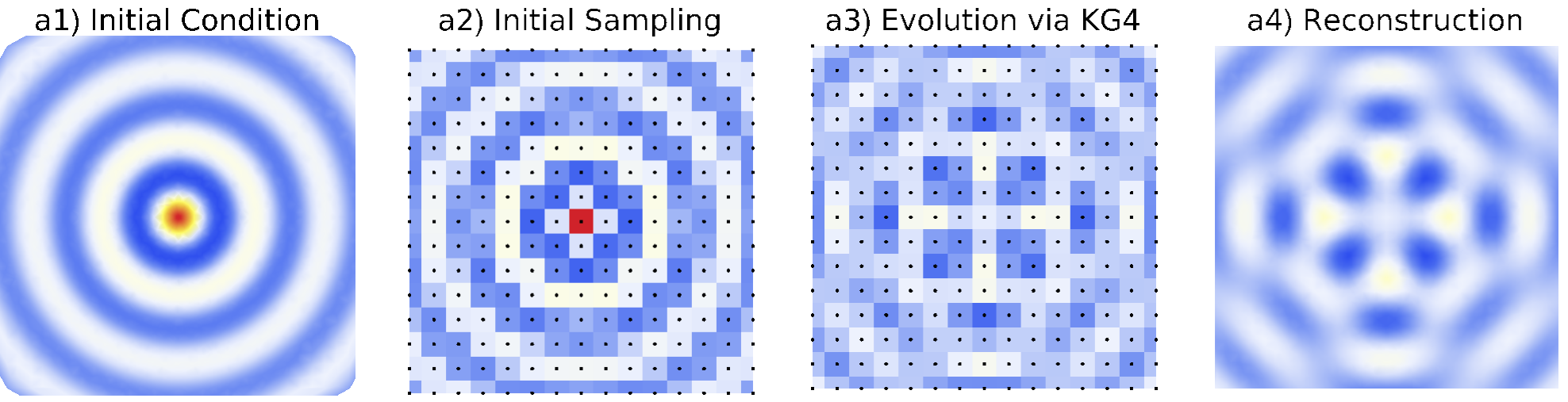}
\centering
\includegraphics[width=0.9\textwidth]{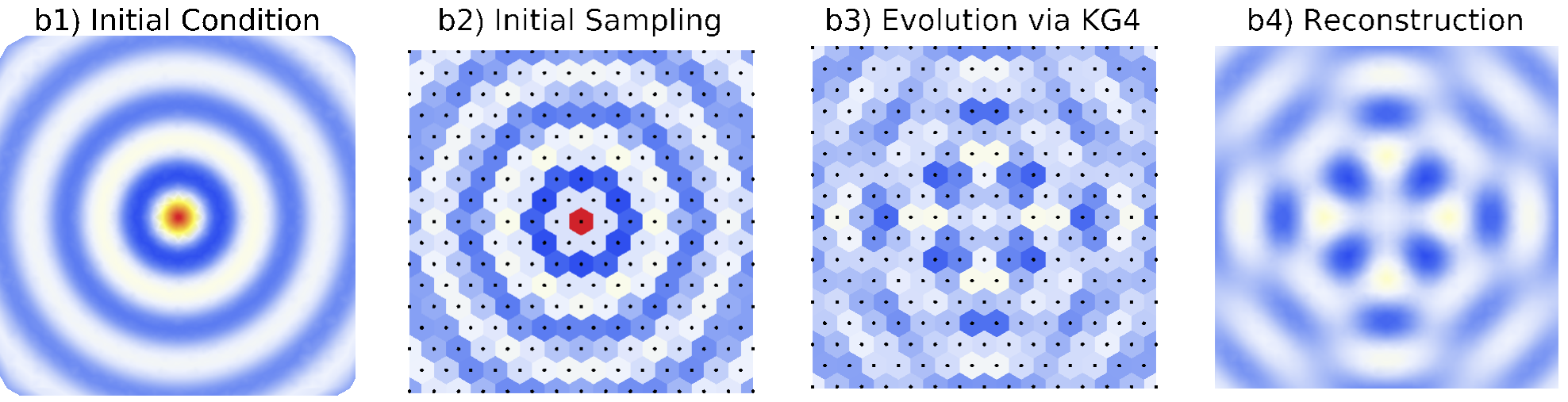}
\centering
\includegraphics[width=0.9\textwidth]{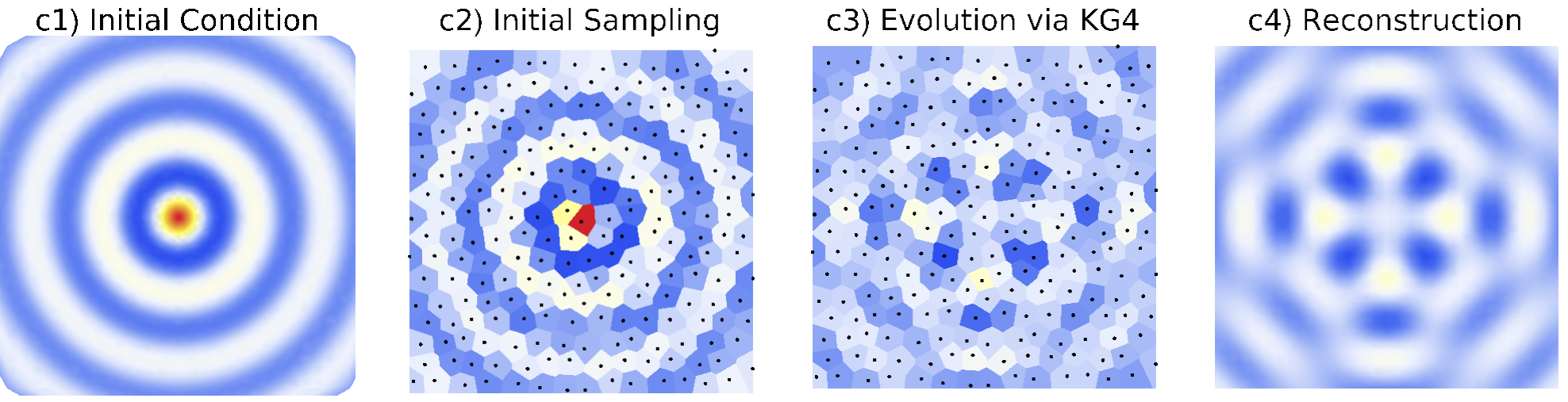}
\caption{The dynamics of KG4 is here shown being carried out in a variety of lattice representations. In the left most column the initial condition is shown in its bandlimited representation, given by Eq.~\eqref{InCond}. In the rightmost column the final evolved state is shown in its bandlimited representation. Here the evolution time is $t=2$ and the field mass is $\mu=0$. This state can be found in four different ways. Firstly by applying the bandlimited dynamics Eq.~\eqref{DKG4bandlimited} to the bandlimited initial condition. The other three ways are shown in the three rows of this figure. The first row shows the initial condition being sampled onto a square lattice. This is then evolved forward in time via Eq.~\eqref{KG4Long}. The bandlimited representation of the final state is then recovered through the methods discussed in Sec.~\ref{SecSamplingTheory}. The second and third rows show the same process carried out on a hexagonal lattice and an irregular lattice. Notice that the final state has a 4-fold symmetry regardless of how the dynamics is represented. Notice that the final state is the same regardless of how the dynamics is represented. We can represent any bandlimited dynamics on any (sufficiently dense) lattice.}\label{FigEvolutionKG4}
\end{figure*}

\begin{figure*}[t]
\centering
\includegraphics[width=0.9\textwidth]{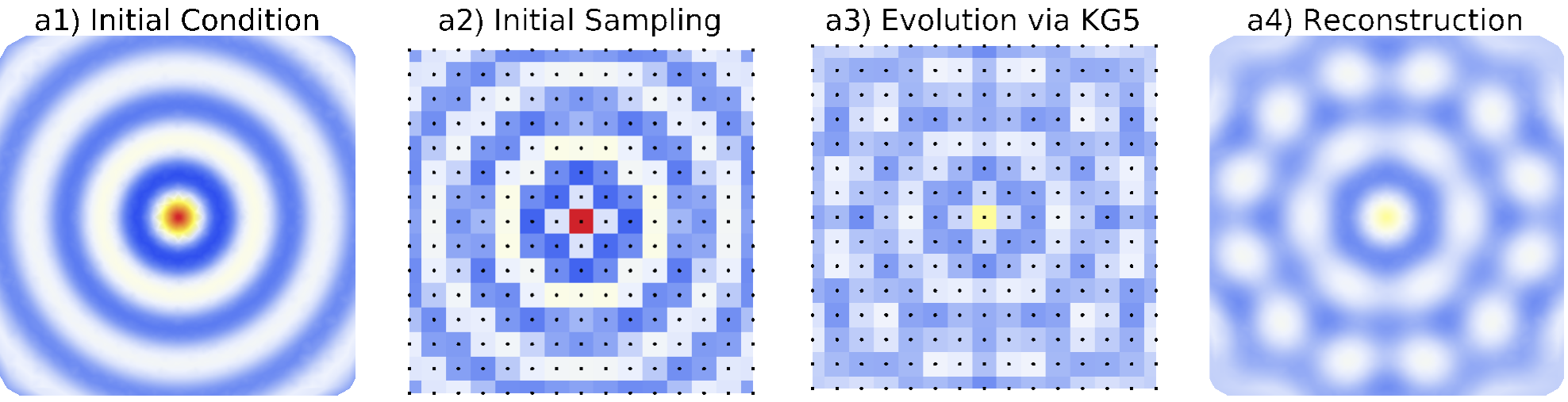}
\centering
\includegraphics[width=0.9\textwidth]{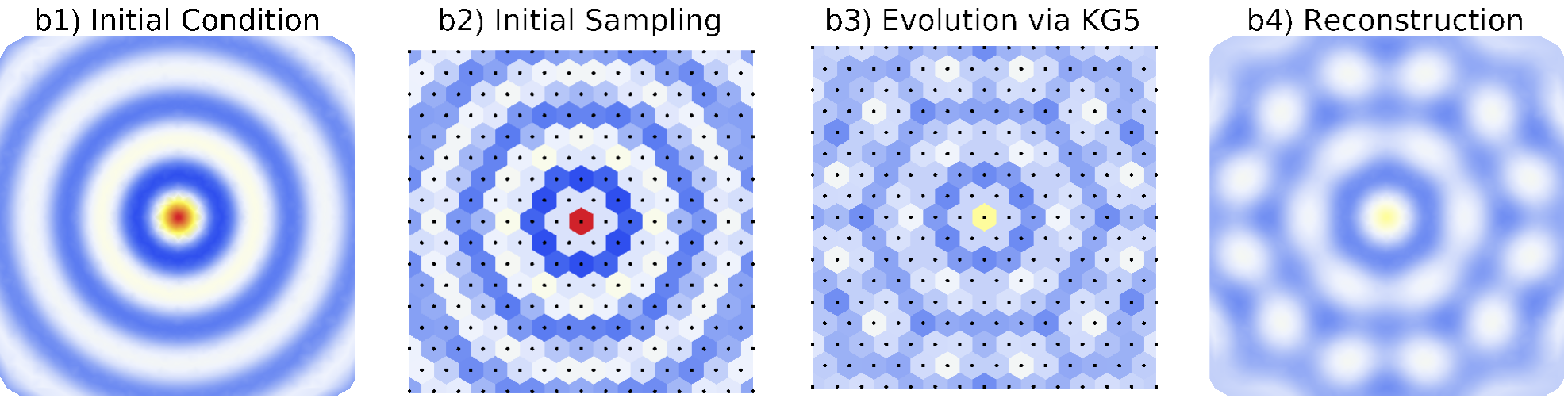}
\centering
\includegraphics[width=0.9\textwidth]{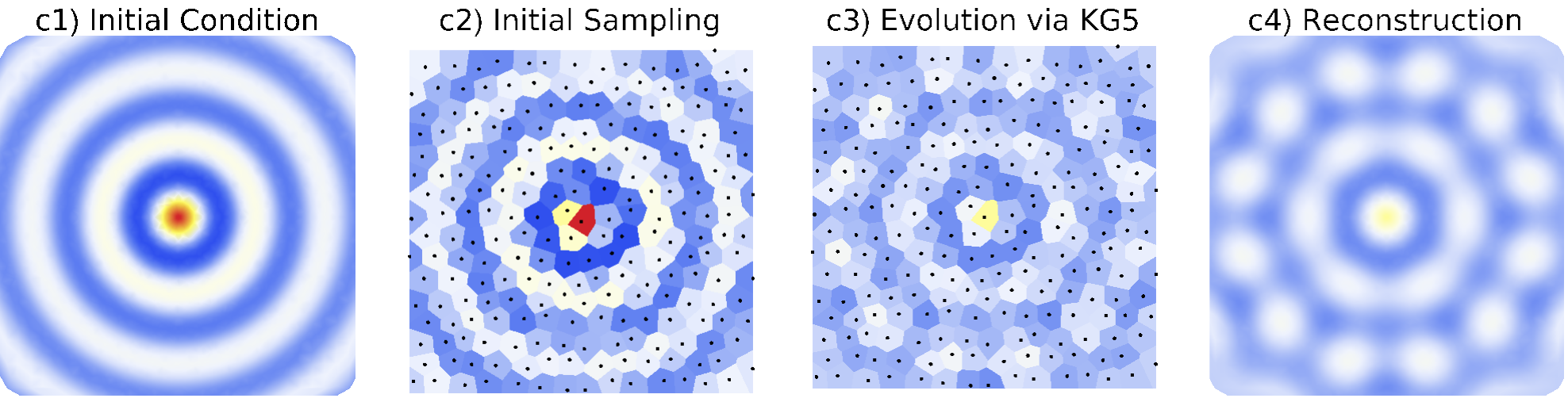}
\caption{The dynamics of KG5 is here shown being carried out in a variety of lattice representations. In the left most column the initial condition is shown in its bandlimited representation, given by Eq.~\eqref{InCond}. In the rightmost column the final evolved state is shown in its bandlimited representation. Here the evolution time is $t=108$ and the field mass is $\mu=0$. This state can be found in four different ways. Firstly by applying the bandlimited dynamics Eq.~\eqref{DKG5bandlimitedSkew} to the bandlimited initial condition. The other three ways are shown in the three rows of this figure. The second row shows the initial condition being sampled onto a hexagonal lattice. This is then evolved forward in time via Eq.~\eqref{KG5Long}. The bandlimited representation of the final state is then recovered through the methods discussed in Sec.~\ref{SecSamplingTheory}. The second and third rows show the same process carried out on a square lattice and an irregular lattice. Notice that the final state has a 6-fold symmetry regardless of how the dynamics is represented. Notice that the final state is the same regardless of how the dynamics is represented. We can represent any bandlimited dynamics on any (sufficiently dense) lattice.}\label{FigEvolutionKG5}
\end{figure*}

\begin{figure*}[t]
\centering
\includegraphics[width=0.9\textwidth]{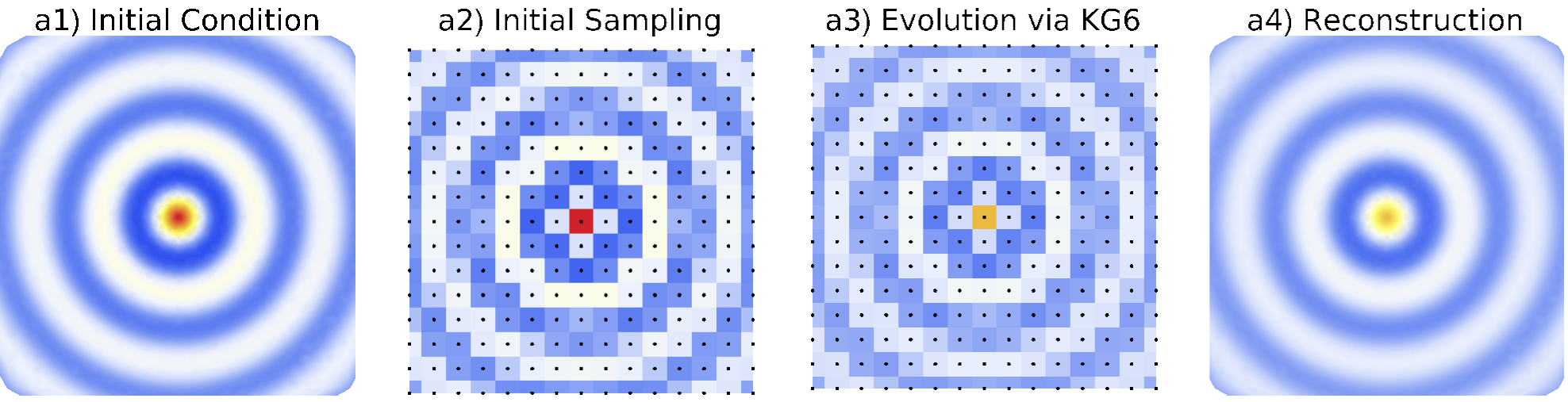}
\centering
\includegraphics[width=0.9\textwidth]{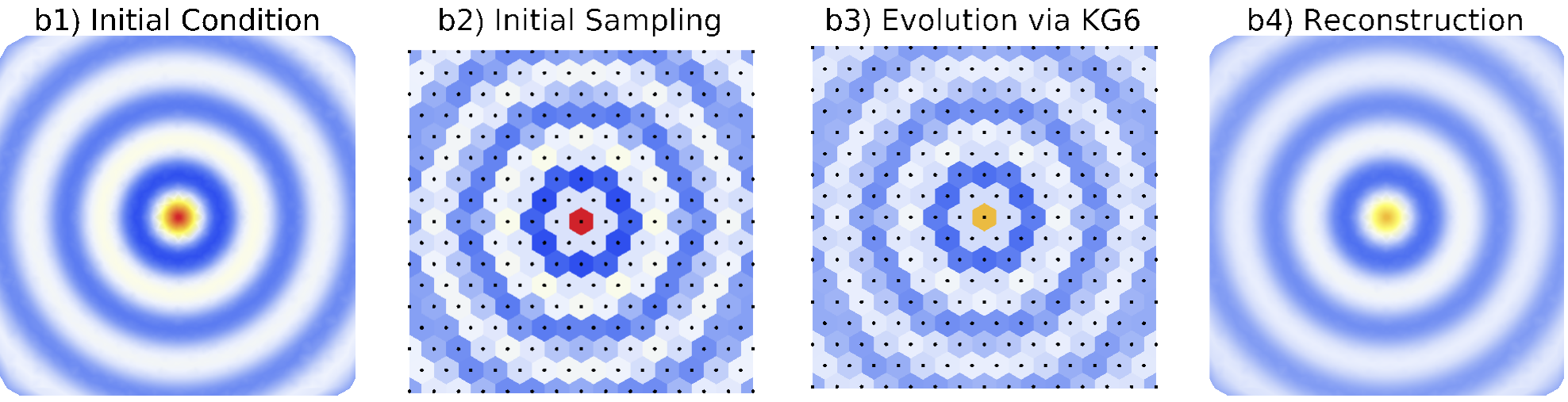}
\centering
\includegraphics[width=0.9\textwidth]{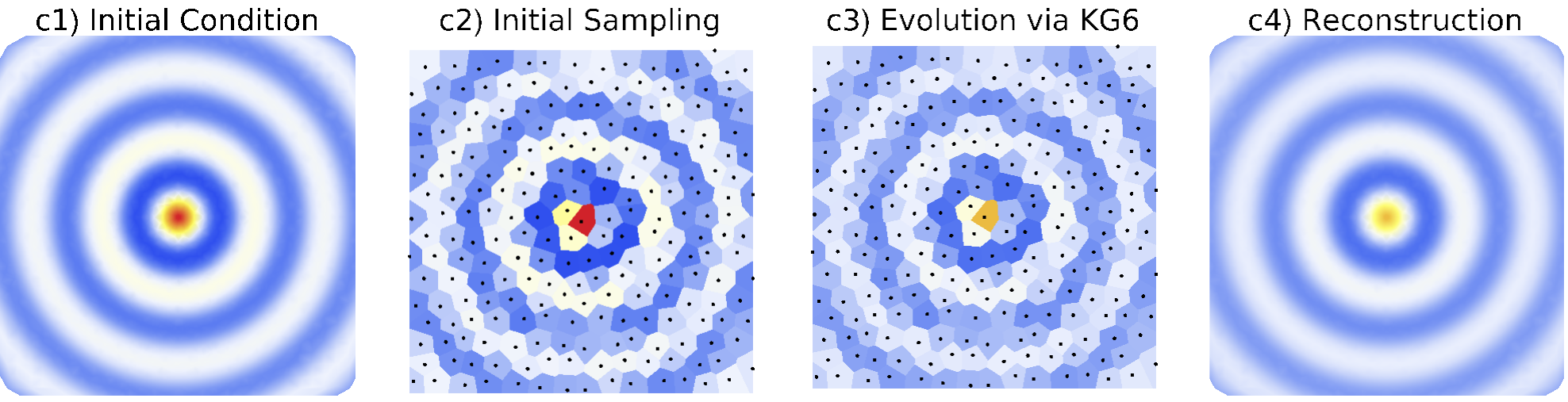}
\caption{The dynamics of KG6 is here shown being carried out in a variety of lattice representations. In the left most column the initial condition is shown in its bandlimited representation, given by Eq.~\eqref{InCond}. In the rightmost column the final evolved state is shown in its bandlimited representation. Here the evolution time is $t=6$ and the field mass is $\mu=0$. This state can be found in four different ways. Firstly by applying the bandlimited dynamics Eq.~\eqref{DKG6bandlimited} to the bandlimited initial condition. The other three ways are shown in the three rows of this figure. The first row shows the initial condition being sampled onto a square lattice. This is then evolved forward in time via Eq.~\eqref{DKG6}. The bandlimited representation of the final state is then recovered through the methods discussed in Sec.~\ref{SecSamplingTheory}. The second and third rows show the same process carried out on a hexagonal lattice and an irregular lattice. Notice that the final state is rotation invariant regardless of how the dynamics is represented. Notice that the final state is the same regardless of how the dynamics is represented. We can represent any bandlimited dynamics on any (sufficiently dense) lattice.}\label{FigEvolutionKG6}
\end{figure*}

In this subsection I will provide some more explicit demonstrations of these theories' symmetries. In particular, I will explicitly demonstrate the fact that the symmetries of some theory's dynamics has nothing to do with the symmetries of the lattice it is represented on. Just as we can represent any bandlimited state on any sufficiently dense lattice, so too can we represent any bandlimited dynamics on any sufficiently dense lattice. To see this consider Figs. \ref{FigEvolutionKG4}, \ref{FigEvolutionKG5} and \ref{FigEvolutionKG6}. 

In each of these figures we begin from some initial state with a bandlimited representation at $t=0$ of,
\begin{align}\label{InCond}
\phi_\text{B}(0,x,y)=\frac{J_1(\pi r_2)}{\pi r_2}
+\frac{J_0(\pi r_2)-J_2(\pi r_2)}{2}
\end{align}
where $J_n(r)$ is the $n^\text{th}$ Bessel function and $r=\sqrt{x^2+y^2}$ and $r_2=r/\sqrt{2}$. This function is shown in the first columns of Figs. \ref{FigEvolutionKG4}, \ref{FigEvolutionKG5} and \ref{FigEvolutionKG6}. Note that this function is rotationally invariant.

This function is bandlimited with bandwidth of $K=\pi/\sqrt{2}$. We can therefore represent this function with sampling on a square 2D lattice with $a=1<\sqrt{2}$. We could equivalently represent this function on a hexagonal 2D lattice or even an irregular 2D lattice. For each of KG4, KG5 and KG6=KG7 Such representations are shown in the second columns of Figs. \ref{FigEvolutionKG4}, \ref{FigEvolutionKG5} and \ref{FigEvolutionKG6}. 

For each of these theories, we then have a choice of which representation to carry out the dynamics in. I here consider four options: as a bandlimited function, as samples on a square lattice, as samples on a hexagonal lattice or as samples on an irregular lattice. The various options for KG4, KG5 and KG6=KG7 are shown in Figs. \ref{FigEvolutionKG4}, \ref{FigEvolutionKG5} and \ref{FigEvolutionKG6} respectively. 

Let's begin with the dynamics of KG5 represented on a hexagonal lattice. This is shown in the middle row of Fig.~\ref{FigEvolutionKG5}. The bandlimited representation of the initial distribution is shown Fig.~\ref{FigEvolutionKG5}b1). The initial sample points on the hexagonal lattice are shown Fig.~\ref{FigEvolutionKG5}b2). These can be evolved forward in time using Eq.~\eqref{KG5Long}. The resulting time-evolved sample points are shown in Fig.~\ref{FigEvolutionKG5}b3). From these we can reconstruct a bandlimited representation for the state using the techniques discussed in Sec.~\ref{SecSamplingTheory}. The resulting reconstruction is shown in Fig.~\ref{FigEvolutionKG5}b4).

Alternatively, we could have carried out this evolution with no lattice representation at all. That is, we could have skipped from Fig.~\ref{FigEvolutionKG5}b1) directly to Fig.~\ref{FigEvolutionKG5}b4). We could do this by applying the dynamics Eq.~\eqref{DKG5bandlimitedSkew} directly to the bandlimited initial condition Eq.~\eqref{InCond}. It is in this sense that the bandlimited and discrete representations of our dynamics are equivalent.

The first and third rows of Fig.~\ref{FigEvolutionKG5} show the exact same evolution via KG5 represented on different lattices, namely a square lattice and an irregular lattice. In the first row the evolution is carried out by a resampled version of Eq.~\eqref{KG5Long}, namely Eq.~\eqref{DKG5Skew}. In the third row the evolution is carried out by whatever resampling of Eq.~\eqref{KG5Long} corresponds to this irregular lattice. 

Notice that the final state has a 6-fold symmetry regardless of how the dynamics is represented. Moreover, notice that the final state is the same regardless of how the dynamics is represented. Just as we can represent any bandlimited state on any lattice, so too can we represent any bandlimited dynamics on any lattice.

Fig.~\ref{FigEvolutionKG4} makes the same demonstration for KG4. Notice that the final state has a 4-fold symmetry regardless of how the dynamics is represented. Notice that the final state is the same regardless of how the dynamics is represented.

Likewise, Fig.~\ref{FigEvolutionKG6} makes the same demonstration for KG6. Notice that the final state is rotation invariant regardless of how the dynamics is represented. Notice that the final state is the same regardless of how the dynamics is represented. 

These figures demonstrate clear as can be that a theory's lattice structure has nothing to do with its dynamical symmetries. We can represent any bandlimited dynamics on any lattice.

\subsection{Bandlimited General Covariance}\label{SecFullGenCov}
As the discussion throughout Sec.~\ref{SecKlein3} has shown, giving our discrete theory a bandlimited representation has had many of the same benefits one expects from a generally covariant formulation. Namely, we have exposed certain parts of our theory as merely representational artifacts and in the process we have come to a better understanding of our theory's symmetries and background structures. This is the work of the titular discrete analog of general covariance. This analogy was originally introduced in \cite{DiscreteGenCovPart1} and has here been expanded to a Lorentzian context. The details of this analogy will be spelled out in detail in Sec.~\ref{SecDisGenCov}.

Before that, however, I will show how to combine this discrete analog with our usual continuum notion of general covariance. Note that on this interpretation KG1-KG7 are ultimately continuum spacetime theories of the sort we are used to interpreting (albeit ones which consider only bandlimited fields). Hence we can apply to these theories the standard techniques of general covariance. (For a review see Appendix A of \cite{DiscreteGenCovPart1}).

As I discussed at the beginning of Sec.~\ref{SecKlein3} on this third interpretation KG1-KG7 are about a bandlimited field \mbox{$\phi_\text{B}:\mathcal{M}\to \mathcal{V}$} with a value space $\mathcal{V}=\mathbb{R}$ and a manifold $\mathcal{M}\cong\mathbb{R}^d$ with $d=2$ for KG1-KG3 and $d=3$ for KG4-KG7. As such, in either case we have access to global coordinate systems for $\mathcal{M}$. The fields $\phi_\text{B}$ considered by this interpretation are bandlimited in the following sense. There exists a fixed special diffeomorphism \mbox{$d_\text{coor.}:\mathbb{R}^d\to\mathcal{M}$} which giving us a fixed special global coordinate system for $\mathcal{M}$. In these special coordinates the field is bandlimited. That is, $\phi_\text{B}\circ d_\text{trans.}\in F^K$. It is in these special coordinates that the fields $\phi_\text{B}$ obey the dynamical equations Eq.~\eqref{DKG1bandlimited}-Eq.~\eqref{DKG7bandlimited}.

Of course, we know from the lessons of continuum general covariance that one ought to be suspicious of any ``special coordinates'' appearing in a supposedly fundamental formulation of a theory. This section will remove any reference to these (or any) coordinates from our bandlimited formulation of KG1-KG7. As is expected of such a generally covariant reformulation, this will reveal these theories' underlying geometric background structures.

For simplicity, however, I will just consider KG4 and KG6 here. To start let us first write the continuum theory KG0 in the coordinate-free language of differential geometry. After substantial work (see \cite{Pooley2015} or Appendix A of \cite{DiscreteGenCovPart1}) one can rewrite Eq.~\eqref{KG0} in the coordinate-free language of differential geometry as follows:
\begin{align}\label{KG0GenCov}
\text{KG0:}\qquad\text{KPMs:}\quad&\langle\mathcal{M},\eta_\text{ab},\varphi\rangle\\
\nonumber
\text{DPMs:}\quad&(\eta^\text{ab}\nabla_\text{a}\nabla_\text{b}-M^2)\,\varphi = 0,   
\end{align}
The geometric objects used in this formulation are as follows. $\mathcal{M}$ is a 2+1 dimensional differentiable manifold. $\eta_\text{ab}$, is a fixed Lorentzian metric field with signature $(-1,1,1)$. Here $\nabla_\text{a}$ is the unique covariant derivative operator compatible with the metric, (i.e. with $\nabla_\text{c}\,\eta_\text{ab}=0$). $\varphi:\mathcal{M}\to\mathbb{R}$ is a dynamical real scalar field. 

Mathematical structures satisfying the above conditions (independent of whether it satisfies the dynamics) are the theory's kinematically possible models (KPMs). This theory's dynamically possible models (DPMs) are the subset of the KPMs which additionally satisfy the theory's dynamics.

Next let's consider KG4. Rewritten in the coordinate-free language of differential geometry, KG4 becomes:
\begin{align}\label{DKG4GenCov}
\text{KG4:}\ \text{KPMs:}\ &\langle \mathcal{M}, \eta_\text{ab},T^\text{a},X^\text{a},Y^\text{a},\phi_\text{B}\rangle\\
\nonumber
\text{DPMs:}\ &
F(T^\text{a}\nabla_\text{a})\phi_\text{B}
= \big[-\mu^2+F(X^\text{b}\nabla_\text{b})\\
\nonumber
&\qquad\qquad\qquad\qquad \ +F(Y^\text{c}\nabla_\text{c})\big]\,\phi_\text{B}.
\end{align}
where 
\begin{align}
F(z)=2\text{cosh}(a\,z)-2.
\end{align}
At the level of dynamics, KG4 and KG0 are radically different. Note that the dynamics of KG0 is local, only involving finite order of derivative. By contrast, the dynamics of KG4 is highly non-local.

These theories are also substantially different at the level of KPMs. The three key differences between KG4 and the continuum theory KG0 are as follows. Firstly KG4 resticts us to manifolds $\mathcal{M}\cong\mathbb{R}^3$ whereas KG0 does not. Secondly, the field $\phi_\text{B}\in F^K$ in KG4 is bandlimited with bandwidth $\omega,k_1,k_2\in[-K,K]$ whereas the field $\psi$ in KG0 is not. I owe the reader a geometric explanation of what this means.

Finally, KG4 has three extra pieces of spacetime structure which KG0 lacks. Namely, $T^\text{a}$, $X^\text{a}$ and $Y^\text{a}$ are three fixed constant unit vectors which are orthogonal to each other. $T^\text{a}$ is timelike whereas $X^\text{a}$ and $Y^\text{a}$ are spacelike. That is, 
\begin{align}
\nabla_\text{a}T^\text{b}&=0 &
\nabla_\text{a}Y^\text{b}&=0 &
\nabla_\text{a}Y^\text{b}&=0\\
\nonumber
\eta_\text{ab} T^\text{a}T^\text{b}&=-1 &
\eta_\text{ab} X^\text{a}X^\text{b}&=1 &
\eta_\text{ab} Y^\text{a}Y^\text{b}&=1\\
\nonumber
\eta_\text{ab} T^\text{a}X^\text{b}&=0 &
\eta_\text{ab} X^\text{a}Y^\text{b}&=0 &
\eta_\text{ab} Y^\text{a}T^\text{b}&=0
\end{align}
Roughly, $X^\text{a}$ and $Y^\text{a}$ here serve to pick out the directions for the rotational anomalies appearing in Fig.~\ref{FigEvolutionKG4}. $T^\text{a}$ picks out a standardized way of moving forward in time (i.e, translation generated by $T^\text{a}\nabla_\text{a}$). That is, $T^\text{a}$ provides a rest frame.

In terms of these spacetime structures, what does it here mean to say that $\phi_\text{B}\in F^K$ is  ``bandlimited with bandwidth $\omega,k_1,k_2\in[-K,K]$''. Consider the following eigen-problem for functions $h:\mathcal{M}\to\mathbb{R}$ on the manifold: 
\begin{align}
T^\text{a}T^\text{b}\nabla_\text{a}\nabla_\text{b} h = -\omega^2 h\\
\nonumber
X^\text{a}X^\text{b}\nabla_\text{a}\nabla_\text{b} h = -k_1^2 h\\
\nonumber
Y^\text{a}Y^\text{b}\nabla_\text{a}\nabla_\text{b} h = -k_2^2 h
\end{align}
A function $f:\mathcal{M}\to\mathbb{R}$ is bandlimited if and only if it is the sum of these eigensolutions over a compact range of Fourier space. The extent of this range is its bandwidth. In these terms we can state the requirement that $\phi_\text{B}$ is bandlimited with bandwidth $\omega,k_1,k_2\in[-K,K]$.

Let's move on to KG6. Rewritten in the coordinate-free language of differential geometry, KG6 becomes:
\begin{align}\label{DKG6GenCov}
\text{KG6:}\qquad\text{KPMs:}\quad&\langle \mathcal{M}, \eta_\text{ab},T^\text{a},X^\text{a},Y^\text{a},\phi_\text{B}\rangle\\
\nonumber
\text{DPMs:}\quad&(\eta^\text{ab}\nabla_\text{a}\nabla_\text{b}-\mu_0^2)\,\phi_\text{B} = 0
\end{align}
with spacetime structures as defined following Eq.~\eqref{KG0GenCov}, $\mathcal{M}\cong\mathbb{R}^3$ and $\phi_\text{B}\in F^K$ in the same sense as KG4. 

At the level of dynamics, KG6 and KG0 are nearly identical. The dynamics for KG0 has a field mass of $M$ whereas KG6 has $\mu_0=\mu/a$. This can be disregarded by setting $\mu_0=M$.

However, at the level of KPMs, KG6 and KG0 are substantially different. The three key differences between KG6 and the continuum theory KG0 are as follows. Firstly KG6 resrticts us to manifolds $\mathcal{M}\cong\mathbb{R}^3$ whereas KG0 does not. Secondly, the field $\phi_\text{B}$ in KG6 is bandlimited with bandwidth $\omega,k_1,k_2\in[-K,K]$ in the sense defined above whereas the field $\psi$ in KG0 is not. Finally, KG6 has the same extra spacetime structures $T^a$, $X^a$ and $Y^a$ that KG4 does. Indeed, at the level of KPMs KG6 and KG4 are identical.

One might notice that these extra pieces of spacetime structure don't play any role in the dynamics of KG6. Why then are they there? $T^a$, $X^a$, and $Y^a$ are needed to spell out what it means for $\bm{\phi}_\text{B}$ to be bandlimited with bandwidth $\omega,k_1,k_2\in[-K,K]$. This region in Fourier space is not Poincar\'e invariant, we need some extra structure to point out in which directions the corners go.

This is a very strange sort of spacetime structure. It doesn't participate in the dynamics, all it does is allows us to articulate a certain restriction on the space of KPMs. As I will discuss in Sec.~\ref{PerfectLorentz}, it is best to think of KG6 as being the part of an extended theory which is visible to us when we restrict our set of representational tools. On this view, while $X^a$ and $Y^a$ are real spacetime structures for KG4, for KG6 they are merely representational artifacts. 

Before that, however, allow me to spell out in detail the discrete analogs of general covariance which have been developed throughout this paper.

\section{Two Discrete Analogs of General Covariance}\label{SecDisGenCov}
Three lessons have been repeated throughout this paper. Each of these lessons is visible in both our second and third attempts at interpreting KG1-KG7. Combined these lessons give us a rich analogy between lattice structures and coordinate systems: Lattice structure is rather less like a fixed background structure and rather more like a coordinate system, i.e., merely a representational artifact.

These three lessons run counter to the three first intuitions one is likely to have regarding lattice structure discussed in Sec.~\ref{SecIntro}. Namely, that lattices and lattice structure: restrict our symmetries, distinguish our theories, and are fundamentally ``baked-into'' the theory. As we have seen, they do not restrict our symmetries, they do not distinguish our theories and they are merely representational not fundamental. In particular, we have learned the following three lessons.

Our first lesson was that taking the lattice structure seriously as a fixed background structure or as the underlying spacetime manifold systematically under predicts the symmetries that discrete spacetime theories can and do have. Indeed, discrete theories can have significantly more symmetries than our first intuitions might allow for. As Sec.~\ref{SecKlein2} and Sec.~\ref{SecKlein3} have shown each of KG1-KG7 has a continuous translation symmetry despite being introduced with explicit lattice structures. Moreover, KG6 and KG7 even have a continuous rotational symmetry and a (limited) Lorentzian boost symmetry. As I will discuss in Sec.~\ref{PerfectLorentz}, these theories can be seen as part of a larger lattice theory which is perfectly Lorentzian. Thus, the fact that a lattice structure was used in the initial statement of these theories' dynamics does not in any way restrict their symmetries. There is no conceptual barrier to having a theory with continuous symmetries represented on a discrete lattice.

In light of the proposed analogy between lattice structure and coordinate systems this first lesson is not mysterious. Coordinate systems are neither background structure nor a fundamental part of the manifold. The use of a certain coordinate system does not in any way restrict a theory's symmetries. Indeed, it is a familiar fact that there is no conceptual barrier to having a rotationally invariant theory formulated on a Cartesian coordinate system.

Our second lesson was that discrete theories which are initially presented to us with very different lattice structures may nonetheless turn out to be completely equivalent theories. Indeed, as we have seen, two of our discrete theories (KG6 and KG7) have a nice one-to-one correspondence between their solutions. This despite the fact that these theories were initially presented to us with different lattice structures (i.e., a square lattice in space and a hexagonal lattice in space respectively). However, despite this nice correspondence, when in Sec.~\ref{SecKlein1} we took these lattice structures seriously as a fixed background structure, we found that KG6 and KG7 were inequivalent; namely, they were here judged to have different symmetries. 

Only in Sec.~\ref{SecKlein2} and Sec.~\ref{SecKlein3} when stopped taking the lattice structure so seriously did we ultimately see KG6 and KG7 as having the same symmetries. Indeed, in these later two interpretations KG6 and KG7 were seen to be completely equivalent. They are simply re-descriptions of a single theory with slightly different limitations in representational capacities. In Sec.~\ref{SecKlein2} this re-description is a change of basis in the theory's value space, whereas in Sec.~\ref{SecKlein3} this re-description is merely a change of coordinates. What we thought were distinct lattice theories are really just different sample points being used to describe one-and-the-same bandlimited field theory.

In light of the proposed analogy between lattice structure and coordinate systems this second lesson is not mysterious. Unsurprisingly, continuum theories presented to us in different coordinate systems may turn out to be equivalent. Moreover, we can always re-describe any continuum theory in any coordinates we wish.

Our third lesson was that, in addition to being able to switch between lattice structures, we can also reformulate any discrete theory in such a way that it has no lattice structure whatsoever. I have shown two ways of doing this. In Sec.~\ref{SecKlein2} this was done by internalizing the lattice structure into the theory's value space. In Sec.~\ref{SecKlein3} this was done by embedding the discrete theory onto a continuous manifold using bandlimited functions. Adopting a lattice structure and switching between them was then handled using Nyquist-Shannon sampling theory discussed in Sec.~\ref{SecSamplingTheory}.

In light of the proposed analogy between lattice structure and coordinate systems this third lesson is not mysterious. This is analogous to the familiar fact that any continuum theory can be written in a generally covariant (i.e., coordinate-free) way. Thus, the two above-discussed ways of reformulating a discrete theory to be lattice-free are each analogous to reformulating a continuum theory to be coordinate-free (i.e., a generally covariant reformulation). Thus we have not one but two discrete analogs of general covariance. See Fig.~\ref{FigTwoAnalogies}.

\begin{figure}[t!]
\begin{flushleft}
\text{{\bf Internal Discrete General Covariance:}}
\end{flushleft}
$\begin{array}{rcl}
\text{Coordinate Systems} & \!\!\leftrightarrow\!\! & \text{Lattice Structure}\\
\text{Changing Coordinates} & \!\!\leftrightarrow\!\! & \text{Changing Lattice Structures}\\
\ & \ & \text{by changing basis in value space}\\
\text{Gen. Cov. Formulation} & \!\!\leftrightarrow\!\! & \text{Basis-Free Formulation}\\
\text{(i.e., coordinate-free)} & \ & \text{(i.e., lattice-free)}\\
\end{array}$\\
\begin{flushleft}
\text{{\bf External Discrete General Covariance:}}
\end{flushleft}
$\begin{array}{rcl}
\text{Coordinate Systems} & \!\!\leftrightarrow\!\! & \text{Lattice Structure}\\
\text{Changing Coordinates} & \!\!\leftrightarrow\!\! & \text{Changing Lattice Structures}\\
\ & \ & \text{by Nyquist-Shannon resampling}\\
\text{Gen. Cov. Formulation} & \!\!\leftrightarrow\!\! & \text{Bandlimited Formulation}\\
\text{(i.e., coordinate-free)} & \ & \text{(i.e., lattice-free)}\\
\end{array}$
\caption{A schematic of the two notions of discrete general covariance introduced \cite{DiscreteGenCovPart1}. The internal strategy is applied to KG1-KG7 in Sec.~\ref{SecKlein2} whereas the external strategy is applied to KG1-KG7 in Sec.~\ref{SecKlein3}. These are compared in Sec.~\ref{SecDisGenCov}. [\textit{Reproduced with permission from} \cite{DiscreteGenCovPart1}.]}
\label{FigTwoAnalogies}
\end{figure}

Before contrasting these two analogies, let's recap what they agree on. In either case, as one would hope, our discrete analog helps us to disentangle a discrete theory's substantive content from its merely representational artifacts. In particular, in both cases, lattice structure is revealed to be non-substantive and merely representational as is the lattice itself. Lattice structure is no more attached or baked-into to our discrete spacetime theories than coordinate systems are to our continuum theory. In either case, getting clear about this has helped us to expose our discrete theory's hidden continuous symmetries.

What distinguishes these two notions of discrete general covariance is how they treat the lattice structure after it has been revealed as being coordinate-like and so merely representational. The approach in Sec.~\ref{SecKlein2} was to internalize the lattice structure into the theory's value space. By contrast, the approach in Sec.~\ref{SecKlein3} was to keep the lattice structure external, but to flesh it out into a continuous manifold such that it is no longer fundamental. Let us therefore call these two notions of discrete general covariance internal and external respectively.

These two approaches pick out very different underlying manifolds for our discretely-representable spacetime theories. As a consequence, they license very different conclusions about locality and symmetries. In particular, while for KG1-KG7 these internal and external approaches have more-or-less agreed as to what symmetries there are, they have disagreed about how they are to be classified.

In each of these differences I find reason to favor the external approach. To briefly overview my feelings: It is more natural for the continuous Poincar\'e symmetries of KG1-KG7 to be classified as external. Moreover, keeping the lattice structure external as a part of the manifold, allows us to draw intuitions about locality from it. In particular, the external approach has a better way  of resolving some puzzles about locality and convergence in the continuum limit. However, neither of these reasons are decisive and I think either the internal or external approach is likely to be fruitful for further investigation/use.

\section{A Perfectly Lorentzian Lattice Theory}\label{PerfectLorentz}
As I will now discuss, we can leverage this analogy between lattice structures and coordinate systems to construct a perfectly Lorentz invariant lattice theory.

As discussed in Sec.~\ref{SecFullGenCov} KG6 is not Lorentz invariant in an unqualified way due to its KPMs not being closed under boosts. Stating the condition that $\phi_\text{B}$ be bandlimited with $\omega,k_1,k_2\in[-K,K]$ requires the use of two space-like fields $X^a$ and $Y^a$ and a time-like field $T^a$ which break Lorentz invariance.

As discussed in Sec.~\ref{SecKlein3}, we can see KG6 as a part of an extended theory. In terms of rotations, we can see KG6 as a part of an larger rotation invariant theory defined over $F^{\sqrt{2}K}_\text{rot.inv.}$ that is with $\sqrt{k_1^2+k_2^2}<\sqrt{2}K$. As I will now discuss, we can do the same thing here with Poincar\'e invariance. Namely, we can find a Poincar\'e invariant extension of KG6.

In particular, we can expand $F^K$ by adding to it every state reachable from $F^K$ by any Poincar\'e transformation and then closing under addition. This defines $F^K_\text{Poin.inv.}$ defined as follows: \mbox{$f(t,x,y)\in F^K_\text{Poin.inv.}$} if and only if 
\begin{enumerate}
    \item[1)] there exists some compact region $\mathcal{K}(f)$ where the Fourier transform of $f$ is supported and 
    \item[2)] this compact region $\mathcal{K}(f)$ itself is contained within the portion of Fourier space satisfying \mbox{$- d\, K^2 \leq w^2 - \vert k\vert^2 \leq K^2$}
\end{enumerate}
with $d$ being the number of spacial dimensions and $\vert k\vert$ being the norm of the wavenumber, i.e., $d=2$ and \mbox{$\vert k\vert=\sqrt{k_1^2+k_2^2}$}. The first condition comes from the fact that Poincar\'e transformations map compact regions in Fourier space to compact regions in Fourier space. The second condition comes from first closing $F^K$ under rotations in space and then Lorentz boosts then finite sums.

To get away from coordinates, let's now define $F^K_\text{Poin.inv.}$ geometrically in terms of the metric $\eta_\text{ab}$. This definition comes in two parts. First let's define the space $F_\text{B}$ which satisfies the first constraint. Consider the eigen-problem for functions $h:\mathcal{M}\to\mathbb{R}$
\begin{align}\label{FBdef}
\eta^\text{ab} \nabla_\text{a}\nabla_\text{b} h= -\lambda h
\end{align}
A function $f:\mathcal{M}\to\mathbb{R}$ is in $F_\text{B}$ if and only if it is the sum of these eigensolutions over a compact range of $\lambda\in\mathbb{R}$ of Fourier space. A function is further within $F^K_\text{Poin.inv.}$ if and only if this compact range of $\lambda$s is contained within \mbox{$- d\, K^2 \leq \lambda \leq K^2$}.

We can thus define following perfectly Lorentz invariant lattice theory:
\begin{align}\label{KGPoininvKGenCov}
\text{KG}_\text{Poin.inv.}^K\qquad\text{KPMs:}\quad&\langle\mathcal{M},\eta_\text{ab},\phi_\text{B}\rangle\\
\nonumber
\text{DPMs:}\quad&(\eta^\text{ab}\nabla_\text{a}\nabla_\text{b}-M^2)\,\phi_\text{B} = 0,
\end{align}
with spacetime structures as defined following Eq.~\eqref{KG0GenCov}, $\mathcal{M}\cong\mathbb{R}^3$ and $\phi_\text{B}\in F^K_\text{Poin.inv.}$ in the sense discussed above. Note that KG6 is a subtheory of $\text{KG}_\text{Poin.inv.}^K$. 

How does $\text{KG}_\text{Poin.inv.}^K$ differ from KG0? Their only difference is that we are here limited to manifolds $\mathcal{M}\cong\mathbb{R}^3$ and that the field $\phi_\text{B}$ in $\text{KG}_\text{Poin.inv.}^K$ is bandlimited in $F_\text{Poin.inv.}^K$, whereas the field $\psi$ in KG0 is not. In fact, for either theory the dynamics guarantees that if the field amplitude starts off bandlimited it will stay bandlimited. Thus this restriction of the allowed dynamical fields is really just a restriction on the allowed initial conditions. Thus, ultimately the only difference between $\text{KG}_\text{Poin.inv.}^K$ and KG0 is a restriction on the initial conditions.

As innocent as this restriction on initial conditions may seem, I believe it has serious implications for the nature of the underlying manifold. As I will discuss later, $\text{KG}_\text{Poin.inv.}^K$ has a nice sampling property and therefore has access to discrete descriptions which make no reference to the manifold. By contrast, the manifold in KG0 seems descriptively essential. Moreover, as discussed in \cite{DiscreteGenCovPart1}, this restriction on initial conditions has serious implications for counterfactual reasoning and locality.

But what makes $\text{KG}_\text{Poin.inv.}^K$ a lattice theory? Note that because every compact region is bounded in a large enough cube, every $f\in F_\text{B}$ is representable by a fine-enough cubic lattice. Doing so one would find its sample values on this lattice obey our original discrete equation for KG6, Eq.~\eqref{DKG6}, with some rescaling of $\mu$. Hence, as each state in this theory is representable on some lattice, this is a perfectly Lorentzian lattice theory. 

This, should be compared with compared with the Lorentz-covariant sampling theory for fields developed in~\cite{Pye2022}. There, only the second condition defining $\text{KG}_\text{Poin.inv.}^K$ is applied to the fields. Let us call this a Lorentz-covariant bandlimitation. Lorentz-covariant UV cutoffs have been considered across the physics literature recently, see~\cite{Kempf2004a,Kempf2004b,Kempf_2010,Kempf2013,Pye2015,Kempf2018,PyeThesis,BEH_2020,Pye2022} among others.

Since \mbox{$- d\, K^2 \leq w^2 - \vert k\vert^2 \leq K^2$} is not a compact region in Fourier space, the resulting fields are not bandlimited in the usual sense; as their support in Fourier space may be non-compact. Rather these functions are Lorentz-bandlimited. Such functions don't have the sampling property discussed in Sec.~\ref{SecSamplingTheory}, but may have a new Lorentz sampling property. Indeed, \cite{Pye2022} develops a new Lorentz-covariant sampling theory. Here none of that is necessary as our fields are each bandlimited in the original sense. Each field can be represented exactly on some lattice in the standard way discussed in Sec.~\ref{SecSamplingTheory}.

One may still complain that this should not be called a lattice theory. For each state here we need a different lattice to represent it, there is no single ontologically significant lattice on which we can represent all of our states. As I will soon discuss, there is in fact such an all-representing lattice (albeit a non-fundamental one), but first let me argue against the spirit of this complaint.

Given the deflationary attitude taken in this paper towards lattice structures, the lack of an all-representing lattice wouldn't bother me. As I have argued throughout this paper, lattice structures are merely representational artifacts playing no substantive role in our theories: they do not limit our symmetries, they do not distinguish our theories, and they are not fundamentally ``baked-into'' our theories. Indeed, as I have argued (here and in \cite{DiscreteGenCovPart1}), there is a rich analogy between the lattice structures appearing in our discrete spacetime theories and the coordinate systems appearing in our continuum spacetime theories. We wouldn't refer to our continuum spacetime theories as ``coordinate theories'', they are rather coordinate-representable theories. Analogously, I claim that properly understood, there are no such thing as lattice-fundamental theories, rather there are only lattice-representable theories. Hence, the above theory is a lattice theory in the strongest sense available.

Accepting this, one might still feel there ought to be an all-representing lattice (albeit a non-fundamental one). As I will discuss, there is one but if there weren't this wouldn't be an issue. Note that we are often forced to represent our continuum theories with multiple overlapping complementary coordinate systems. That is, we often lack an all-representing coordinate system. Given the central analogy of this paper, why should the lack of an all-representing lattice bother us?

Returning to answering the complaint directly however, there is such an all-representing lattice. Consider a cubic lattice with some fixed spacing. Consider another cubic lattice with twice the resolution, offset from the first lattice by some rational amount. Consider an infinite sequence of lattices each doubling in resolution with some rational offset. At some point any function $f\in F_\text{B}$ will be representable on one of these lattices. Consider the \textit{deep lattice} which results from taking the infinite union of all of these lattices (but not their closure under the limit). Every function $f\in F_\text{B}$ is representable on this deep lattice\footnote{Indeed any $f\in F_\text{B}$ can be represented on some finite length initial segment on this.}. Moreover, note that the deep lattice has a countable number of lattice sites, by construction it is a subset of the rationals $\mathbb{Q}^3$.

One final comment on $F_\text{Poin.inv.}^K$. Note that this space is closed under (finite) addition and scalar multiplication and is hence a vector space. As the deep lattice shows, this is a vector space with a countably infinite dimension. Thus, given its translation, rotation and Lorentz boost invariances $F_\text{Poin.inv.}^K$ supports a countably infinite dimensional representation of the Poincar\'e group.

Viewing KG6 as a part of $\text{KG}_\text{Poin.inv.}^{K}$ sheds some light on its failure to be Lotentz invariant in an unqualified way. To see how, it is instructive to restate this in terms of sample lattices. KG6 is a theory about all of the bandlimited functions which are representable on a fixed cubic sample lattice with spacing $a$. By contrast, $\text{KG}_\text{Poin.inv.}^{K}$ is a theory about\footnote{Technically, $\text{KG}_\text{Poin.inv.}^{K}$ is about this collection of bandlimited functions closed under addition.} all of the bandlimited functions representable on a fixed cubic sample lattice with spacing $a$ \textit{or any Poincar\'e transformation thereof.}

Thus, KG6 is the part of $\text{KG}_\text{Poin.inv.}^{K}$ which is visible to us when we restrict ourselves to only a certain set of representational tools. We can thus see KG6's failure to be Lorentz invariant in an unqualified way as a problem of representational capacity and not of physics. Sure, if we limit ourselves to functions which are representable on a fixed cubic lattice, we lose rotation symmetry. But what physical reason do we have to limit our representational tools in this way?

Consider an analogous situation involving coordinate systems. Almost all manifolds cannot be covered with a single global coordinate system. For instance, the 2-sphere needs at least two coordinate systems to cover it. Consider the sphere under arbitrary rotations. Note that no coordinate system is closed under these transformations. If we limit ourselves to functions which are supported entirely within the scope of a single fixed coordinate system, we lose rotation symmetry. Would it be right to say that no theory which is set on a sphere can be rotation invariant? Of course not, this just reflects the fact that each of our coordinate systems individually have a limited representational capacity. We can regain rotation invariance by using multiple representational tools, i.e., multiple coordinate systems. 

Applying this lesson to KG6, we ought to view it as being able to represent a Lorentz non-invariant part of a wider Lorentz invariant theory. As discussed above, making use of multiple sample lattices (i.e., all Poincar\'e transformed cubic lattices with spacing $a$) we can describe a rotation invariant theory $\text{KG}_\text{Poin.inv.}^{K}$ over $F^{K}_\text{Poin.inv.}$.

The equivalence between KG6 and KG7 can also be seen this way. Note that the coordinate transformation which maps KG7 onto KG6, namely Eq.~\eqref{CoorKG7KG6}, maps a cubic 3D lattice to a hexagonal 3D lattice maintaining the lattice spacing $a$, see Fig.~\ref{FigSkew}. Thus, KG7 is a theory about all bandlimited functions representable on a hexagonal 3D sample lattice with spacing $a$. The subsets of KG6 and KG7's KPMs which are equivalent to each other ($F^K_\text{KG6}\cong F^K_\text{KG7}$) are exactly those parts which are representable on both a square and hexagonal lattice. All of KG6 and KG7's DPMs are representable in both ways, but not all of their KPMs. We can here see KG6 and KG7 as parts of a larger unified theory viewed in two ways with differently limited representational capacities.

Note that even the extended theory $\text{KG}_\text{Poin.inv.}^{K}$ has an artificially limited representational capacity. What space do we find if we use all of our representational tools (i.e., every possible sample lattice)? The answer is $F_\text{B}$ defined above.

Contrast $F_\text{B}$ with $F(\mathbb{R}^3)$ the space of all functions $f(t,x,y)$. The Gaussian distribution is in $F(\mathbb{R}^3)$ but not $F_\text{B}$. One might gloss this difference as follows: $F(\mathbb{R}^3)$ contains actually infinite frequencies, whereas $F_\text{B}$ contains arbitrarily large but always finite frequencies. That is, $F_\text{B}$ contains only a potential infinity of frequencies. Note that while $F_\text{B}$ is closed under finite sums it is not closed under infinite sums.

Using $F_\text{B}$ we can thus define following perfectly Lorentz invariant lattice theory:
\begin{align}\label{KGBGenCov}
\text{KG}_\text{B}\text{:}\qquad\text{KPMs:}\quad&\langle\mathcal{M},\eta_\text{ab},\phi_\text{B}\rangle\\
\nonumber
\text{DPMs:}\quad&(\eta^\text{ab}\nabla_\text{a}\nabla_\text{b}-M^2)\,\phi_\text{B} = 0,
\end{align}
with spacetime structures as defined following Eq.~\eqref{KG0GenCov}, $\mathcal{M}\cong\mathbb{R}^3$ and $\phi_\text{B}\in F_\text{B}$ in the sense defined following Eq.~\eqref{FBdef}.

We can even give $\text{KG}_\text{B}$ a diffeomorphism invariant reformulation as:
\begin{align}\label{SR2like}\text{KG}_\text{B}'\text{:}\qquad\text{KPMs:}\quad&\langle\mathcal{M},g_\text{ab},\phi_\text{B}\rangle\\
\nonumber
\text{DPMs:}\quad&(g^\text{ab}\nabla_\text{a}\nabla_\text{b}-M^2)\,\phi_\text{B} = 0\\
\nonumber
&R^\text{a}{}_\text{bcd}=0
\end{align}
Here $\mathcal{M}\cong\mathbb{R}^3$. The fixed Lorentzian metric field, $\eta_\text{ab}$, has been replaced with a dynamical metric field, $g_\text{ab}$. The dynamical metric field varies from model to model. $\nabla_\text{a}$ is still the unique covariant derivative operator compatible with the metric, (i.e. with $\nabla_\text{c}\,g_\text{ab}=0$). The dynamical metric field obey a new dynamical equation, $R^\text{a}{}_\text{bcd}=0$, where $R^\text{a}{}_\text{bcd}$ is the Riemann tensor associated with the unique compatible derivative $\nabla_\text{a}$. The dynamical field $\phi_\text{B}$ is bandlimited in the same sense as above (i.e., $\phi_\text{B}\in F_\text{B}$) with two differences. Firstly, $g_\text{ab}$ replaces $\eta_\text{ab}$ in Eq.~\eqref{FBdef}. Secondly, this condition ought to be viewed as a compatibility constraint acting jointly on the metric $g_\text{ab}$, the covariant derivative $\nabla_\text{a}$, and the field $\phi_\text{B}$. Comparing with \cite{Pooley2015}, $\text{KG}_\text{B}$ is like SR1 and $\text{KG}_\text{B}'$ is like SR2. 

Note that the above discussion is suggestive of a possible procedure by which any spacetime theory can be given a bandlimited formulation and so gain the sampling property discussed in Sec.~\ref{SecSamplingTheory}. 

Perhaps surprisingly, all of the above claims about $\text{KG}_\text{Poin.inv.}^K$ hold for $\text{KG}_\text{B}$ (and relatedly $\text{KG}_\text{B}'$). Its state space $F_\text{B}$ is closed under Poincar\'e transformation. Moreover, it is closed under generic diffeomorphisms. Every function $f\in F_\text{B}$ is representable on some fine-enough cubic lattice. Doing so one would find its sample values on this lattice obey Eq.~\eqref{DKG6} with some rescaling of $\mu$.

Note that this space is closed under (finite) addition and scalar multiplication and is hence a vector space. Moreover, $F_\text{B}$ has a deep lattice just like $F^K_\text{Poin.inv.}$ does. Indeed, it has exactly the same deep lattice. Hence $F_\text{B}$ is a countably infinite dimensional vector space. It supports a countably infinite dimensional representation of the Poincar\'e group.

As I see it, the proper view of KG6 (even beginning from $\phi_\ell$) is the one given by Eq.~\eqref{DKG6GenCov} understood as a part of the maximally extended theory $\text{KG}_\text{B}$ (or $\text{KG}_\text{B}'$ depending on taste). In particular, KG6 is the part of $\text{KG}_\text{B}$ which is visible to us if we restrict ourselves to only certain representational tools.

How does $\text{KG}_\text{B}$ differ from the continuum theory KG0? KG0 assents to the existence of actually infinite frequencies whereas $\text{KG}_\text{B}$ does not. To me $\text{KG}_\text{B}$ appears to be on better ground empirically speaking than KG0: I have never and will never see a literally infinite frequency photon. As small as the difference is between $\text{KG}_\text{B}$ and KG0, I expect there to be radical differences in views on spacetime, but this is a topic for another paper.

\section{Conclusion}\label{SecConclusion}
This paper has given an exhaustive study of the seven discrete Klein Gordon equations presented in Sec.~\ref{SecSevenKG}. These theories were initially formulated in terms of a real valued function over lattice sites, $\phi_\ell\in F_L$. Next, as an infinite dimensional vector, $\bm{\Phi}\in \mathbb{R}^L$. And finally, as a bandlimited field $\phi_\text{B}\in F^K$ over a continuous spacetime manifold $\mathcal{M}$. Each of these redescriptions was carried out by a vector space isomorphism, $F_L\cong\mathbb{R}^L\cong F^K$.

Along the way, we have learned three substantial lessons about the role that lattice structures play in our discrete spacetime theories. These lessons serve to undermine the three first intuitions about lattice structure laid out in Sec.~\ref{SecIntro}. As I have shown, lattice structures don't restrict symmetries, they don't distinguish our theories, and they are not fundamentally ``baked-into'' these theories. As I have discussed, these lessons lay the foundation for a rich analogy between the lattice structures which appear in our discrete spacetime theories and the coordinate systems which appear in our continuum spacetime theories. Indeed, my analysis has shown that lattice structure is rather less like a fixed background structure or a fundamental part of some underlying manifold and rather more like a coordinate system, i.e., merely a representational artifact. This extends the results of \cite{DiscreteGenCovPart1} to a Lorentzian setting.

Based upon this analogy, we have two discrete analogs of general covariance (see Fig.~\ref{FigTwoAnalogies}) each of which are useful for exposing hidden symmetries and background structures in our discrete spacetime theories (i.e., lattice theories). In either case, as hoped, when applied to such theories this discrete analog helps us disentangle the theory's substantive content from its representational artifacts. 

These results are significant as they tell strongly against the intuitions laid out in Sec.~\ref{SecIntro}. One might have an intuition that the world could be fundamentally set on a lattice. This lattice might be square or hexagonal and we might discover which by probing the world at the smallest possible scales looking for violations of rotational symmetry, or other lattice artifacts. Many serious efforts at quantum gravity assume that the world is set on something like a lattice at the smallest scales (although these are often substantially more complicated than the lattices assumed here). However, as this paper clearly demonstrates this just cannot be the case. 

The world cannot be ``fundamentally set on a square lattice'' (or any other lattice) any more than it could be ``fundamentally set in a certain coordinate system''. Like coordinate systems, lattice structures are just not the sort of thing that can be fundamental; they are both thoroughly merely representational. Spacetime cannot be a lattice (even when it might be representable as such). Specifically, I claim that properly understood, there are no such things as lattice-fundamental theories, rather there are only lattice-representable theories.

The primary inspiration for this paper has been the work of mathematical physicist Achim Kempf~\cite{Kempf_1997,UnsharpKempf,Kempf2000b,Kempf2003,Kempf2004a,Kempf2004b,Kempf2006,Martin2008,Kempf_2010,Kempf2013,Pye2015,Kempf2018} among others~\cite{PyeThesis,Pye2022,BEH_2020}. For an overview of Kempf's works on this topic and his closely related thesis that ``Spacetime could be simultaneously continuous and discrete, in the same way that information can be''~\cite{Kempf_2010}, see~\cite{Kempf2018}. My thesis here inserts a ``representable as'' and then analyzes which parts of these representations should be thought of as substantial. Lattice structures turn out to be coordinate-like representational artifacts.

For a discussion of the impact I anticipate these results having on the philosophy of space and time generally, see the conclusion of \cite{DiscreteGenCovPart1}. This includes some discussion of the dynamical versus geometric debate regarding how we should view spacetime structures~\cite{EarmanJohn1989Weas,TwiceOver,BelotGordon2000GaM,BrownPooley1999,Nonentity,HuggettNick2006TRAo,StevensSyman2014Tdat,DoratoMauro2007RTbS,Norton2008,Pooley2013}. As we are now in a Lorentzian context, I will focus my attention on the aspects of this work which relate to quantum gravity and Lattice QFT.

People often have arguments (e.g., from quantum gravity) that spacetime must be a lattice of some sort, are such arguments in disagreement with my conclusion? This would need to be checked on a case-by-case basis, but there is room for compatibility here. Upon closer investigation, such arguments may only show that spacetime must be \textit{representable as} a lattice. In this case there is no disagreement with my conclusion. For instance, bandlimited functions can always be represented as a lattice of sample values; it's just that upon closer philosophical investigations these lattice representations cannot possibly be fundamental, as they are merely coordinate-like representational tools. However, if their arguments carry ontological force as well as representational force, then there is disagreement.

What does my conclusion mean for those who model spacetime as a lattice (e.g., some approaches to quantum gravity)? There is no issue per se with modeling spacetime as a lattice: e.g., as I just mentioned bandlimited physics is discretely representable. My conclusion only speaks to how such models should be interpreted. Indeed, those who describe their physics on a lattice for reasons of approximation (e.g. the Lattice QFT community, people studying crystalline solid state systems) are free to ignore my conclusion. However, they ought not ignore the central analogy or the three lessons which support it. 

For them my lesson is this: ``lattice artifacts'' arise under the following two conditions. They arise when 1) we represent our continuum physics on a lattice \textit{2) and then modify the dynamics to be simple in this representation}. As this paper has shown, the blame is to be placed entirely on the second step. Square-shaped lattice artifacts do not come from using a square lattice, see Fig.~\ref{FigEvolutionKG4}. These artifacts come from using dynamics which are relatively simple (i.e., nearest neighbor interactions) when represented on a square lattice. One can define rotation invariant dynamics on a square lattice, Fig.~\ref{FigEvolutionKG6}, but not with nearest neighbor interactions.

So how should lattice-based approaches to quantum gravity be interpreted then? Namely, what about loop quantum gravity (LQG) and causal set theory~\cite{Surya2019}? I am not an authority on either of these approaches, but I can offer some comments. The conclusion of this paper does not directly apply to LQG. This is because the spacetime considered in LQG consists not only of lattice sites but also links/edges between lattice sites. In order to more directly apply, the results of this paper would need to be extended to more complicated gauge theories first.

Regarding causal set theory~\cite{Surya2019}, its proponents often note that no fixed spacetime lattice is Poincar\'e invariant. This (apparently) spells big trouble for any lattice-based Lorentzian theories. They, however, avoid this issue by considering instead a random Poisson sprinkling of lattice points which does not pick out any preferred direction and hence does not explicitly break Poincar\'e symmetry, at least on average. However, given the deflationary position this paper takes towards lattices, I claim there is no issue to be avoided. Like coordinate systems, lattice structures are just a representational tool for helping us express our theory in a certain way. There is no need for the symmetries of our representational tools to latch onto the symmetries of the thing being represented. Cartesian coordinates are distorted under Lorentz boosts, but we can still use them to describe our Lorentzian theories without issue. Indeed, in Sec.~\ref{PerfectLorentz} I have presented a perfectly Lorentzian lattice theory.

\begin{acknowledgments}
The authors thanks James Read, Achim Kempf, Jason Pye, and Nick Menicucci and the Barrio RQI for their helpful feedback.
\end{acknowledgments}

\bibliographystyle{apsrev4-1}
\bibliography{references}

\appendix

\appendix
\section{Analysis of Symmetries for KG1-KG7}\label{AppA}
This appendix will identify the symmetries for KG1-KG7 under the three interpretations put forward in the main text. However, it is convenient to do this in the reverse order in which these interpretations were introduced in the main text.

\subsection{Symmetries in the Third Interpretation}
Let's begin by analyzing the symmetries of KG1-KG7 in our third attempt at interpreting them, in Sec.~\ref{SecKlein3}.

For KG6 (and relatedly KG7) we can look to their generally covariant formulation Eq.~\eqref{DKG6GenCov} to determine their symmetries. For KG6 we find the same symmetries as we did for the continuum Klein Gordon equation, KG0. Namely, the 2+1 dimensional Poincar\'e group plus reflections, global linear rescalings and certain local affine rescalings. Technically, the space of KPMs for KG6 is not closed under rotations and Lorentz boosts. This has been discussed at length in Secs.\ref{SecKlein3} and \ref{SecFullGenCov}. This is an issue of representational capacity, not of physics.

Similarly, for KG4 we can look to its generally covariant formulation Eq.~\eqref{DKG4GenCov} to identify its symmetries. Its extra spacetime structures $T^a$, $X^a$ and $Y^a$ appearing in its dynamics restrict its rotation symmetries to just quarter turns and forbid Lorentz boosts. Applying the same analysis to KG5 one would find we are restricted to one-sixth turns.

We could also cast KG1-KG3, (namely Eq.~\eqref{DKG1bandlimited}-\eqref{DKG3bandlimited}) into a generally covariant form as well. Doing so we would find their symmetries are the 1+1D Poincar\'e group plus reflections, global linear rescalings and certain local affine rescalings.

None of this is mysterious.

\subsection{Symmetries in the Second Interpretation}
Let us next analyze the symmetries of KG1-KG7 in our second attempt at interpreting them, in Sec.~\ref{SecKlein2}.

Before this, note that our redescription of KG1-KG7 in terms of $\phi_\ell\in F_L$, $\bm{\Phi}\in \mathbb{R}^L$, and $\phi_\text{B}\in F^K$ were each mediated by a vector space isomorphism, \mbox{$F_L\cong\mathbb{R}^L\cong F^K$}. This gives us a nice solution-preserving one-to-one correspondence between our three interpretations.

Thus, given any transformation on one interpretation we can always find the equivalent transformation on the other interpretations. However, as we learned following Eq.~\eqref{GaugeVR} and Eq.~\eqref{BandlimitedSymmetries} this does not mean that these theories have the same class of possible symmetries in each interpretation. While we have a one-to-one correspondence between a generic transformation in one interpretation and another, what counts as a \textit{symmetry} transformation is interpretation dependent. The scope of possible symmetry transformation varies from interpretation to interpretation.

Thus every symmetry revealed by our third interpretation gives us a candidate symmetry for our other two interpretations, but more must be done. In particular, we need to check whether these transformations are of the forms allowed in Eq.~\eqref{Permutation} and Eq.~\eqref{GaugeVR}.

All of the symmetries revealed by our third interpretation are also symmetries on the second interpretation. However, as discussed following Eq.~\eqref{BandlimitedSymmetries}, our second interpretation has a wider scope of possible symmetries than our third interpretation does. In fact, as I will soon discuss it includes local Fourier rescalings among other things.

Let's begin however, with the symmetries shared by our second and third interpretations. As revealed in Sec.~\ref{SecSamplingTheory} translation of a bandlimited function $f_\text{B}(x)\to f_\text{B}(x-\epsilon)$ is represented in terms of its vector of sample values $\bm{f}$ as $\bm{f}\to T^\epsilon_\text{B}\bm{f}$, see Eq.~\eqref{TBdef}. Moreover, as discussed following Eq.~\eqref{DBMatrix}, the operator $T_\text{B}^\epsilon$ in Sec.~\ref{SecSamplingTheory} is numerically identical to the operator $T^\epsilon$ appearing in Sec.~\ref{SecKlein2}. 

Thus for KG1-KG3 our candidate symmetry for continuous translation is $\bm{\Phi}\to T^\epsilon_\text{j}\bm{\Phi}$ are $\bm{\Phi}\to T^\epsilon_\text{n}\bm{\Phi}$. Likewise for KG4-KG7 the candidate symmetries are $\bm{\Phi}\to T^\epsilon_\text{j}\bm{\Phi}$, $\bm{\Phi}\to T^\epsilon_\text{n}\bm{\Phi}$ and $\bm{\Phi}\to T^\epsilon_\text{m}\bm{\Phi}$. These are all of the form Eq.~\eqref{GaugeVR} and are thus viable symmetries under our second interpretation. Indeed, this is a symmetry of KG1-KG7 under our second interpretation.

Similar considerations apply for the reflection symmetries and for the linear and affine rescalings. It is easy to find the symmetry candidates here and check that they are of the form Eq.~\eqref{GaugeVR}. The only other non-trivial symmetry to transfer over is the continuous rotational symmetry and the Lorentz symmetry. To see this we need the following facts. 

For functions $h(t,x,y)$ rotations are generated through the derivative as 
\mbox{$h(R(t,x,y))= \exp(\theta (x \partial_y - y \partial_x))h(x,y)$}. Suppose that $h=h_\text{B}$ is bandlimited and we sample it in two ways. Once on some cubic lattice, and once on another lattice identical to the first but rotated in space around $(j,n,m)=(0,0)$ by an angle $\theta$. Sometimes (but not always) the state will be representable on both of these lattices. If so then its sample values in these two case, $\bm{h},\bm{h}'\in\mathbb{R}^\mathbb{Z}\otimes\mathbb{R}^\mathbb{Z}\otimes\mathbb{R}^\mathbb{Z}$ are related as $\bm{h}'=R_\text{B}^\theta \bm{h}$ where
\begin{align}
R_\text{B}^\theta \coloneqq \exp(-\theta (N_\text{n} D_\text{B,m}-N_\text{m} D_\text{B,n}))
\end{align}
with $\theta\in\mathbb{R}$ and where $N_\text{n}$ and $N_\text{m}$ are position operators which return the second and third index. Finally, since $D_\text{B}$ is numerically identical to $D$ we have $R_\text{B}^\theta=R^\theta$ as defined in Eq.~\eqref{RthetaDef}. 

This transformation is of the form Eq.~\eqref{GaugeVR} and is thus a viable symmetry under our second interpretation. Indeed, this is a symmetry of KG6 under our second interpretation (at least restricted to $\mathbb{R}^L_\text{rot.inv.}$). Moreover, with a slight change of basis (namely, Eq.~\eqref{SkewKG6KG7}) it is for KG7 as well (at least restricted to $\mathbb{R}^L_\text{rot.inv.}$ post-transformation).

The same holds for Lorentzian boosts. Allow me to discuss only KG3 here with KG6 and KG7 being treated analogously. For functions $h(t,x)$ Lorentz boosts are generated through the derivative as 
\mbox{$h(\Lambda(t,x))= \exp(-w (t \partial_x + x \partial_t))h(t,x)$}. Suppose that $h=h_\text{B}$ is bandlimited and we sample it in two ways. Once on some square lattice, and once on another lattice identical to the first but boosted around $(n,m)=(0,0)$ by a boost parameter $w$. Sometimes (but not always) the state will be representable on both of these lattices, see Fig.\ref{FigKleinSpaceTime}. If so then its sample values in these two case, $\bm{h},\bm{h}'\in\mathbb{R}^\mathbb{Z}\otimes\mathbb{R}^\mathbb{Z}$ are related as $\bm{h}'=\Lambda_\text{B}^w \bm{h}$ where
\begin{align}
\Lambda_\text{B}^w \coloneqq \exp(-w (N_\text{j} D_\text{B,n}+N_\text{n} D_\text{B,j}))
\end{align}
Finally, since $D_\text{B}$ is numerically identical to $D$ we have $\Lambda_\text{B}^\theta=\Lambda^\theta$ as defined in Eq.~\eqref{LambdaWDef}. 

This transformation is of the form Eq.~\eqref{GaugeVR} and is thus a viable symmetry under our second interpretation. Indeed, this is a symmetry of KG3 under our second interpretation (at least restricted to a finite-sized region around $\omega,k_1,k_2=0$ and $w=0$). Similarly, for KG6 and KG7.

But what about the local Fourier rescaling symmetry present in our second interpretation. It's straightforward to check that this is a symmetry, but how do we know there aren't other symmetries of the form Eq.~\eqref{GaugeVR} missed by our third interpretation?

Let's begin the search. Firstly note our possibility for affine transformations $\bm{c}$ has already been accounted for above. Thus, the only place extra symmetries could be realized is as $\bm{\Phi}\mapsto\Lambda\,\bm{\Phi}$ where $\Lambda$ is some invertible linear map. This will be a symmetry if and only if $\Lambda$ commutes with the relevant operator (call it $\Delta_\text{KG1-KG7}^2$) appearing in the theory's dynamical equation: namely, Eq.~\eqref{DKG1}, Eq.~\eqref{DKG2}, Eq.~\eqref{DKG3} and Eqs.~\eqref{DKG4}-\eqref{DKG7}. In each case, $\Delta_\text{KG1-KG7}^2$ is diagonal in the relevant discrete Fourier basis: $\bm{\Phi}(\omega,k)$ for KG1-KG3 and $\bm{\Phi}(\omega,k_1,k_2)$ for KG4-KG7. 

There are roughly two ways for $\Lambda$ to commute with $\Delta_\text{KG1-KG7}^2$ it could either be diagonal in the discrete Fourier basis itself, or it could operate within a degenerate subspaces of $\Delta_\text{KG1-KG7}^2$. The transformations $\Lambda$ which are diagonal in the discrete Fourier basis are exactly the local Fourier rescalings which we have already discussed.

What then are the degenerate subspaces of $\Delta_\text{KG1-KG7}^2$? The spectrum of these operators are given by the dispersion relations with $\lambda\in\mathbb{R}$ replacing $\mu^2>0$. For KG1-KG3 only one of these degenerate subspaces contains solutions, the one with $\lambda=\mu^2$. These are plotted in Fig.~\ref{FigKleinDisp}.  For relatively small $\omega$ and $k$ these degenerate subspaces trace out approximate hyperbolas in Fourier space. For KG3 this degenerate subspace remains exactly hyperbolic as we increase our wavenumbers, but for KG4 and KG5 these become increasingly distorted. In every case, for sufficiently high wavenumbers the degenerate subspaces begin to touch the sides of the region $\omega,k\in[-\pi,\pi]$. Recall that we have a $2\pi$ periodic identification of planewaves on our second interpretation. 

Thus, these solution-containing degenerate subspaces trace out a compact figure in Fourier space. Collecting all of these degenerate Fourier modes together into a vector space we have something isomorphic to $\mathbb{R}^\mathbb{Z}$. The dynamical symmetries within each of these degenerate subspaces is $\text{GL}(\mathbb{R}^\mathbb{Z})$, i.e., tremendously large. For instance, permuting these degenerate planewaves in a discontinuous way is a symmetry here.

If this feels excessive, one can add in some structure to Fourier space to allow such discontinuous permutations. Indeed, throughout this paper I have talked about compact regions in Fourier space. We must therefore have some native topology on Fourier space. Moreover, let's give Fourier space a differentiable structure. We can then restrict our attention to symmetries generated by diffeomorphisms in Fourier space. As I mentioned above, the degenerate Fourier spaces for KG1-KG3 trace out closed curves in Fourier space. Our symmetries are then smooth maps which flow along these degeneracy curves. This may still seem excessive. On this view, KG1 and KG2 have something like Lorentz boost symmetries (smoothly moving along their non-hyperbolic degenerate spaces). Although it should be noted that these pseudo-boosts to do not fit together with these theories' translation symmetries to form a Poincar\'e group. 

One might further restrict the symmetries in this second interpretation by giving its Fourier space a Minkowski metric structure. The only symmetries then are generated by metric preserving transformations of Fourier space (i.e., Lorentz boosts). This forbids KG1 and KG2's pseudo-rotations but allows for KG3's authentic Lorentz boosts.

I leave an analogous analysis of KG4-KG7 degenerate subspaces to the reader.

\subsection{Symmetries in the First Interpretation}
Let us now check whether the above discussed symmetries are still symmetries in our first interpretation. To do this we just need to check which are understandable in terms of permutations of the lattice site and time reparametrizations. That is, when are the above-discussed linear transformations $\Lambda$ permutations?

It is not hard to check when the translations, rotations, and Lorentz boosts reduce to permutations of lattice sites for KG1-KG7. Ultimately, this shows the symmetries under the first interpretation are just those claimed in Sec.~\ref{SecKlein1}.

\end{document}